\documentclass[10pt]{article}
\usepackage{amssymb}
\usepackage{axodraw}
\usepackage{rotate}
\usepackage{epsfig}
\usepackage{psfig}
\usepackage{graphicx}
\usepackage{wrapfig}
\usepackage{floatflt}
\begin{document}
\newcommand{\eqnzero}{\setcounter{equation}{0}} 

\newcommand{\bq}{\begin{equation}}
\newcommand{\eq}{\end{equation}}
\newcommand{\bqa}{\begin{eqnarray}}
\newcommand{\eqa}{\end{eqnarray}}
\newcommand{\baa}[1]{\begin{array}{#1}}
\newcommand{\eaa}{\end{array}}
\newcommand{\nll}{\nonumber\\}

\newcommand{\Litwo}{\mbox{${\rm{Li}}_{2}$}}
\newcommand{\alem}{\alpha_{em}}
\newcommand{\alsS}{\alpha^2_{_S}}
\newcommand{\ds }{\displaystyle}
\newcommand{\sss}[1]{\scriptscriptstyle{#1}}
\newcommand{\eps}{\varepsilon^*}
\newcommand{\dprop}{\overline\Delta}
\newcommand{\dpropi}[1]{d_{#1}}
\newcommand{\sla}[1]{/\!\!\!#1}
\def\mgn{mgn}
\def\mw {M_{\sss{W}}}
\def\mws{M_{\sss{W}^2}}
\def\mz {M_{\sss{Z}}}
\def\mh {M_{\sss{H}}}
\def\men{m_{\nu_e}}
\def\mel{m_e}
\def\mup{m_u}
\def\mdn{m_d}
\def\mmn{m_{\nu}}
\def\mmo{m_{\mu}}
\def\mch{mch}
\def\mst{mst}
\def\mtn{mtn}
\def\mta{mta}
\def\mtp{m_t}
\def\mbt{m_b}
\def\mp{mp}
\def\mf{m_f}
\def\mv{M_{\sss{V}}}
\def\srt{\sqrt{2}}
\newcommand{\vpa}[2]{\sigma_{#1}^{#2}}
\newcommand{\vma}[2]{\delta_{#1}^{#2}}
\newcommand{\af}{I^3_f}
\newcommand{\sqs}{\sqrt{s}}

\newcommand{\stw}{s_{\sss{W}}  }
\newcommand{\ctw}{c_{\sss{W}}  }
\newcommand{\stws}{s^2_{\sss{W}}}
\newcommand{\stwf}{s^4_{\sss{W}}}
\newcommand{\ctws}{c^2_{\sss{W}}}
\newcommand{\ctwf}{c^4_{\sss{W}}}

\newcommand{\siw }{s_{\sss{W}}}           
\newcommand{\cow }{c_{\sss{W}}}
\newcommand{\siws}{s^2_{\sss{W}}}
\newcommand{\cows}{c^2_{\sss{W}}}
\newcommand{\siwc}{s^3_{\sss{W}}}
\newcommand{\cowc}{c^3_{\sss{W}}}
\newcommand{\cowsc}{c^6_{\sss{W}}}
\newcommand{\siwf}{s^4_{\sss{W}}}
\newcommand{\cowf}{c^4_{\sss{W}}}

\newcommand{\bff}[4]{B_{#1}\big( #2;#3,#4\big)}             
\newcommand{\fbff}[4]{B^{F}_{#1}\big(#2;#3,#4\big)}        
\newcommand{\scff}[1]{C_{#1}}             
\newcommand{\sdff}[1]{D_{#1}}                 
\newcommand{\dffp}[6]{D_{0} \big( #1,#2,#3,#4,#5,#6;}       
\newcommand{\dffm}[4]{#1,#2,#3,#4 \big) }       
\newcommand{\tHmus}{\mu^2}
\newcommand{\epsh}{\hat\varepsilon}
\newcommand{\epsb}{\bar\varepsilon}


\newcommand{\chapt}[1]{Chapter~\ref{#1}}
\newcommand{\chaptsc}[2]{Chapter~\ref{#1} and \ref{#2}}
\newcommand{\eqn}[1]{Eq.~(\ref{#1})}
\newcommand{\eqns}[2]{Eqs.~(\ref{#1})--(\ref{#2})}
\newcommand{\eqnss}[1]{Eqs.~(\ref{#1})}
\newcommand{\eqnsc}[2]{Eqs.~(\ref{#1}) and (\ref{#2})}
\newcommand{\eqnst}[3]{Eqs.~(\ref{#1}), (\ref{#2}) and (\ref{#3})}
\newcommand{\eqnsf}[4]{Eqs.~(\ref{#1}), 
          (\ref{#2}), (\ref{#3}) and (\ref{#4})}
\newcommand{\eqnsv}[5]{Eqs.(\ref{#1}), 
          (\ref{#2}), (\ref{#3}), (\ref{#4}) and (\ref{#5})}
\newcommand{\tbn}[1]{Table~\ref{#1}}
\newcommand{\tabn}[1]{Tab.~\ref{#1}}
\newcommand{\tbns}[2]{Tabs.~\ref{#1}--\ref{#2}}
\newcommand{\tabns}[2]{Tabs.~\ref{#1}--\ref{#2}}
\newcommand{\tbnsc}[2]{Tabs.~\ref{#1} and \ref{#2}}
\newcommand{\fig}[1]{Fig.~\ref{#1}}
\newcommand{\figs}[2]{Figs.~\ref{#1}--\ref{#2}}
\newcommand{\figsc}[2]{Figs.~\ref{#1} and \ref{#2}}
\newcommand{\sect}[1]{Section~\ref{#1}}
\newcommand{\sects}[2]{Sections~\ref{#1} and \ref{#2}}
\newcommand{\subsect}[1]{Subsection~\ref{#1}}
\newcommand{\appendx}[1]{Appendix~\ref{#1}}

\def\itf{I^{(3)}_f}
\def\thmn{\vartheta_{u\nu}}
\def\thmo{\vartheta_{d l}}
\def\thle{\vartheta_{l}}
\def\mml{m_l}
\def\mmf{m_f}
\newcommand{\ip }[1]{u\left({#1}        \right)}  
\newcommand{\iap}[1]{{\bar{v}}\left({#1}\right)}  
\newcommand{\op }[1]{{\bar{u}}\left({#1}\right)}  
\newcommand{\oap}[1]{v\left({#1}\right)}          
\def\betaf{\beta_f}
\def\betap{\beta_{+}}
\def\betam{\beta_{-}}
\setcounter{page}{0}
\thispagestyle{empty}

$\,$
\vspace*{-1cm}

\begin{flushright}
{{\tt hep-ph/0411186} \\
      November 2004
}
\end{flushright}
\vspace*{\fill}
\begin{center}

{\LARGE\bf SANCscope -- v.1.00}
\vspace*{1.5cm}

{\bf A.~Andonov, A.~Arbuzov$^*$, D.~Bardin, S.~Bondarenko$^{*}$, \\[1mm]
P.~Christova, L.~Kalinovskaya, G.~Nanava$^{**}$,
and W.~von~Schlippe$^{***}$}
\vspace*{13mm}

{\normalsize
{\it Dzhelepov Laboratory for Nuclear Problems, JINR,      \\
$^{*}$ Bogoliubov Laboratory of  Theoretical Physics, JINR,\\ 
       ul. Joliot-Curie 6, RU-141980 Dubna, Russia,        \\
$^{**}$ on leave from IHEP, TSU, Tbilisi, Georgia,         \\
$^{***}$ Petersburg Nuclear Physics Institute,             \\ 
        Gatchina, RU-188300 St. Petersburg, Russia   }}
\vspace*{13mm}

\end{center}

\begin{abstract}
\noindent
In this article we have summarized the status of the system {\tt SANC} version {\tt 1.00}.
We have implemented theoretical predictions for many high energy interactions of
fundamental particles at the one-loop precision level for up to 4-particle processes.
In the present part of our {\tt SANC} description we place emphasis on  
an extensive discussion of an important first step of calculations
of the one-loop amplitudes of 3- and 4-particle processes in QED, QCD and EW theories.

{\tt SANC} version {\tt v1.00} is accessible from  servers at Dubna 
{\it http://sanc.jinr.ru/ (159.93.74.10)} and CERN 
{\it http://pcphsanc.cern.ch/ (137.138.180.42)}.
\end{abstract}

\vspace*{5mm}

\centerline{\it (To be published in Computer Physics Communications)}

\vfill

\vspace*{1mm}
\bigskip
\footnoterule
\noindent
{\footnotesize \noindent
Work supported in part by INTAS grant $N^{o}$ 03-51-4007.
\\
E-mail: sanc@jinr.ru}
\clearpage
\tableofcontents
\clearpage
\listoffigures
\listoftables    
\clearpage

\begin{center}
{\bf PROGRAM SUMMARY}
 \begin{itemize}

 \item {\it Title of program}: {\tt SANC}
 \item {\it Catalogue identifier}:
 \item {\it Program obtainable from}: Internet sites at DLNP, JINR, Dubna, Russia, \\
       {\it http://sanc.jinr.ru/ (159.93.74.10)} and at CERN, 
       {\it http://pcphsanc.cern.ch (137.138.180.42)}
 \item {\it Computers for which the program is designed and others on which it has been 
            tested:\\
            Designed for:} platforms on which Java and FORM3 are available\\
       {\it Tested on:} Intel-based PC's
 \item {\it Operating systems}: Linux, Windows
 \item {\it Programming languages used}: Java, FORM3, PERL, FORTRAN
 \item {\it Memory required to execute with typical data}: 10 Mb
 \item {\it No. of bits in a word}: 32
 \item {\it No. of processors used}: 1 on {\tt SANC} server, 1 on {\tt SANC} client
 \item {\it Distribution format}: gzipped tar archive
 \item {\it Keywords}: Feynman diagrams, Perturbation theory, Quantum field theory,
                       Standard Model, Electroweak interactions, QCD, QED,
                       One-loop calculations, Monte Carlo generators.
 \item {\it Nature of physical problem}: Automatic calculation of pseudo-- and realistic
            observables for various processes and decays in the Standard Model of Electroweak
            interactions, QCD and QED at one-loop precision level. Form factors and 
            helicity amplitudes free of UV divergences are produced. For exclusion of
            IR singularities the soft photon emission is included.
 \item {\it Method of Solution}: Numerical computation of analytical formulae 
                                 of form factors and helicity amplitudes.  
                                 For simulation of two fermion 
                                 radiative decays of Standard Model bosons $(W^{\pm},Z)$
                                 and the Higgs boson a Monte Carlo technique is used. 
 \item {\it Restrictions on the complexity}: In the current version of {\tt SANC}
                                             there are 3 and 4 particle processes 
                                             and decays available at one-loop precision level.
                                               
 \item {\it Typical Running time}: The running time depends on the selected process.
                                   For instance, the symbolic calculation of 
                                   form factors (with precomputed building blocks)
                                   of Bhabha scattering 
                                   in the Standard Model takes about 15 sec,
                                   helicity amplitudes --- about 30 sec,
                                   and bremsstrahlung --- 10 sec. The numerical 
                                   computation of cross-section for this process 
                                   takes about 5 sec (CPU 3GHz IP4, RAM 512Mb, L2 1024 KB).
 \end{itemize}
\end{center}

\clearpage

\section{Introduction}

\hspace*{4mm}
\underline{\bf Project motivation}
\vspace{1mm}

The main goal is the creation of a computer system for semi-automatic calculations of realistic 
and pseudo-observables for various processes of elementary particle interactions 
``from the SM Lagrangian to event distributions''
at the one-loop precision level for the present and future colliders --
TEVATRON (Runs II and III), LHC, electron Linear Colliders (ISCLC, CLIC), muon factories
and others.

Furthermore, the {\tt SANC} system, even at the level which is reached already,
may be used for educational purposes by students specializing in high energy physics.
With its help, it is easy to follow all steps of calculations at the one-loop precision level
for $W,Z,H\to f\bar{f}(\gamma),\;H\to\gamma\gamma,Z\gamma,ZZ,WW,\;t\to bW$ decays, and 
many other processes.
Moreover, all the calculations are realized in the spirit of the book \cite{Bardin:1999ak}
which makes the {\tt SANC} system particularly appealing for pedagogical purposes.

\underline{\bf Historical overview}
\vspace*{1mm}

The {\tt SANC} project has been started in early 2001.
During the first phase of the project (2001--2003), the {\tt SANC} group demonstrated the 
workability of the computer system which is being developed \cite{Bardin:2002gs}.
The {\tt version 0.01}, from 03/28/2001, was already able to compute one-loop Feynman 
diagrams for all SM $1\to 2$ decays and $2f\to 2f$ processes (in $R_{\xi}$ and unitary gauges,
including QCD, accessing thereby all one-loop diagrams needed for the processes considered 
by the Dubna group in the past 
\cite{Bardin:1999yd}--\cite{Bardin:1997nc}
in connection with theoretical support of experiments at CERN and DESY.
The {\tt FORM} codes (at present {\tt FORM3} \cite{Vermaseren:2000nd} is being used)
computing the one-loop ultraviolet finite scalar form factors of  
the amplitudes of the decays $Z(H,W)\to f\bar{f}$ were unified and put into 
a special program environment, written in {\tt JAVA}. 
This version was used for a revision of Atomic Parity Violation \cite{Bardin:2001ii},
 and for a calculation of the one-loop electroweak radiative corrections 
for the process $e^{+}e^{-}\to f\bar{f}$ \cite{Andonov:2002xc} and neutrino 
DIS~\cite{Arbuzov:2004zr}.

In the second phase of the project (2004--2006), we extend automatic calculations of 
such a kind to a large number of HEP processes, with emphasis  on LHC physics.

\underline{\bf Present status}
\vspace*{1mm}

The present level of the system is realized in the version {\tt v1.00}. 
This  version has a fresh new layout and is more user friendly than 
earlier versions.

New in version {\tt 1.00} are Compton scattering and several other $ffbb$ processes.
By our philosophy we treat them  as building blocks for future calculations 
of $5\to 0$ processes (fully massive case).

For the last year we substantially enhanced our computer system compared to 
the status presented in the years 2002--2003 at large-scale international conferences, 
such as ACAT2002 
\cite{Andonov:2003xe}--\cite{Nanava:2003xh},
ICHEP2002 \cite{Andonov:2002jg}, RADCOR2002 \cite{Bardin:2003zd} and Workshops at Saint-Malo 
\cite{BardinK:SM}, CERN \cite{BardinK:MK}, Montpelier \cite{BardinKA:MP} 
and Paris~\cite{Arbuzov:2004xxx}.

\underline{\bf How to get started and use {\tt SANC}}
\vspace*{1mm}

To learn more about available {\tt SANC} servers look at our home pages at Dubna
{\it http://sanc.jinr.ru} or CERN {\it http://pcphsanc.cern.ch}.

{\tt SANC} may be accessed via the so called {\tt SANC} client --- a software free to download.
The user will always get the latest (updates) versions from either of the two above addresses.

\underline{\bf Levels of the calculations}
\vspace*{1mm}

{\tt SANC} is subdivided into  three logical levels, each with a specific purpose. 

\vspace*{-2mm}
\begin{itemize} 
\item{Level 1, Analytic}
\end{itemize} 
\vspace*{-2mm}

The analytical application includes enhanced tools of {\tt FORM} procedures.
{\tt SANC} has three types of procedures: {\it specific, intrinsic} and {\it special}. 
The specific procedures are used by {\tt FORM} source codes, typically only once; they
are always visible (see Section~\ref{precomputation}) and can be modified by the user.
The action of intrinsic procedures is uniquely specified by their arguments, therefore
they may be used by many {\tt FORM} codes. 
Their bodies are not accessible to the users. 
Finally, special procedures are used only a few number of times to perform some special action
in a given {\tt FORM} code. Normally, they have very simple arguments like field indices. 

The Figures~\ref{PrecQED}--\ref{PrecEW} show fully open menus for ``Precomputation''
and for available ``Processes'' in the QED part, Fig~\ref{ProcQED},
and ``Processes'' in the EW part, Fig.~\ref{ProcEW}.
In this article we explain in detail the process of ``Precomputation'' .

Entering your chosen process, you are in an active session and receive the
analytical result for scalar Form Factors (FF), Helicity Amplitudes (HA) and the accompanying 
Bremsstrahlung contributions (BR). 

For the calculation of the HAs we use techniques of Vega--Wudka \cite{Vega:1995cc}
and Kleiss--Stirling \cite{Kleiss:1985yh}.
 
As a main example of the description of the calculation of FF and HA for processes
$f_1\bar{f}_1\to f\bar{f}$ we mention Ref.~\cite{Andonov:2002xc}.
All calculations of FF$\to $HA$\to $BR on the {\tt SANC} tree are realized in the same job stream.

\vspace*{-2mm}
\begin{itemize} 
\item{Level 2, Numerical}
\end{itemize} 
\vspace*{-2mm}

The analytic results are transferred to the second level where they are 
analyzed by a software package {\tt s2n.f} ({\it symbols-to-numbers}), written in {\tt PERL}.
The {\tt s2n.f} package automatically generates {\tt FORTRAN} codes for subsequent numerical 
computations of decay rates and process cross sections.
The calculational flow inside levels 1 and 2 and the exchange of data between them is fully 
automatized and is governed by selecting corresponding items in menus.

The {\tt s2n} part of the {\tt SANC} system is completely fixed for all available decays (besides
($Z\to W^{+}W^{-}$); for NC $4f\to 0$ processes we have {\tt s2n} for FF$\to $HA$\to $SoftBR parts,
and for CC processes $f_1\bar{f}'_1\to f'_1\bar{f}''_1$
we have {\tt s2n} for FF$\to $HA$\to $ SoftBR+HardBR parts
(here $f_1$ with or without primes denotes massless fermions,
{\it e.g.} of the 1st generation).
We have performed many high-precision comparisons of the numerical results derived with the aid 
of {\tt s2n.f} with the results of the alternative systems
{\tt FeynArts}~\cite{Hahn:2000kx} and {\tt GRACE}~\cite{Fleischer:2002nn}
and the code {\tt topfit}~\cite{Fleischer:2002xxx}.

\vspace*{-2mm}
\begin{itemize}
\item{ Level 3, MC generators}
\end{itemize}
\vspace*{-2mm}
In {\tt version 1.00}, MC generators are available only for $B\to f\bar{f}$ decays together 
with relevant graphic interface. The results can be presented in a variety of histograms. 
The user may ``play with the parameters'' of histograms in the window.

A first Monte Carlo generator for decays $Z(H,W)\to f\bar{f}\gamma$ 
is created in close contact with members of the KK collaboration,  
see papers \cite{Andonov:2002mx,Nanava:2003cg}. 
The MC generator is also accessible via menus, therefore, for the case of decays we are able 
to demonstrate how {\it the full chain of calculations} works out within our integrated system. 
The MC event generators are supposed to be usable also in a ``stand alone'' mode ready 
to be incorporated into the software of experiments.
\vspace*{2mm}

This paper is organized as follows:

In Section 2 we describe amplitudes for all available in {\tt version 1.00}  3- and 4-leg 
processes.

Section 3, the main section of this paper, is fully devoted to the concept of precomputation;
a comprehensive description of the {\tt SANC} precomputation tree and its modules is given.
We assume that while reading this part the reader will be looking inside the corresponding 
modules.
For one example, namely precomputation of photonic vacuum polarization, we demonstrate the whole
process of calculations by presenting intermediate results after each procedure is called.

In Section 4 we describe a part of {\tt SANC} procedures, mostly those which are used by 
precomputation modules.

In Section 5 we briefly describe the {\tt SANC} trees of processes implemented for the time 
being.

Although this paper is mostly devoted to the {\tt SANC} precomputation, in 
Section 6 we give a short {\bf User Guide} for {\tt version 1.00}. 
A more detailed description of the {\tt SANC} processes trees and computer aspects of 
{\tt SANC} system will be given elsewhere.

\clearpage

\section{Amplitude Basis, Scalar Form Factors, Helicity Amplitudes}
\subsection{Introduction}
In this section we present a collection of formulae for the amplitudes of basic SM $1\to 2$ decays
and $2\to 2$ processes available in {\tt SANC v.1.00}.
The {\em covariant one-loop amplitude} (CA) corresponds to a result of the straightforward standard
calculation by means of {\tt SANC} programs and procedures of {\em all} diagrams contributing 
to a given process at the tree (Born) and one-loop levels.
It is represented in a certain {\em basis}, made of strings of Dirac matrices and/or external 
momenta ({\em structures}), contracted with polarization vectors of vector bosons, if any. 
We usually omit Dirac spinors. The amplitude also contains kinematical factors and
coupling constants
and is parameterized by a certain number of form factors, which we denote by ${\cal F}$, 
in general with an index labeling the corresponding structure.
The number of FFs is equal to the number of structures. If there is only one FF, we normally
do not label it. For the processes with non zero tree-level amplitudes the FFs have the form
\bqa
{\cal F} = 1 + k {\tilde{\cal F}}\,,
\eqa 
where ``1'' is due to the Born level and the term ${\tilde{\cal F}}$ with the factor
\bqa
k=\frac{g^2}{16\pi^2}\,,
\label{kaen}
\eqa 
is due to the one-loop level. We also use various coupling constants
\bqa
Q_f\,,\quad I^{(3)}_f\,,\quad \sigma_f = v_f + a_f\,,\quad \delta_f = v_f - a_f\,,
\quad \stw=\frac{e}{g}\,,\quad 
\ctw=\frac{\mw}{\mz}\,,\quad \mbox{\it etc.}
\eqa

Given a CA parameterized by a certain number of FFs, {\tt SANC} computes a set
of HAs, denoted by ${\cal H}_{\lambda_1 \lambda_2 \lambda_3\dots}$, 
where $\lambda_1 \lambda_2 \lambda_3\dots$ denote the signs of particle spin projections onto 
a quantization axis as will be explained in the following sections.

In the representation of massive HAs, the following notation is very useful:
\bqa
P^{\pm}\left(I,M_1,M_2\right) = \sqrt{\big|I-\left(M_1\pm M_2 \right)^2\big|}\,.
\eqa

\subsection{The 3-leg processes ${\boldmath B(Q)\to f(p_1)+\bar{f}(p_2)}$}
In this section we present amplitudes for $1\to 2$ decays involving one vector boson
and two fermions.
For all $B\to f\bar{f}$ decays, except Higgs boson decay, the three $\lambda_i$ in 
${\cal H}_{\lambda_1\lambda_2\lambda_3}$ denote the signs of the boson, fermion and antifermion
spin projections, respectively. Fermion spins are projected onto their momenta,
the boson spin is projected onto the fermion momentum. For example, ${\cal H}_{++-}$ 
corresponds to the following spin configuration:
\vspace*{-20mm}

\begin{figure}[!h]
\[
\begin{picture}(125,86)(0,0)
    \DashLine(5,40)(115,40){5}
    \ArrowLine(114,40)(115,40)
\Text(117,35)[lb]{ $Z$}
    \Line(40,25)(80,25)
    \ArrowLine(79,25)(80,25)
    \Vertex(60,10){1.5}
    \Line(25,10)(100,10)
    \ArrowLine(99,10)(100,10)
    \ArrowLine(25,10)(24,10)
    \Line(30,-15)(50,-15)
    \ArrowLine(51,-15)(52,-15)

    \Line(70,-15)(90,-15)
    \ArrowLine(89,-15)(90,-15)

\Text(-2,20)[lb]  { Spin $B$}
\Text(23,5)[lt]  { $p_2$   }
\Text(84,5)[lt]  { $p_1$   }
\Text(15,-32)[lb]{ Spin ${\bar{ f}}$}
\Text(60,-32)[lb]{ Spin $ f$}
\end{picture}
\]
\end{figure}
\vspace*{-2cm}

\clearpage

\begin{itemize}

\item{$H\to f+\bar{f}$}

Its CA is described by one FF only: 
\bqa
{\cal A}_{\sss Hff}= \left( -\frac{g}{2} \frac{\mf}{\mw} \right) {\cal F}_{\sss S}\,. 
\eqa
Correspondingly, there is only one independent HA:
\bqa
{\cal H}_{++}&=& {\cal H}_{--}= \frac{g}{2}\frac{\mf}{\mw}P^+_{\sss Hff} {\cal F}_{\sss S}\,,
\nll
{\cal H}_{+-}&=& {\cal H}_{-+}= 0,
\eqa
here
\bqa
P^+_{\sss Hff}=P^{+}\left(\mh^2,\mf,\mf\right)=\sqrt{\mh^2-4\mf^2}\,.
\eqa

\item{$V\to f+\bar{f}$}

\noindent{The CA of the decay of a heavy vector particle ($V$) contains two FFs:}
\bqa
{\cal A}_{\sss Aff} = 
e Q_f \left[ i \gamma_\mu {\cal F}_{\sss Q} + \mf D_\mu {\cal F}_{\sss D} \right]\epsilon_\mu(Q)\,,
\eqa
here and below in this section, until stated otherwise, $D_\mu=(p_2-p_1)_\mu$.

\noindent{The two independent HAs are:}
\bqa
{\cal H}_{++-}  &=&{\cal H}_{--+} = \srt\, e\, Q_f \mv {\cal F}_{\sss Q}\,,
\nll
{\cal H}_{0++}  &=&{\cal H}_{0--}  = 
 e\,\mf Q_f\left[\left(P^{+}_{\sss Vff}\right)^2{\cal F}_{\sss D}+2{\cal F}_{\sss Q}\right],
\nll
{\cal H}_{+\pm+}&=&{\cal H}_{\pm--}={\cal H}_{0+-}={\cal H}_{0-+}={\cal H}_{-+\pm}=0\,,
\eqa
\bqa
P^+_{\sss Vff}=P^{+}\left(\mv^2,\mf,\mf\right)=\sqrt{\mv^2-4\mf^2}\,.
\eqa

\item{$Z\to f+\bar{f}$}

\noindent{In this case, the CA is described by three FFs:}
\bqa 
{\cal  A}_{\sss Zff} &=&
 \frac{g}{2\ctw} 
      \left[  i \gamma_\mu \gamma_6 {I^{(3)}_f} {\cal F}_{\sss L}              
            + i \gamma_\mu            \vma{f}{} {\cal F}_{\sss Q}              
            + \mf D_\mu {I^{(3)}_f} {\cal F}_{\sss D} \right]\epsilon_\mu(Q)\,,
\eqa
where $\gamma_{6,\,7}=1\pm\gamma_5$.
\noindent{The three independent HAs look as follows:}
\bqa
{\cal H}_{++-}  &=& \frac{g}{\srt \ctw}  \left[\itf (\mz -P^+_{\sss Zff}) {\cal F}_{\sss L}
 + \vma{f}{}  \mz  {\cal F}_{\sss Q} \right],
\nll
{\cal H}_{--+}  &=& \frac{g}{\srt \ctw}  \left[\itf (\mz +P^+_{\sss Zff}) {\cal F}_{\sss L}
 + \vma{f}{} \mz  {\cal F}_{\sss Q} \right],
\nll
{\cal H}_{ 0--} &=& {\cal H}_{0++}  
= \frac{g\mf}{\ctw} \left[ \itf {\cal F}_{\sss L} + \vma{f}{} {\cal F}_{\sss Q}
+\frac{1}{2}\itf\left(P^+_{\sss Zff}\right)^2{\cal F}_{\sss D}\right],
\nll
{\cal H}_{+\pm+}&=&  
{\cal H}_{\pm--}={\cal H}_{0+-}={\cal H}_{0-+}={\cal H}_{-++}={\cal H}_{-+-}=0\,,
\eqa
\bqa
P^+_{\sss Zff}=P^{+}\left(\mz^2,\mf,\mf\right)=\sqrt{\mz^2-4\mf^2}\,.
\eqa

\item{$W^{+}\to u+\bar{d}$ $(W^{-}\to \bar{u}+d)$}

\noindent{The CA of this decay is described by four FFs:}
\bqa
{\cal A}_{ W u \bar{d}} &=&
          \frac{g}{2\srt} \Bigl[                                 
          i \gamma_\mu \gamma_6 {\cal F}_{\sss L}                    
        + i \gamma_\mu \gamma_7 {\cal F}_{\sss R}                     
        + D_\mu \gamma_6 {\cal F}_{\sss LD}                 
        + D_\mu \gamma_7 {\cal F}_{\sss RD} \Bigr] \epsilon_\mu(Q)\,.
\label{Wdecay}
\eqa
\noindent{The corresponding HAs read:}
\bqa
{\cal H}_{--+} &=& \frac{g}{2}\Bigl[
       \left( P^-_{{\sss W}ud} + P^+_{{\sss W}ud}\right) {\cal F}_{\sss L}   
      +\left( P^-_{{\sss W}ud} - P^+_{{\sss W}ud}\right) {\cal F}_{\sss R}\Bigr],   
\nll
{\cal H}_{++-} &=& \frac{g}{2}\Bigl[
       \left( P^-_{{\sss W}ud} - P^+_{{\sss W}ud}\right) {\cal F}_{\sss L}
      +\left( P^-_{{\sss W}ud} + P^+_{{\sss W}ud}\right) {\cal F}_{\sss R}\Bigr],
\nll
{\cal H}_{0++} &=& \frac{g}{2 \srt} \frac{1}{\mw}
\Bigl[
 \left(P^-_{{\sss W}ud} m_{u+d} + P^+_{{\sss W}ud} m_{u-d}\right) {\cal F}_{\sss L}   
+\left(P^-_{{\sss W}ud} m_{u+d} - P^+_{{\sss W}ud} m_{u-d}\right) {\cal F}_{\sss R}    
\nll &&
+ P^+_{{\sss W}ud} P^-_{{\sss W}ud}
  \left[P^+_{{\sss W}ud} + P^-_{{\sss W}ud}\right] {\cal F}_{\sss LD}
+ P^+_{{\sss W}ud} P^-_{{\sss W}ud}
  \left[P^+_{{\sss W}ud} - P^-_{{\sss W}ud}\right] {\cal F}_{\sss RD}    
\Bigr],
\nll
{\cal H}_{0--} &=& \frac{g}{2 \srt} \frac{1}{\mw} 
\Bigl[
 \left(P^-_{{\sss W}ud} m_{u+d} - P^+_{{\sss W}ud} m_{u-d}\right) {\cal F}_{\sss L} 
+\left(P^-_{{\sss W}ud} m_{u+d} + P^+_{{\sss W}ud} m_{u-d}\right) {\cal F}_{\sss R}     
\nll &&
+ P^+_{{\sss W}ud} P^-_{{\sss W}ud} 
  \left( P^+_{{\sss W}ud} - P^-_{{\sss W}ud}\right) {\cal F}_{\sss LD}
+ P^+_{{\sss W}ud} P^-_{{\sss W}ud}
  \left( P^+_{{\sss W}ud} + P^-_{{\sss W}ud}\right) {\cal F}_{\sss RD}
\Bigr],
\nll
{\cal H}_{+\pm+} &=&  {\cal H}_{\pm--} = {\cal H}_{0+-} 
= {\cal H}_{0-+} = {\cal H}_{-+\pm} = 0\,,
\eqa
where
\bq
P^\pm_{{\sss W}ud}=P(\mw^2,\mup,\pm\mdn)=\sqrt{\mw^2-m^2_{u\pm d}}\,,
\qquad m_{u\pm d}=m_{u}\pm m_{d}\,.
\eq

\item{$t\to b+W^{+}$}

\noindent{The CA has the same structure as in Eq.~(\ref{Wdecay}),
 but here with $D_\mu=(p_1+p_2)_\mu$.}\\
\noindent{The four HAs are:}
\bqa
{\cal H}_{+-+} &=&
 \frac{g}{2} \left[ \left(P^+_{t{\sss W}b}+P^-_{t{\sss W}b}\right) {\cal F}_{\sss L} 
                   -\left(P^+_{t{\sss W}b}-P^-_{t{\sss W}b}\right) {\cal F}_{\sss R}\right],
\nll[2mm] 
{\cal H}_{-+-} &=&
-\frac{g}{2} \left[ \left(P^+_{t{\sss W}b}-P^-_{t{\sss W}b}\right) {\cal F}_{\sss L}
                   -\left(P^+_{t{\sss W}b}+P^-_{t{\sss W}b}\right) {\cal F}_{\sss R}\right],
\nll[2mm]
{\cal H}_{++0} &=&
\frac{g}{2 \srt\mw}\Bigl[
      \left( P^+_{t{\sss W}b} m_{t-b} - P^-_{t{\sss W}b} m_{t+b} \right){\cal F}_{\sss L}
     -\left (P^+_{t{\sss W}b} m_{t-b} + P^-_{t{\sss W}b} m_{t+b} \right){\cal F}_{\sss R}
\nll &&
+ P^+_{t{\sss W}b}P^-_{t{\sss W}b}\left(P^-_{t{\sss W}b}+P^+_{t{\sss W}b}\right){\cal F}_{\sss LD}
- P^+_{t{\sss W}b}P^-_{t{\sss W}b}\left(P^-_{t{\sss W}b}-P^+_{t{\sss W}b}\right){\cal F}_{\sss RD}
\Bigr],
\nll[2mm]
{\cal H}_{--0} &=&-
\frac{g}{2 \srt\mw}\Bigl[
      \left( P^+_{t{\sss W}b} m_{t-b} + P^-_{t{\sss W}b} m_{t+b} \right) {\cal F}_{\sss L}
     -\left( P^+_{t{\sss W}b} m_{t-b} - P^-_{t{\sss W}b} m_{t+b} \right) {\cal F}_{\sss R}
\nll[2mm] &&
+ P^+_{t{\sss W}b}P^-_{t{\sss W}b}\left(P^-_{t{\sss W}b}-P^+_{t{\sss W}b}\right){\cal F}_{\sss LD}
- P^+_{t{\sss W}b}P^-_{t{\sss W}b}\left(P^-_{t{\sss W}b}+P^+_{t{\sss W}b}\right){\cal F}_{\sss RD}
\Bigr],
\nll[2mm] 
{\cal H}_{++\pm}&=&{\cal H}_{+-0}={\cal H}_{+--}={\cal H}_{-++}={\cal H}_{-+0}={\cal H}_{--\pm}=0,
\eqa
where
\bq
P^{\pm}_{t{\sss W}b}=P(\mw^2,\mtp\pm\mbt)=\sqrt{m^2_{t\pm b}-\mw^2}\,,
\qquad m_{t\pm b}=m_{t}\pm m_{b}\,.
\eq

\item{$\bar{t}\to \bar{b}+W^{-}$}

The amplitudes of this process are similar to the previous one (though not identical).
Their explicit expressions can be found in the relevant module of the {\tt SANC} tree.
\end{itemize}

\clearpage

\subsection{The 3-leg processes $B(Q)\to V(p_1)+V(p_2),~~V=\gamma, Z, W$}
In this section we just list CAs and HAs for basic three-boson decays
in the SM. Note, the first two decays do not proceed at the tree level, this is why their FFs
do not start with ``1''. 

\begin{itemize}

\item{$H \to \gamma(p_1)+\gamma(p_2)$}
\bqa
{\cal  A}_{\sss H \gamma\gamma} &=& i\,k\,
g\stws\left(\delta_{\mu\nu}+2\frac{p_{1\mu}p_{2\nu}}{\mh^2}\right)\epsilon_\nu(p_1)\epsilon_\mu(p_2)
{\tilde{\cal F}}\,,
\eqa
\bqa
{\cal H}_{++} &=& {\cal H}_{--} = k\,g\stws {\tilde{\cal F}}\,,
\nll
{\cal H}_{+-} &=& {\cal H}_{-+} = 0\,.
\eqa

\item{$H \to Z(p_1)+\gamma(p_2)$} 
\bqa
{\cal  A}{\sss H\gamma Z} &=& i\,k\,
g\stw\left[\left(1-\frac{\mz^2}{\mh^2}\right)\delta_{\mu\nu}+2\frac{p_{1\mu}p_{2\nu}}{\mh^2}\right]
                                      \epsilon_\nu(p_1)\epsilon_\mu(p_2){\tilde{\cal F}}\,,
\eqa
\bqa
{\cal H}_{++} &=& {\cal H}_{--} =k\,g\,\stw\left(1-\frac{\mz^2}{\mh^2}\right){\tilde{\cal F}}\,,
\nll
{\cal H}_{+-} &=& {\cal H}_{0\pm}={\cal H}_{-+} = 0\,.
\eqa

\item{$H \to Z(p_1)+Z(p_2)$}
\bqa
{\cal A}_{\sss HZZ} = \left( -\frac{g\mz}{\ctw} \right)
\left(\delta_{\mu\nu} {\cal F}_{\sss D}+\frac{p_{1\mu} p_{2\nu}}{\mh^2} {\cal F}_{\sss P} \right)
                                      \epsilon_\nu(p_1)\epsilon_\mu(p_2)\,,
\eqa
\bqa
{\cal H}_{++}  &=& {\cal H}_{--} = \left( -\frac{g\mz}{\ctw} \right){\cal F}_{\sss D}\,,
\nll
{\cal H}_{00}    &=& \left( -\frac{g\mz}{\ctw} \right)
                 \left[\left(1-\frac{1}{2}\frac{\mh^2}{\mz^2}\right) {\cal F}_{\sss D}
                     - \left(1-\frac{1}{4}\frac{\mh^2}{\mz^2}\right) {\cal F}_{\sss P}\right],
\nll
{\cal H}_{\pm0} &=& {\cal H}_{+-} = {\cal H}_{0\pm} = {\cal H}_{-+} = 0\,.
\eqa

\item{$H \to W(p_1)+W(p_2)$}
\bqa
{\cal A}_{\sss HWW} = \left( -g \mw \right)    
\left(\delta_{\mu\nu}{\cal F}_{\sss D}+\frac{p_{1\mu} p_{2\nu}}{\mh^2} {\cal F}_{\sss P}\right)
                                     \epsilon_\nu(p_1)\epsilon_\mu(p_2)\,,
\eqa
\bqa
{\cal H}_{++} &=& {\cal H}_{--} = \left( -g \mw \right){\cal F}_{\sss D},
\nll 
{\cal H}_{00} &=& \left( -g \mw \right)  
                \left[\left(1-\frac{1}{2}\frac{\mh^2}{\mw^2}\right){\cal F}_{\sss D} 
                     -\left(1-\frac{1}{4}\frac{\mh^2}{\mw^2}\right){\cal F}_{\sss P}\right],
\nll
{\cal H}_{+0} &=& {\cal H}_{+-} = {\cal H}_{0+} 
 = {\cal H}_{0-} = {\cal H}_{-+} = {\cal H}_{-0} = 0\,.
\eqa

\item{$Z(Q)\to W(p_1)+W(p_2)$}
\bqa
{\cal A}_{\sss ZWW} &=& i\,g\,\ctw \Bigl(p_{1\mu}p_{2\nu}D_\alpha {\cal F}_{{\sss D}_{12}} 
+ \delta_{\alpha \nu} p_{1\mu} {\cal F}_{{\sss D}_1} 
+ \delta_{\alpha \mu} p_{2\nu} {\cal F}_{{\sss D}_2} 
\nll
&&+\delta_{\mu\nu} D_\alpha {\cal F}_{{\sss D}d}
+ \epsilon_{\beta\alpha\mu\nu} D_\beta {\cal F}_{{\sss D}{\epsilon}}^{\rm fer}\Bigr)
                      \epsilon_\alpha(Q)\epsilon_\nu(p_1)\epsilon_\mu(p_2)\,.
\eqa
The HAs for this decay are not implemented in {\tt SANC v.1.00}.
\end{itemize}

\subsection{The 4-leg NC processes $f_1\bar{f}_1\to(\gamma Z)\to f\bar{f}$\label{proc-ffff}}
Here we present the CAs and HAs for any $f_1\bar{f}_1 f\bar{f}\to 0$ NC 
process at any channel $s$, $t$ or $u$. 
Here $0$ stands for {\em vacuum}, and by  $f_1$ we mean a first 
generation fermion with field index 11,12,13,14, whose mass is neglected everywhere except in 
arguments of $\log$s (mass singularities) and by $f$ we mean any fermion with field indices 
from 11 to 22 (see Section~\ref{Uguide} for definition of field indices).
For such a case, the Higgs and $\phi^{0}$ boson interactions with the $f_1$ current are also 
neglected.

The covariant one-loop amplitude of the $2f \to 2f$ process
\vspace*{-2mm}

\begin{figure}[!h]
\[
\begin{picture}(125,86)(0,0)
  \Photon(25,43)(100,43){3}{10}
    \Vertex(100,43){5}
  \ArrowLine(125,86)(100,43)
  \ArrowLine(100,43)(125,0)
  \ArrowLine(0,0)(25,43)
    \Vertex(25,43){5}
  \ArrowLine(25,43)(0,86)
\Text(6,75)[lb]  { $\bar{f}_1(p_1)$}
\Text(6,12)[lt]  { $f_1(p_2)$    }
\Text(60,25)[bc] { $(Z,\gamma)$  }
\Text(125,77)[lb]{ $\bar{f}(p_4)$}
\Text(125,12)[lt]{ $f(p_3)$      }
\end{picture}
\]
\vspace*{-6mm}

\caption[$f_1\bar{f}_1\to f\bar{f}$ process] 
        {$f_1\bar{f}_1\to f\bar{f}$ process.}
\label{fig:oneloopP}
\end{figure}
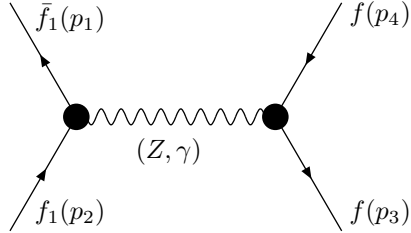

\noindent
is described in terms of six form factors: $LL,QL,LQ,QQ,LD$ and $QD$,
corresponding to six Dirac structures (${\cal A}_{\gamma}$ is also described by a $QQ$ structure,
it is separated out for convenience; $\alpha(s)=\alpha(0) {\cal F}_{\gamma}(s)$).
Note that all 4-momenta are incoming and the usual Mandelstam invariants in 
our metric (i.e. $p^2=-m^2$) are:
\bq
\left( p_1+p_2 \right)^2 =-s,
\qquad
\left( p_2+p_3 \right)^2 =-t,
\qquad
\left( p_2+p_4 \right)^2 =-u.
\eq

The $\gamma$ and $Z$ exchange amplitudes are: 
\bqa
{\cal A}_{\gamma}(s) &=& 
i \,e^2\,\frac{\ds Q_{f_1} Q_{f}}{ \ds s} {\cal F}_{\gamma}(s)
     \gamma_\mu \otimes \gamma_\mu\,\,,
\label{ggNC}
\\[1mm]
{\cal A}_{\sss{Z}}(s)&=&
i \,e^2\,\frac{\chi_{\sss{Z}}(s)}{s}
\nll [1mm] &&
\times \Biggl\{
 I^{(3)}_{f_1}I^{(3)}_{f}\gamma_\mu\left(1+\gamma_5\right)\otimes\gamma_\mu\left(1+\gamma_5\right)
                   {\cal F}_{\sss{LL}}(s,t,u)
+ \delta_{f_1} I^{(3)}_{f} \gamma_\mu \otimes \gamma_\mu \left( 1 + \gamma_5 \right)
                   {\cal F}_{\sss{QL}}(s,t,u)
\nll [1mm] &&
+ I^{(3)}_{f_1}\delta_{f}
\gamma_\mu \left( 1 + \gamma_5 \right) \otimes\gamma_\mu \,
                   {\cal F}_{\sss{LQ}}(s,t,u)
+\delta_{f_1}\delta_{f}
\gamma_\mu\otimes\gamma_\mu \,
                   {\cal F}_{\sss{QQ}}(s,t,u)
\label{zzNC}
\\ [1mm] &&
+ I^{(3)}_{f_1}   I^{(3)}_{f} 
\gamma_\mu{\left( 1 + \gamma_5 \right)}\otimes\left(- i m_f D_{\mu} \right)
                   {\cal F}_{\sss{LD}}(s,t,u)
+\delta_{f_1} I^{(3)}_{f}
\gamma_\mu \otimes \left( - i m_f D_{\mu} \right) {\cal F}_{\sss{QD}}(s,t,u)\hspace*{-.3mm}
\Biggr\}.
\nonumber
\eqa
Here and below in this and in the next sections, $D_\mu=(p_4-p_3)_\mu$ and $\chi_{\sss{Z}}(s)$ 
is the $Z/\gamma$ propagator ratio
\bq
\chi_{\sss{Z}}(s)=\frac{1}{4\stw^2\ctw^2}\,\frac{s}{s-\mz^2+i\mz\Gamma_{\sss Z}}\,.
\eq
Symbol $\gamma_\mu\otimes\gamma_\mu$ is used in the following short-hand notation:
\bq
\gamma_\mu\otimes\gamma_\nu=\iap{p_1}\gamma_{\mu}\ip{p_2}\op{p_3}\gamma_{\nu}\oap{p_4}.
\eq
For more details see Ref.~\cite{Andonov:2002xc}.

If the $f_1$ mass is neglected, we have six corresponding HAs.
They depend on kinematical variables, coupling constants and our six scalar form factors: 
\bqa
{\cal H}_{-++-} &=& - e^2\left( 1+\cos\vartheta \right) 
 \Bigg( Q_{f_1} Q_f {\cal F}_{\gamma}(s)
 +{\chi_{\sss Z}(s)}\delta_{f_1}\Bigg[\left(1-\frac{P^+}{\sqrt{s}}\right)I^{(3)}_f{\cal F}_{\sss QL}
        +\delta_f {\cal F}_{\sss QQ}\Bigg]\Bigg),
\nll
{\cal H}_{-+-+} &=& - e^2\left( 1-\cos\vartheta\right) 
 \Bigg(Q_{f_1} Q_f {\cal F}_{\gamma}(s)
 +{\chi_{\sss Z}(s)}\delta_{f_1}\Bigg[\left(1+\frac{P^+}{\sqrt{s}}\right)I^{(3)}_f{\cal F}_{\sss QL}
        +\delta_f {\cal F}_{\sss QQ} \Bigg] \Bigg),
\nll
{\cal H}_{-+--} &=& {\cal H}_{-+++}
                   =  e^2\,\frac{2\mf}{\sqs} \sin\vartheta
 \Bigg(Q_{f_1} Q_f {\cal F}_{\gamma}(s)
\nll &&
     +{\chi_{\sss Z}(s)}\delta_{f_1}\Bigg[ I^{(3)}_f {\cal F}_{\sss QL}+\delta_f{\cal F}_{\sss QQ}
     +\frac{1}{2} (P^+)^2  I^{(3)}_f  {\cal F}_{\sss QD}\Bigg] \Bigg),
\nll
{\cal H}_{+-++} &=& {\cal H}_{+---} 
                  = - e^2\,\frac{2\mf}{\sqs} \sin\vartheta
 \Bigg(Q_{f_1} Q_f {\cal F}_{\gamma}(s)
+{\chi_{\sss Z}(s)}\Bigg[2I^{(3)}_{f_1}I^{(3)}_f{\cal F}_{\sss LL}
                     +2I^{(3)}_{f_1}\delta_f {\cal F}_{\sss LQ}
\nll &&
 +\delta_{f_1} I^{(3)}_f {\cal F}_{\sss QL}
                                 +\delta_{f_1} \delta_f {\cal F}_{\sss QQ}
  +\frac{1}{2} (P^+)^2 I^{(3)}_f
       \Big( 2 I^{(3)}_{f_1} {\cal F}_{\sss LD}+\delta_{f_1} {\cal F}_{\sss QD}\Big) \Bigg] \Bigg),
\nll
{\cal H}_{+-+-} &=& - e^2\left( 1-\cos\vartheta \right) 
  \Bigg( Q_{f_1} Q_f {\cal F}_{\gamma}(s)
+{\chi_{\sss Z}(s)} \Bigg[ \left( 1 - \frac{P^+}{\sqrt{s}} \right) I^{(3)}_f
    \Big( 2 I^{(3)}_{f_1} {\cal F}_{\sss LL}+\delta_{f_1} {\cal F}_{\sss QL}\Big)
\nll &&
+\delta_f\left(2I^{(3)}_{f_1}{\cal F}_{\sss LQ}+\delta_{f_1}{\cal F}_{\sss QQ}\right)\Bigg] \Bigg),
\nll
{\cal H}_{+--+} &=& - e^2\left( 1+\cos\vartheta \right) 
   \Bigg(
  Q_{f_1} Q_f {\cal F}_{\gamma}(s)
+{\chi_{\sss Z}(s)} \Bigg[ \left( 1 + \frac{P^+}{\sqrt{s}} \right) I^{(3)}_f
    \Big( 2 I^{(3)}_{f_1}   {\cal F}_{\sss LL}+\delta_{f_1} {\cal F}_{\sss QL}\Big)
\nll &&
             +\delta_f  \Big( 2 I^{(3)}_{f_1} {\cal F}_{\sss LQ}
             +\delta_{f_1}  {\cal F}_{\sss QQ} \Big) \Bigg] \Bigg),
\nll
{\cal H}_{+++\pm} &=&  {\cal H}_{++-\pm}  = {\cal H}_{--+\pm}  = {\cal H}_{---\pm} = 0;
\eqa
helicity indices, for example, 
$+--+$ denote the signs of the fermion spin projections onto their momenta $p_1,p_2,p_3,p_4$, 
respectively. 

Moreover,
\bqa
P^+=P^+(s,\mf,\mf)=\sqrt{s-4\mf^2}\,,
\eqa
and the scattering angle $\cos\vartheta$ is connected to the invariant $t$:
\bqa
\cos\vartheta &=& \left( t-\mf^2+\frac{s}{2} \right) \frac{2}{s \beta_f}\,,
\eqa
where 
\bq
\beta_f=\ds\sqrt{1-\frac{4\,m_f^2}{s}}\,.
\label{betaf}
\eq
\clearpage

\subsection{The 4-leg CC processes $f_1\bar{f}'_1\to(W)\to f\bar{f'}$}
In version {\tt 1.00} we have implemented particular $2f\to 2f$ CC processes, having 
in mind their application to Drell--Yan type CC processes at hadron colliders as well
as for 3-particle top decays.
\begin{itemize}
\item $\bar{u} + d\to l^- + \bar{\nu}_l$
\bqa
{\cal A}_{\sss W^-} &=& i\,e^2\,\frac{\chi_{\sss W}(s)}{4\,s}\Big[
    \gamma_\mu \left( 1+\gamma_5 \right) \otimes \gamma_\mu \left( 1+\gamma_5 \right)
                           {\cal F}_{\sss{LL}}(s,t,u)
\nll
&&+ \gamma_\mu \left( 1+\gamma_5 \right) \otimes \left( 1+\gamma_5 \right)(-i D_\mu)  
                           {\cal F}_{\sss{LD}}(s,t,u)\Big].
\eqa

There are only two non-zero HAs for the case when only one mass (lepton)
is not neglected:
\bqa
{\cal H}_{+--+}  &=&  - e^2\,\left( 1 + \cos\thmo  \right) \frac{P^+}{\sqrt{s}}\,\chi_{\sss W}(s)
                           {\cal F}_{\sss LL}(s,t,u)\,,
\nll
{\cal H}_{+-++}  &=&  - e^2\,\sin\thmo P^+\, \chi_{\sss W}(s)
                \left[ \frac{\mml}{s}  {\cal F}_{\sss LL}(s,t,u)  
               +\left(  1 - \frac{\mml^2}{s} \right)  {\cal F}_{\sss LD}(s,t,u) \right],
\nll[3mm]
{\cal H}_{+++\pm} &=& {\cal H}_{++-\pm} = {\cal H}_{+-\pm-} = {\cal H}_{-++\pm} =
{\cal H}_{-+-\pm} = {\cal H}_{--+\pm} = {\cal H}_{---\pm} = 0\,,
\eqa
here and below
\vspace*{-7mm}

\bqa
P^+=P^+(s,\mml,0)=\sqrt{s-\mml^2}\,.
\eqa
\vspace*{-7mm}

\item $\bar{d} + u \to l^+ + \nu_l$
\bqa
{\cal A}_{\sss W^+} &=& i\,e^2\,\frac{\chi_{\sss W}(s)}{4\,s}\Big[
    \gamma_\mu \left( 1+\gamma_5 \right) \otimes \gamma_\mu \left( 1+\gamma_5 \right)
                           {\cal F}_{\sss{LL}}(s,t,u)
\nll
&&+ \gamma_\mu \left( 1+\gamma_5 \right) \otimes \left( 1-\gamma_5 \right)(-i D_\mu) 
                           {\cal F}_{\sss{RD}}(s,t,u)\Big],
\eqa
\bqa
  {\cal H}_{+--+} &=& - e^2\,\left( 1 + \cos\thmn \right) \frac{P^+}{\sqrt{s}}\,\chi_{\sss W}(s) 
                      {\cal F}_{\sss LL}(s,t,u)\,, 
\nll
  {\cal H}_{+---} &=& - e^2\,\sin\thmn P^+\,\chi_{\sss W}(s) 
                         \left[         \frac{\mml}{s} {\cal F}_{\sss LL}(s,t,u) 
                        +\left(1-\frac{\mml^2}{s} \right) {\cal F}_{\sss RD}(s,t,u)\right],
\nll[3mm]
{\cal H}_{+++\pm} &=& {\cal H}_{++-\pm} = {\cal H}_{+-+\pm} = {\cal H}_{-++\pm} =
{\cal H}_{-+-\pm} = {\cal H}_{--+\pm} = {\cal H}_{---\pm} = 0\,.
\eqa
Note our angle convention: $\thmo$ and $\thmn$ are chosen to be the angles between 
{\em particle momenta} in the initial and final states in the cms reference frame,
and here
\bq
\chi_{\sss W}(s)=\frac{s}{2\stw^2}\,\frac{1}{s-\mw^2+i\mw\Gamma_{\sss W}}\,.
\eq

\item $t(p_2) \to b(p_1) + l^+(p_4) + \nu_l(p_3)$

For this case there are four different structures and scalar form factors 
if the mass of the $b$ quark is not neglected. The CA reads:
\bqa
{\cal A}_{t} &=&i\,e^2\,\frac{d_{\sss W}(s)}{4}\Big[
    \gamma_\mu \left( 1+\gamma_5 \right) \otimes \gamma_\mu \left( 1+\gamma_5 \right)
                       {\cal F}_{\sss{LL}}(s,t)\,
\nll
&&+ \gamma_\mu \left( 1-\gamma_5 \right) \otimes \gamma_\mu \left( 1+\gamma_5 \right)
                       {\cal F}_{\sss{RL}}(s,t)\,
\nll
&&+ \left( 1+\gamma_5 \right) \otimes \gamma_\mu \left( 1+\gamma_5 \right)(-i D_\mu) 
                       {\cal F}_{\sss{LD}}(s,t)\,
\nll
&&+ \left( 1-\gamma_5 \right) \otimes \gamma_\mu \left( 1+\gamma_5 \right)(-i D_\mu) 
                       {\cal F}_{\sss{RD}}(s,t)\Big].
\eqa
Here $D_\mu$ and 4-momentum conservation read:
\bq
D_\mu=(p_1+p_2)_\mu\,,\quad p_2=p_1+p_3+p_4\,,
\eq
while the invariants are
\bq
s=-(p_3+p_4)^2,\quad t=-(p_1+p_4)^2, 
\eq
and
\bq
d_{\sss W}(s)=\frac{1}{2\stw^2}\,\frac{1}{s-\mw^2+i\mw\Gamma_{\sss W}}\,.
\eq

The four non-zero HAs are:
\bqa
  {\cal H}_{++-+} &=&+\frac{1}{2}\,e^2\,{d_{\sss W}(s)}\sin\thle\Big\{
   \left(P^{+} m_{t-b}-P^{-}m_{t+b}\right){\cal F}_{\sss LL}
  -\left(P^{+} m_{t-b}+P^{-}m_{t+b}\right){\cal F}_{\sss RL}
\nll
&&\hspace*{16mm}
  -P^{+}P^{-}\Big[\left(P^{+}+P^{-}\right){\cal F}_{\sss LD}
                 +\left(P^{+}-P^{-}\right){\cal F}_{\sss RD}\Big]\Big\}\,,
\nll
  {\cal H}_{---+} &=&-\frac{1}{2}\,e^2\,{d_{\sss W}(s)}\sin\thle\Big\{
   \left(P^{+} m_{t-b}+P^{-}m_{t+b}\right){\cal F}_{\sss LL}
  -\left(P^{+} m_{t-b}-P^{-}m_{t+b}\right){\cal F}_{\sss RL}
\nll
&&\hspace*{16mm}
  +P^{+}P^{-}\Big[\left(P^{+}-P^{-}\right){\cal F}_{\sss LD}
                 +\left(P^{+}+P^{-}\right){\cal F}_{\sss RD}\Big]\Big\}\,,
\nll
  {\cal H}_{+--+} &=&+\frac{1}{2}\,e^2\,{d_{\sss W}(s)}\left(1-\cos\thle\right)\sqrt{s}\Big\{
   \left(P^{+} - P^{-}\right){\cal F}_{\sss LL}
  -\left(P^{+} + P^{-}\right){\cal F}_{\sss RL}\Big\}\,,
\nll
  {\cal H}_{-+-+} &=&-\frac{1}{2}\,e^2\,{d_{\sss W}(s)}\left(1+\cos\thle\right)\sqrt{s}\Big\{
   \left(P^{+} + P^{-}\right){\cal F}_{\sss LL}
  -\left(P^{+} - P^{-}\right){\cal F}_{\sss RL}\Big\}\,.
\eqa
Here
\bq
P^{\pm}=\sqrt{(m_t\pm m_b)^2-s}\,.
\eq
Finally, $\thle$ is the angle between leptonic 4-momentum in R-frame ($\vec{p}_3+\vec{p}_4=0$)
and the z-axis is chosen along $p_1$ momentum in the rest frame of decaying top.
It is related to $t$ invariant by
\bq
t=m^2_{b}+\frac{1}{2}\left[m^2_{t}-m^2_{b}-s-\sqrt{\lambda(s,m^2_{t},m^2_{b})}\cos\thle\right],
\eq
where $\lambda(x,y,z)$ is the ordinary kinematical function
\bq
\lambda(x,y,z)=x^2+y^2+z^2-2xy-2xz-2yz\,.
\eq

\item $\bar{t} \to \bar{b} + l^- + \bar{\nu}_l$

This case is similar, although not identical to the previous one. For exact expressions see 
relevant module in the {\tt SANC} tree:\\
{\bf EW $\to$ Processes $\to$ 4 legs $\to$ Charged Current $\to$ t$\to$ b~l~nu (HA)}.

\end{itemize}

\clearpage

\subsection{Bhabha scattering}
The CA for Bhabha scattering can be derived from Eqs.(\ref{ggNC}--\ref{zzNC}) as
follows 
(if the electron mass is neglected):
\bqa
{\cal A}_{\rm{Bhabha}}&=&{\cal A}_{\gamma}(s)+{\cal A}_{\sss{Z}}(s)
                -\left[{\cal A}_{\gamma}(t)+{\cal A}_{\sss{Z}}(t)\right]
\nll
&=& i\,e^2 \left[ 
 \gamma_\mu \otimes \gamma_\mu \frac{{\cal F}_{\gamma}(s)}{s} 
-\gamma_\mu \otimes \gamma_\mu \frac{{\cal F}_{\gamma}(t)}{t} \right]
\nll[1mm]
& &+i\,e^2\,\frac{\chi_{\sss{Z}}(s)}{s}
\nll [1mm] &&
\times \biggl\{\hspace*{-.3mm} \left(I^{(3)}_{e}\right)^2\hspace*{-1mm}
\gamma_\mu\left(1+\gamma_5\right)\otimes\gamma_\mu\left(1+\gamma_5\right)
                   {\cal F}_{\sss{LL}}(s,t,u)
+ \delta_{e} I^{(3)}_{e} \gamma_\mu \otimes \gamma_\mu \left( 1 + \gamma_5 \right)
                   {\cal F}_{\sss{QL}}(s,t,u)
\nll [1mm] &&
+ I^{(3)}_{e}\delta_{e}
\gamma_\mu \left( 1 + \gamma_5 \right) \otimes\gamma_\mu \,
                   {\cal F}_{\sss{LQ}}(s,t,u)
+\delta^2_{e}\gamma_\mu\otimes\gamma_\mu \,
                   {\cal F}_{\sss{QQ}}(s,t,u)
\biggr\}\qquad\quad
\nll [1mm] 
& &-i\,e^2\,\frac{\chi_{\sss{Z}}(t)}{t}
\nll [1mm] &&
\times \biggl\{\hspace*{-.3mm} \left(I^{(3)}_{e}\right)^2\hspace*{-1mm}
\gamma_\mu\left(1+\gamma_5\right)\otimes\gamma_\mu\left(1+\gamma_5\right)
                   {\cal F}_{\sss{LL}}(t,s,u)
+ \delta_{e} I^{(3)}_{e} \gamma_\mu \otimes \gamma_\mu \left( 1 + \gamma_5 \right)
                   {\cal F}_{\sss{QL}}(t,s,u)
\nll [1mm] &&
+ I^{(3)}_{e}\delta_{e}
\gamma_\mu \left( 1 + \gamma_5 \right) \otimes\gamma_\mu \,
                   {\cal F}_{\sss{LQ}}(t,s,u)
+\delta^2_{e}
\gamma_\mu\otimes\gamma_\mu \,
                   {\cal F}_{\sss{QQ}}(t,s,u)
\biggr\}.
\eqa
It is described by the electromagnetic running coupling constant and four FFs with exchanged
arguments $s$ and $t$.

There are six non-zero HAs:
\bqa
{\cal H}_{++++} &=&
 -2 e^2\,\frac{s}{t} \Bigl\{ {\cal F}^\gamma_{\sss QQ}(t,s,u) - {\chi_{\sss Z}}(t)\vma{e}{} 
\left[{\cal F}^{\sss Z}_{\sss QL}(t,s,u)-\vma{e}{}{\cal F}^{\sss Z}_{\sss QQ}(t,s,u)\right]\Bigr\},
\nll
{\cal H}_{----} &=&  
 -2 e^2\,\frac{s}{t} \Bigl\{ {\cal F}^\gamma_{\sss QQ}(t,s,u) - {\chi_{\sss Z}}(t)\vma{e}{}    
\left[{\cal F}^{\sss Z}_{\sss LQ}(t,s,u)-\vma{e}{}{\cal F}^{\sss Z}_{\sss QQ}(t,s,u)\right]\Bigr\},
\nll
{\cal H}_{+-+-} &=& -e^2\,\left(1-\cos\theta\right)\Bigl\{ {\cal F}^\gamma_{\sss QQ}(s,t,u)
 -{\chi_{\sss Z}}(s)\vma{e}{}\left[{\cal F}^{\sss Z}_{\sss LQ}(s,t,u)
                      -\vma{e}{}{\cal F}^{\sss Z}_{\sss QQ}(s,t,u)\right]\Bigr\},
\nll
{\cal H}_{-+-+} &=& -e^2\,\left(1-\cos\theta\right)\Bigl\{ {\cal F}^\gamma_{\sss QQ}(s,t,u) 
 -{\chi_{\sss Z}}(s)\vma{e}{}\left[{\cal F}^{\sss Z}_{\sss QL}(s,t,u)
                      -\vma{e}{}{\cal F}^{\sss Z}_{\sss QQ}(s,t,u)\right]\Bigr\},
\nll     
{\cal H}_{+--+} &=& -e^2\,\left( 1+\cos \theta\right) \Bigl\{ {\cal F}^\gamma_{\sss QQ}(s,t,u)
\nll &&\hspace*{15mm}
    +{\chi_{\sss Z}}(s)\left( {\cal F}^{\sss Z}_{\sss LL}(s,t,u)
            -\vma{e}{} \left[ {\cal F}^{\sss Z}_{\sss LQ}(s,t,u)
                             +{\cal F}^{\sss Z}_{\sss QL}(s,t,u)
                    -\vma{e}{}{\cal F}^{\sss Z}_{\sss QQ}(s,t,u) \right] \right)
\nll &&\hspace*{-14mm}
 +\frac{s}{t} 
\Big[ 
    {\cal F}^\gamma_{\sss QQ}(t,s,u)
    +{\chi_{\sss Z}}(t)\left( {\cal F}^{\sss Z}_{\sss LL}(t,s,u)
            -\vma{e}{} \left[ {\cal F}^{\sss Z}_{\sss LQ}(t,s,u)
                             +{\cal F}^{\sss Z}_{\sss QL}(t,s,u)
                    -\vma{e}{}{\cal F}^{\sss Z}_{\sss QQ}(t,s,u) \right] \right) 
\Big]
\Bigr\},
\nll 
{\cal H}_{-++-} &=& - e^2\,\left( 1+ \cos \theta \right) \Bigl\{
                    {\cal F}^\gamma_{\sss QQ}(s,t,u) 
 + \vma{e}{2}{\chi_{\sss Z}}(s){\cal F}^{\sss Z}_{\sss QQ}(s,t,u) 
\nll &&\hspace*{20mm}
 +\frac{s}{t} \left[{\cal F}^\gamma_{\sss QQ}(t,s,u) 
 + \vma{e}{2}{\chi_{\sss Z}}(t){\cal F}^{\sss Z}_{\sss QQ}(t,s,u)\right] \Bigr\},
\nll
{\cal H}_{+++-} &=& {\cal H}_{++-\pm} = {\cal H}_{+-++} = {\cal H}_{+---} = 0\,,
\nll 
{\cal H}_{-+++}&=& {\cal H}_{-+--}  = {\cal H}_{--+\pm} = {\cal H}_{---+} = 0\,,
\eqa
however, since for Bhabha scattering 
${\cal F}^{\sss Z}_{\sss LQ}={\cal F}^{\sss Z}_{\sss QL}$, the number of independent HAs
is actually reduced to four as expected.
\clearpage

\subsection{$ffbb\to 0$ processes}
In {\tt SANC v.1.00} we have implemented three classes of $ffbb\to 0$ processes:
$ff\gamma\gamma\to 0$, $ffZ\gamma\to 0$ and $ffH\gamma\to 0$. Due to space shortage,
we limit ourselves in this paper to the process $ffH\gamma\to 0$. 
Moreover, for $ffbb\to 0$ processes the variety of cross channels is more rich than for the case 
of $ffff\to 0$. For example, for process  $ffH\gamma\to 0$
it is worth considering at least three channels: \\
$\,$\hspace*{5mm}$\bullet$ annihilation, $f\bar{f}\to H\gamma$; \\
$\,$\hspace*{5mm}$\bullet$ decay, $H\to f\bar{f}\gamma$;\\
$\,$\hspace*{5mm}$\bullet$ and $H$ production at $\gamma e$ colliders, $\gamma e\to He$.

For all the channels we still can write down almost unique CA. Below we give
it exactly for the annihilation channel, $f(p_2)\bar{f}(p_1)\to H(p_4)\gamma(p_3)$,
but it might be easily rewritten into any other channel.
This is not the case, however, for the HAs. The latter must be recomputed for
any given channel.
 
There are eight structures transverse in photonic 4-momentum, 4 vector and 4 axial ones
\bqa
{\cal A}_{\sss ffH\gamma}
       &=& -e\,g\,\frac{Q_f m_f}{\mw}\,\Biggl\{
           \iap{p_1}\ip{p_2}\left[(U^2+m^2_f)(p_2)_{\nu}-(T^2+m^2_f)(p_1)_{\nu}\right]
                                                 \varepsilon^{\gamma}_{\nu}(p_3)F_{v1}(s,t)
\nll[3mm]
       & &-\iap{p_1}\sla{p_3}\gamma_{\nu}\ip{p_2}\varepsilon^{\gamma}_{\nu}(p_3)F_{v2}(s,t)
\nll[1mm]
       & &-\iap{p_1}\,i\,\left[\sla{p_3}(p_1)_{\nu}+\frac{1}{2}(U^2+m^2_f)\gamma_{\nu}\right]
                                         \ip{p_2}\varepsilon^{\gamma}_{\nu}(p_3)F_{v3}(s,t)
\nll
       & &-\iap{p_1}\,i\,\left[\sla{p_3}(p_2)_{\nu}+\frac{1}{2}(T^2+m^2_f)\gamma_{\nu}\right]
                                         \ip{p_2}\varepsilon^{\gamma}_{\nu}(p_3)F_{v4}(s,t)
\nll[1mm]
       & &+\iap{p_1}\gamma_5\ip{p_2}\left[(U^2+m^2_f)(p_2)_{\nu}-(T^2+m^2_f)(p_1)_{\nu}\right]
                                                 \varepsilon^{\gamma}_{\nu}(p_3)F_{a1}(s,t)
\nll[3mm]
       & &-\iap{p_1}\sla{p_3}\gamma_{\nu}\gamma_5\ip{p_2}\varepsilon^{\gamma}_{\nu}(p_3)F_{a2}(s,t)
\nll[1mm]
       & &-\iap{p_1}\,i\,\left[\sla{p_3}(p_1)_{\nu}+\frac{1}{2}(U^2+m^2_f)\gamma_{\nu}\right]
                                 \gamma_5\ip{p_2}\varepsilon^{\gamma}_{\nu}(p_3)F_{a3}(s,t)
\nll
       & &-\iap{p_1}\,i\,\left[\sla{p_3}(p_2)_{\nu}+\frac{1}{2}(T^2+m^2_f)\gamma_{\nu}\right]
                      \gamma_5\ip{p_2}\varepsilon^{\gamma}_{\nu}(p_3)F_{a4}(s,t)\Biggr\},
\label{StrffHA}
\eqa
each multiplied by the corresponding FF: $ F_{v1\div v4}$ and $ F_{a1\div a4}$.
In above expressions
\bq
T^2=(p_2+p_3)^2,\qquad U^2=(p_2+p_4)^2. 
\eq
 
The eight different HAs are
\bqa
{\cal H}_{+++} &=&-k_0
              \Bigl[s\betaf \left(\betaf F_{v1}(s,t) - F_{a1}(s,t) \right)
                          - \betap\left( F_{v2}(s,t) - F_{a2}(s,t) \right)
\nll &&
                      + \mmf\left( F_{v3}(s,t) -\betaf F_{a3}(s,t) \right) 
                      + \mmf\left( F_{v4}(s,t) +\betaf F_{a4}(s,t) \right)\Bigr],
\nll
{\cal H}_{---} &=& k_0
              \Bigl[s\betaf \left(\betaf F_{v1}(s,t) + F_{a1}(s,t) \right)
                          - \betap\left( F_{v2}(s,t) + F_{a2}(s,t) \right)
\nll &&
                      + \mmf\left( F_{v3}(s,t) +\betaf F_{a3}(s,t) \right) 
                      + \mmf\left( F_{v4}(s,t) -\betaf F_{a4}(s,t) \right)\Bigr],
\nll
{\cal H}_{++-} &=& k_0 
              \Bigl[s\betaf \left(\betaf F_{v1}(s,t) - F_{a1}(s,t) \right)
                          - \betam\left( F_{v2}(s,t) + F_{a2}(s,t) \right)
\nll &&
                      + \mmf\left( F_{v3}(s,t) -\betaf F_{a3}(s,t) \right)
                      + \mmf\left( F_{v4}(s,t) -\betaf F_{a4}(s,t) \right)\Bigr],
\nll
{\cal H}_{--+} &=&-k_0 
              \Bigl[s\betaf\left(\betaf  F_{v1}(s,t) + F_{a1}(s,t) \right)
                          - \betam\left( F_{v2}(s,t) - F_{a2}(s,t) \right)
\nll &&
                      + \mmf\left( F_{v3}(s,t) +\betaf F_{a3}(s,t) \right) 
                      + \mmf\left( F_{v4}(s,t) -\betaf F_{a4}(s,t) \right)\Bigr],
\nll
{\cal H}_{+-+} &=&-k_{+}
       \Bigl[4\frac{\mmf}{s}\left( F_{v2}(s,t) - F_{a2}(s,t) \right)
\nll &&
                  -  \betap \left( F_{v3}(s,t) + \betaf F_{a3}(s,t) \right)
                  -  \betam \left( F_{v4}(s,t) + \betaf F_{a4}(s,t) \right) \Bigr],
\nll
{\cal H}_{-+-} &=&-k_{+}
       \Bigl[4\frac{\mmf}{s}\left( F_{v2}(s,t) + F_{a2}(s,t) \right) 
\nll &&
                  -  \betap \left( F_{v3}(s,t) - \betaf F_{a3}(s,t) \right)
                  -  \betam \left( F_{v4}(s,t) - \betaf F_{a4}(s,t) \right) \Bigr],
\nll
{\cal H}_{+--} &=&-k_{-}
       \Bigl[4\frac{\mmf}{s}\left( F_{v2}(s,t) + F_{a2}(s,t) \right)
\nll &&
                  -  \betam \left( F_{v3}(s,t) + \betaf F_{a3}(s,t)\right)
                  -  \betap \left( F_{v4}(s,t) + \betaf F_{a4}(s,t)\right) \Bigr],
\nll
{\cal H}_{-++} &=&-k_{-}
       \Bigr[4\frac{\mmf}{s}\left( F_{v2}(s,t) - F_{a2}(s,t) \right)
\nll &&
                  -  \betam \left( F_{v3}(s,t) - \betaf F_{a3}(s,t)\right)
                  -  \betap \left( F_{v4}(s,t) - \betaf F_{a4}(s,t)\right) \Bigr]
\eqa
with the coefficients
\bqa
k_0     &=& -e\,g\,\frac{Q_f m_f}{\mw}\,\frac{\sin\vartheta_{\gamma}}{2\sqrt{2}}(s-M^2_{\sss H}),
\nll
k_{\pm} &=& -e\,g\,\frac{Q_f m_f}{\mw}\,\frac{1\pm\cos\vartheta_{\gamma}}{4\sqrt{2}}
               (s-M^2_{\sss H})\sqrt{s}\,.
\eqa
Furthermore, $\beta_f$ as in Eq.~(\ref{betaf}) and 
\bq
\beta_{\pm}=1\pm \beta_f.
\eq
The angle $\vartheta_{\gamma}$ is the cms angle of the produced photon (angle between
$\vec{p}_2$ and $\vec{p}_3$).

For the sake of completeness we also present the amplitude in the Born approximation. 
In terms of structures (\ref{StrffHA}) it reads: 
\bqa
{\cal A}^{\sss Born}_{\sss ffH\gamma}
       &=& -e\,g\,\frac{Q_f m_f}{\mw}\,\frac{1}{(T^2+m^2_f)(U^2+m^2_f)}\,
\nll[1mm]
       & &\times
\iap{p_1}\biggl\{\left[(U^2+m^2_f)(p_2)_{\nu}-(T^2+m^2_f)(p_1)_{\nu}\right]
     +\frac{1}{2}(Q^2+\mh^2)\sla{p_3}\gamma_{\nu}\biggr\}\ip{p_2}\varepsilon^{\gamma}_{\nu}(p_3)\,,
\eqa
where
\bq
Q^2=(p_1+p_2)^2=-s.
\eq

\clearpage

\section{Precomputation\label{precomputation}}
\subsection{Introduction}
This section is devoted to a rather detailed description of {\em precomputation} in {\tt SANC}.
The concept of precomputation is very important for the {\tt SANC} project 
(see Ref.~\cite{Bardin:2002gs}).
The basic idea here is to precompute as many one-loop diagrams and derived
quantities (like renormalization constants, various building blocks {\it etc.})
as possible since the CPU 
time needed is in general quite large for the above quantities making it
impractical to compute them in each {\tt SANC} run.

Recall our particle notation conventions:\\
--- $f$ stands for any fermion (lepton or quark);\\
--- $b$ stands for neutral bosons $A,Z,H$;\\
--- when we need to be more concrete, we use $l$ for leptons instead of $f$,
    and precisely $A,Z,W,H$ for\\ 
\phantom{---} bosons.

\begin{floatingfigure}{75mm}
\hspace*{-6mm}
\includegraphics[width=75mm,height=120mm]{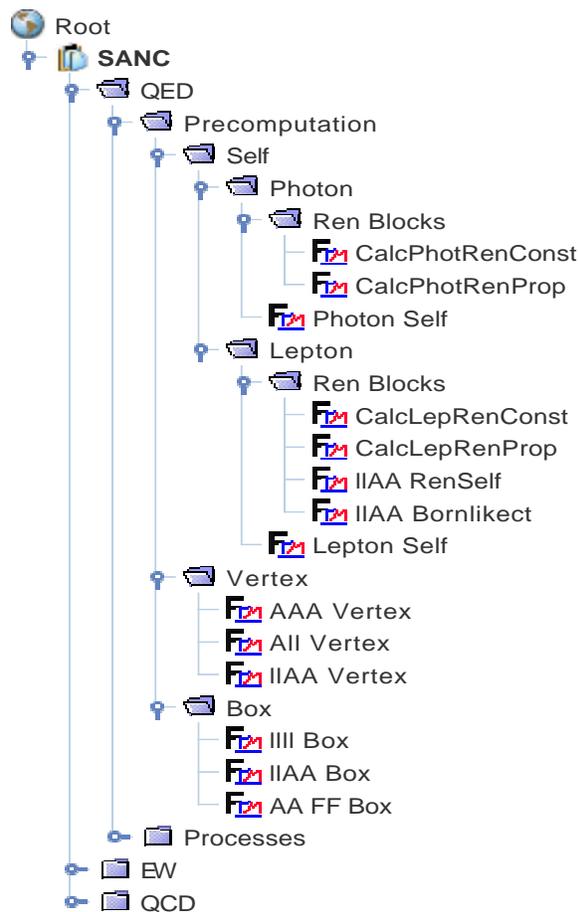}
\caption[Precomputation in QED part]
{Precomputation in QED part.}\label{PrecQED}
\vspace*{4mm}
\end{floatingfigure}

It is worth emphasizing that at the precomputation phase it is not necessary to distinguish the 
process channel. While computing one-loop diagrams  all 4-momenta are considered as incoming.
In the derived expressions (say for the scalar form factors) any required channel is  obtained
by means of an appropriate permutation of arguments (say of Mandelstam variables $s,t,u$).  

The Fig.~\ref{PrecQED} shows the fully open menu for ``Precomputation'' in the QED branch of 
{\tt SANC}.

It consists of {\bf Self} (Energies), {\bf Vertex} and {\bf Box} submenus.
Self energies, in turn, are subdivided into {\bf Photon} and {\bf Lepton} submenus.
They are further subdivided into calculation of diagrams themselves: {\bf Photon Self} and 
{\bf Lepton Self}. Precomputed and stored one-loop diagrams are used for calculation 
of corresponding renormalization constants: {\bf CalcPhot\-RenConst} 
and {\bf Calc\-LepRenConsts},
which are also stored. Altogether they are used for the calculation of renormalized propagators: 
{\bf CalcPhot\-RenProp} and {\bf CalcLepRenProp}. The latter is used to calculate the renormalized
self energy 4-leg diagram for Compton scattering in QED {\bf llAA RenSelf}.
The file {\bf llbb Bornlikect} computes Born-like counterterms (see Section~\ref{fse-ffbb}).

{\bf Vertex} consists of 3-photon-leg {\bf AAA}, photon-2 lepton {\bf All} and 4-leg
vertices for NC $llAA$ Compton-like QED processes 
(any channel, see Section~\ref{vert-ffbb}).

{\bf Box} is represented by the NC 2-photon exchange 4-lepton-leg box (direct and crossed) 
{\bf llll Box} (see Section~\ref{ffff-box}) and by the two topologies (T2 and T4) of boxes
appearing in Compton-like processes {\bf llAA Box} (see Section~\ref{ffbb-box}).
Finally, the file {\bf AA FF Box} transforms results obtained by {\bf llll Box} into the
scalar form factors of a $4l$ process.

\clearpage

The tree of ``Precomputation'' in the EW part, which is shown in Fig.~\ref{PrecEW},
has many more branches.

\begin{floatingfigure}{85mm}
\hspace*{-6mm}
\includegraphics[width=73mm,height=190mm]{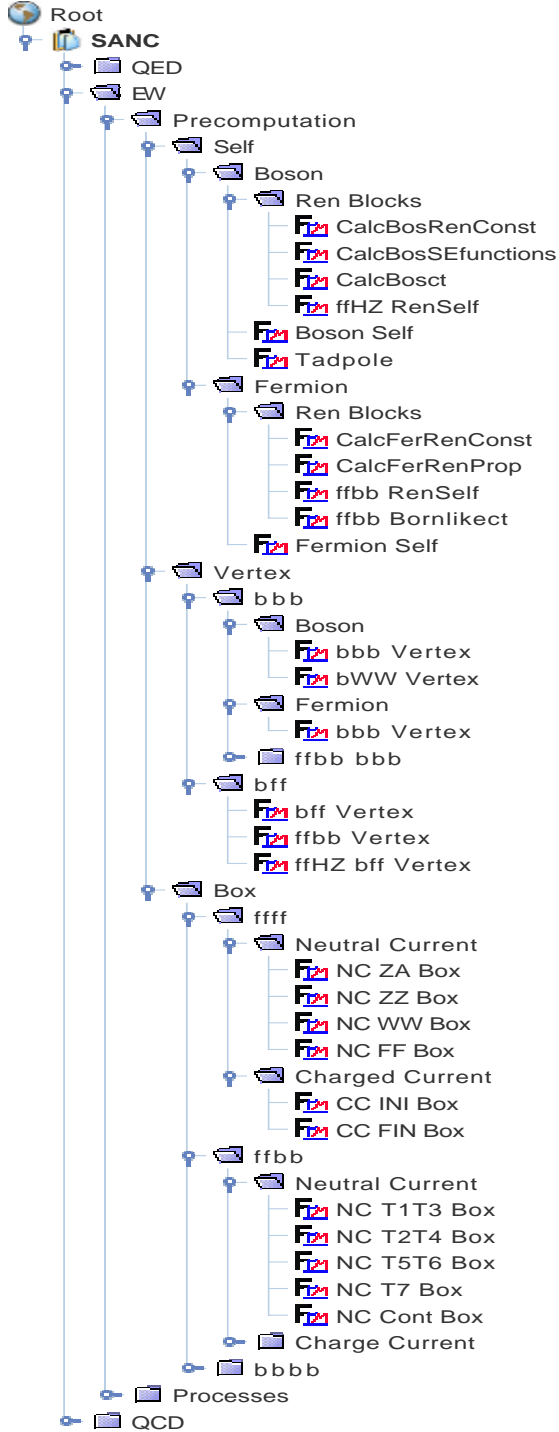}
\caption[Precomputation in EW part]
{Precomputation in EW part.}\label{PrecEW}
\vspace*{-10mm}
\end{floatingfigure}

It also consists of {\bf Self}, {\bf Vertex} and {\bf Box} sub-items.

Self energies are subdivided into {\bf Boson} and {\bf Fermion} submenus.
They are further subdivided into calculation of diagrams themselves: {\bf Boson Self} 
and {\bf Tadpole} (section~\ref{bse}) and {\bf Fermion Self} (section~\ref{fse}). 
Precomputed and stored one-loop diagrams are used to calculate 
the corresponding renormalization constants: {\bf CalcBosRenConst} and 
{\bf CalcFerRenConsts},
which are also stored. They are all used to calculate the ingredients of renormalized 
bosonic propagators: {\bf CalcBosSEfunctions} and {\bf CalcBosct} and renormalized fermi\-onic
propagators {\bf CalcFerRenProp}. The latter is used to calculate the renormalized
self energy 4-leg diagram for $ffbb$ processes by {\bf ffbb RenSelf}; 
{\bf ffbb Bornlikect} computes Born-like counterterms (section~\ref{fse-ffbb}).

{\bf Vertex} consists of 3-boson-leg {\bf bbb} and boson-2-fermion {\bf bff} vertices.
Vertices {\bf bbb} contain {\bf Boson}ic and {\bf Fermion}ic components and {\bf bbb} vertices
for {\bf ffbb} processes (closed here). The former is further 
subdivided into any neutral leg {\bf bbb Vertex} and in particular any neutral boson to $W^+W^-$
{\bf bWW Vertex} (section~\ref{vert-bbb}).
Vertices {\bf bff} are subdivided into any $bff$ 3-leg vertices and 4-leg
vertices for NC $ffbb$ processes (section~\ref{vert-bff}).

{\bf Box} is subdivided into three large classes: {\bf ffff}, {\bf ffbb} and {\bf bbbb} each
of them is subdivided further into NC and CC boxes. The {\bf ffff} class contains a rich collection 
of 4-fermion-leg NC and CC boxes, direct and crossed, (section~\ref{ffff-box}). 
The {\bf ffbb} family is presented by now by seven topologies of boxes ({\bf NC T1--T7 Box})
appearing in NC $ffbb$ processes (section~\ref{ffbb-box}). 

The file {\bf NC FF Box} transforms results obtained by {\bf ffff} into the 
scalar form factors of a $4f$ process, while {\bf NC Cont Box} reali\-zes some further manipulations
with NC $ffbb$ boxes (section~\ref{ffbb-box}).

The files intended for {\bf Charged Current} boxes for $ffbb$ processes and for {\bf bbbb} boxes
exist but are not added for the time being.

\clearpage

\subsection{Self energies}
\subsubsection{Bosonic self energy\label{bse}}
Self energies are the simplest one-loop diagrams.
There are three topologies of bosonic self-energy:
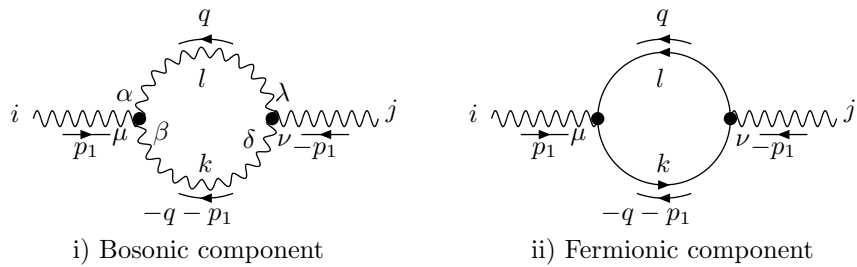
\begin{figure}[!h]
\vspace*{-2mm}
\[
\begin{array}{ccc}
  \vcenter{\hbox{
  \begin{picture}(150,80)(0,0)
  \Photon(0,50)(40,50){3}{7}
  \Photon(90,50)(130,50){3}{7}
  \PhotonArc(65,50)(25,0,180){2}{12}
  \PhotonArc(65,50)(25,-180,0){2}{12}

  \ArrowArc(65,50)(  30,70,110)
  \ArrowArc(65,50)( -30,70,110)

  \ArrowLine(11,44)(29,44)
  \ArrowLine(119,44)(101,44)

\Text( 26.5,40)[lb]{ $\mu $}
\Text( 89,40  )[lb]{ $\nu $}

\Text( 88,57 )[lb]{ $\lambda$}
\Text( 76,39 )[lb]{ $\delta $}
\Text( 28,57 )[lb]{ $\alpha $}
\Text( 42,40 )[lb]{ $\beta  $}

\Text( -12,49)[lb]{ $ i     $}
\Text( 130,49)[lb]{ $ j     $}

\Text( 59,62 )[lb]{ $ l     $}
\Text( 59,30 )[lb]{ $ k     $}

\Text( 38,10 )[lb]{ $-q-p_1 $}
\Text( 59,85 )[lb]{ $ q     $}
\Text( 12,35 )[lb]{ $ p_1   $}
\Text( 98,35 )[lb]{$ -p_1   $}

 \Vertex(90,50){2.5}
 \Vertex(40,50){2.5}
\Text(15,-5)[lb]{i) Bosonic component}
  \end{picture}}}
&&  
  \vcenter{\hbox{
  \begin{picture}(150,80)(0,0)

  \ArrowLine(11,44)(29,44)
  \ArrowLine(119,44)(101,44)

  \ArrowArc(65,50)(25,0,180)

  \ArrowArc(65,50)(  30,70,110)
  \ArrowArc(65,50)( -30,70,110)

  \ArrowArc(65,50)(25,-180,0)

  \Photon(0,50)(40,50){3}{7}
  \Photon(90,50)(130,50){3}{7}

\Text( 26.5,40)[lb]{ $\mu $}
\Text( 89,40  )[lb]{ $\nu $}

\Text( 38,10 )[lb]{ $-q-p_1$}
\Text( 59,85 )[lb]{ $ q    $}
\Text( 59,62 )[lb]{ $ l    $}
\Text( 59,30 )[lb]{ $ k    $}

\Text( -12,49)[lb]{ $ i    $}
\Text( 130,49)[lb]{ $ j    $}

\Text( 12,35)[lb]{ $ p_1   $}
\Text( 98,35 )[lb]{$ -p_1  $}
 \Vertex(90,50){2.5}
 \Vertex(40,50){2.5}
\Text(15,-5)[lb]{ii) Fermionic component}
  \end{picture}}}
\end{array}
\]
\vspace*{-5mm}
\caption[Bosonic self energy, two point diagrams] 
        {Two point diagrams.}
\label{fig:two_bse}
\end{figure}

\begin{figure}[!h]
\vspace*{-6mm}
\[
  \vcenter{ \hbox{
  \begin{picture}(150,80)(0,0)

  \ArrowArc(65,50)(  30,70,110)

  \ArrowLine(11,12)(29,12)
  \ArrowLine(119,12)(101,12)

  \PhotonArc(65,50)(25,0,180){2}{12}
  \PhotonArc(65,50)(25,-180,0){2}{12}

  \Photon(0,19)(65,19){3}{8}
  \Photon(65,19)(130,19){3}{8}

\Text( 70,34 )[lb]{ $\beta  $}
\Text( 48,34 )[lb]{ $\alpha $}

\Text( 70,8 )[lb]{ $\nu $}
\Text( 48,8 )[lb]{ $\mu $}

\Text( -12,16)[lb]{ $ i $}
\Text( 130,16)[lb]{ $ j $}

\Text( 12,3)[lb]{ $ p_1  $}
\Text( 98,3 )[lb]{$ -p_1 $}

\Text( 59,62 )[lb]{ $ l $}
\Text( 59,85 )[lb]{ $ q $}

 \Vertex(67,21){2.5}
\Text(30,-12)[lb]{iii) Bubbles}
  \end{picture}}}
\]
\vspace*{-3mm}
\caption[Bosonic self energy, one point diagrams]
        {One point bosonic diagrams.}
\label{fig:one_bse}
\end{figure}
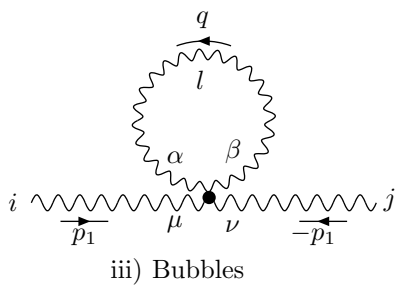

\begin{figure}[!h]
\vspace*{-6mm}
\[
\begin{array}{ccc}
 \vcenter{ \hbox{
  \begin{picture}(150,80)(0,0)

  \ArrowArc(65,50)(  30,70,110)

  \ArrowLine(11 ,-21)(29 ,-21)
  \ArrowLine(119,-21)(101,-21)

  \PhotonArc(65,50)(25,0,180){2}{12}
  \PhotonArc(65,50)(25,-180,0){2}{12}

  \Photon(0 ,-15)(65 ,-15){3}{8}
  \Photon(65,-15)(130,-15){3}{8}

 \Vertex(67, 21){2.5}
 \Vertex(67,-12){2.5}
 \Line(67,-12)(67,-7)
 \Line(67,-2)(67,3)
 \Line(67,8  )(67,12)
 \Line(67,17 )(67,21)

\Text( 72,33 )[lb]{$\beta  $}
\Text( 54,33 )[lb]{$\alpha $}

\Text( 72,-26)[lb]{$\nu $}
\Text( 52,-26)[lb]{$\mu $}

\Text( -9 ,-18)[lb]{$ i $}
\Text( 134,-18)[lb]{$ j $}

\Text( 16,-30)[lb]{$ p_1 $}
\Text( 98,-30)[lb]{$-p_1 $}

\Text( 64,62 )[lb]{$ l $}
\Text( 64,85 )[lb]{$ q $}

\Text(50, 12 )[lb]{$(\delta ) $}
\Text(50,-10 )[lb]{$(\lambda) $}
\Text(15,-50)[lb]{iv) Bosonic component}
  \end{picture} } }
&&
  \vcenter{ \hbox{
  \begin{picture}(150,80)(0,0)

  \ArrowArc(65,50)(  30,70,110)

  \ArrowLine(11 ,-21)(29 ,-21)
  \ArrowLine(119,-21)(101,-21)

   \ArrowArc(65,50)(  25,270,269.999)

  \Photon(0 ,-15)(65 ,-15){3}{8}
  \Photon(65,-15)(130,-15){3}{8}

 \Vertex(67, 24){2.5}
 \Vertex(67,-12){2.5}
 \Line(67,-12)(67,-7)
 \Line(67,-2)(67,3)
 \Line(67,8  )(67,12)
 \Line(67,17 )(67,21)

\Text( 72,-26)[lb]{$\nu $}
\Text( 52,-26)[lb]{$\mu $}

\Text( -9 ,-18)[lb]{$ i $}
\Text( 134,-18)[lb]{$ j $}

\Text( 16,-30)[lb]{$ p_1 $}
\Text( 98,-30)[lb]{$-p_1 $}

\Text( 64,62 )[lb]{$ l $}
\Text( 64,85 )[lb]{$ q $}

\Text(50, 12 )[lb]{$(\delta )$}
\Text(50,-10 )[lb]{$(\lambda)$}
\Text(15,-50)[lb]{v) Fermionic component}
  \end{picture}}}
\end{array}
\]
\vspace*{11mm}
\caption[Bosonic self energy, tadpoles] 
        {Tadpoles.}
\vspace*{-2mm}
\label{fig:tad_bse}
\end{figure}
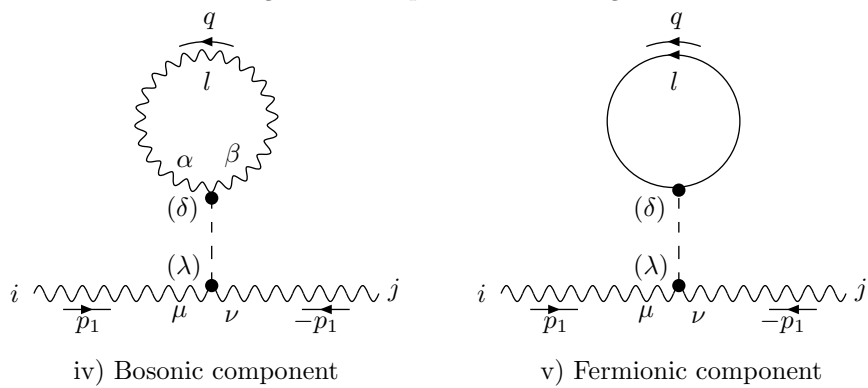

\noindent For the first and the third topologies it is useful to distinguish
{\em bosonic} and {\em fermionic} components.

All these diagrams are precomputed and stored in the {\tt BS.sav} file. 
At user request they may be recomputed by a FORM program
accessible via the menu sequence
{\bf EW $\to$ Precomputation $\to$ Self $\to$ Boson $\to$ Boson Self}.
The five types of self energies are defined in the specific procedure {\sf Diagb(i,j)}
(see the definitions of procedure types in the Introduction).

\noindent
To give an example of what the precomputation really does, we present a simplified version of
a very similar module accessible via the menu sequence
{\bf QED $\to$ Precomputation $\to$ Self $\to$ Photon $\to$ Photon Self}.
It starts with the defining expression {\sf vacpol} for the photonic vacuum polarization, 
Fig.\ref{fig:two_bse}~ii) and continues in five steps of calls to the intrinsic procedures
{\sf (i) FeynmanRules, (ii) Diracizing, (iii) GammaTrace, (iv) Reduction and (v) Scalarizing} 
described in Section \ref{pintr}. After steps 1-4) we show the intermediate results and after
step 5) --- the final result.
\vspace*{-5mm}

\newpage

\begin{verbatim}
#include Declar.h
#call Globals()
 
g vacpol = -Tr*vert(+1,12,-12,mu,ii)*pr(12,q,ii)
              *vert(-1,12,-12,nu,ii)*pr(12,q+p1,ii)*int;
(i) 
#call FeynmanRules(0)

vacpol =
+ den(1,mel,q)*den(1,mel,q+p1)*e^2*qel^2*int*Tr* 
(-gd(ii,mu)*gd(ii,q)*gd(ii,nu)*i_*mel-gd(ii,mu)*gd(ii,q)*gd(ii,nu)*gd(ii,q) 
 -gd(ii,mu)*gd(ii,q)*gd(ii,nu)*gd(ii,p1)+gd(ii,mu)*gd(ii,nu)*mel^2 
 -gd(ii,mu)*gd(ii,nu)*gd(ii,q)*i_*mel-gd(ii,mu)*gd(ii,nu)*gd(ii,p1)*i_*mel );
(ii) 
#call Diracizing(0)

vacpol =
+ den(1,mel,q)*den(1,mel,q+p1)*e^2*qel^2*int*Tr* 
(-2*gd(ii,mu)*gd(ii,al1)*qf(al1)*qf(nu)-gd(ii,mu)*gd(ii,al1)*gd(ii,nu)*qf(al1)*i_*mel
 -gd(ii,mu)*gd(ii,al1)*gd(ii,nu)*gd(ii,p1)*qf(al1)+gd(ii,mu)*gd(ii,nu)*mel^2
 +gd(ii,mu)*gd(ii,nu)*q.q-gd(ii,mu)*gd(ii,nu)*gd(ii,p1)*i_*mel 
 -gd(ii,mu)*gd(ii,nu)*gd(ii,al1)*qf(al1)*i_*mel );
(iii)
#call GammaTrace(1)

vacpol =
+ den(1,mel,q)*den(1,mel,q+p1)*e^2*qel^2*int*
( 4*d_(mu,nu)*mel^2+4*d_(mu,nu)*q.q+4*d_(mu,nu)*q.p1 
 -4*qf(mu)*p1(nu)-8*qf(mu)*qf(nu)-4*qf(nu)*p1(mu) );
(iv)
#call Reduction(0)

vacpol =
+ p1(mu)*p1(nu)*e^2*qel^2*
(-8*b21(p1s,1,mel,1,mel)-8*b1(p1s,1,mel,1,mel))
 +d_(mu,nu)*e^2*qel^2*
(4*a0(1,mel)-8*b22(p1s,1,mel,1,mel)+4*b1(p1s,1,mel,1,mel)*p1s);
(v)
#call Scalarizing(0)

vacpol =
+ p1(mu)*p1(nu)*e^2*qel^2*
(-4/9-8/3*p1s^-1*mel^2-8/3*a0(1,mel)*p1s^-1+4/3*b0(p1s,1,mel,1,mel)
 -8/3*b0(p1s,1,mel,1,mel)*p1s^-1*mel^2 )
+ d_(mu,nu)*e^2*qel^2 * 
( 4/9*p1s + 8/3*mel^2 + 8/3*a0(1,mel) - 4/3*b0(p1s,1,mel,1,mel)*p1s 
 +8/3*b0(p1s,1,mel,1,mel)*mel^2 );
\end{verbatim}

\noindent We continue describing the 
\underline{structure of the basic program {\bf Boson Self}:}\\
The calculation of all bosonic self-energies (diagram-by-diagram) is done by two calls 
to specific procedure {\sf CalcBos}(i,j), with \{$i=3,\,6$, $j=i$\}, corresponding
to the {\sf charged} external bosons $W^\pm$ and $\phi^\pm$,
four calls to {\sf CalcBos}(i,j) with \{$i=1,2,4,5$, $j=i$\}, corresponding to
$\gamma$, $Z$, $H$ and $\phi^0$, and one call to {\sf CalcBos}(i,j) with
\{$i=2$, $j=1$\}, corresponding to external bosons $Z$ and $\gamma$, respectively.

Procedure {\sf CalcBos}(i,j) calls specific procedure {\sf Diagb}(i,j) and
seven {\it intrinsic procedures}:\linebreak
(i) {\sf FeynmanRules}, (ii) {\sf GammaRight}, (iii) {\sf Diracizing}, (iv) 
{\sf GammaTrace}, (v) {\sf Reduction}, (vi) {\sf Sing} and (vii) {\sf Scalarizing}. 
The intrinsic procedures are described in Section \ref{pintr}.

At each call to procedure {\sf CalcBos}(i,j), the diagrams
Figs. \ref{fig:two_bse}, \ref{fig:one_bse}, \ref{fig:tad_bse} are calculated
for twelve virtual fermions field indices $k,l$ running over $11,\,12\,\ldots\,22$ 
(corresponding to 
$\nu_e$, $e$, $u$, $d$, $\nu_\mu$, $\mu$, $c$, $s$, $\nu_\tau$, $\tau$, $t$ and $b$, 
respectively) and for ten virtual bosons $k,l=1,\,2,\,\ldots,\,10$, corresponding to
$\gamma$, $Z$, $W^\pm$, $H$, $\phi^0$, $\phi^\pm$ and four Faddeev--Popov ghosts
$X^{+}$,$X^{-}$, $Y_{\sss Z}$, $Y_{\sss A}$, respectively.

The topology of the self-energy diagrams is specified in procedure {\sf Diagb(i,j)}
in terms of the {\it vertices} and {\it propagators} of the diagrams.
In the class of diagrams available at  present, {\it vertices} are of two kinds:
boson-fermion-fermion (Bff) and three-boson (BBB) vertices.
The diagrams are computed in nested loops over all allowed
field indices of the virtual particles.

Two more FORM programs are accessible via menu sequences\\[.5mm]
{\bf QED $\to$ Precomputation $\to$ Self $\to$ Photon $\to$ Boson Self $\to$ Photon Self} 
and \\[.5mm]
{\bf EW $\to$ Precomputation $\to$ Self $\to$ Boson $\to$ Tadpoles}.\\[.5mm]
They have very similar structures and compute respectively photonic self energy
(vacuum polarization) 
in the QED tree of {\tt SANC} and tadpole diagrams separately from self energies. 
The results are stored in {\tt PSqed.sav} and {\tt TP.sav} files, respectively.

Precomputed bosonic self-energies are used by the other FORM programs which calculate bosonic
counterterms \\[.5mm]
{\bf EW $\to$ Precomputation $\to$ Self $\to$ Boson $\to$ Ren Blocks $\to$ CalcBosRenConst} \\[.5mm]
and the photonic counterterm\\[.5mm]
{\bf QED $\to$ Precomputation $\to$ Self $\to$ Photon $\to$ CalcPhotRenConst} \\[.5mm]
in the QED tree; bosonic self energy functions \\[.5mm] 
{\bf EW $\to$ Precomputation $\to$ Self $\to$ Boson $\to$ Ren Blocks $\to$ 
CalcBosSEFunctions},\\[.5mm]
and the renormalized photonic propagator in the QED branch \\[.5mm]
{\bf QED $\to$ Precomputation $\to$ Self $\to$ Photon $\to$ CalcPhotRenProp}. \\[.5mm]
Their results, in turn, are used by a FORM program which computes
{\em counter term blocks (crosses)}, see Fig. 15 of Ref.~\cite{Andonov:2002xc}.
The latter is accessible via the chain \\
{\bf EW $\to$ Precomputation $\to$ Self $\to$ Boson $\to$ Ren Blocks $\to$ CalcBosct}.

\subsubsection{Fermionic self energy\label{fse}}
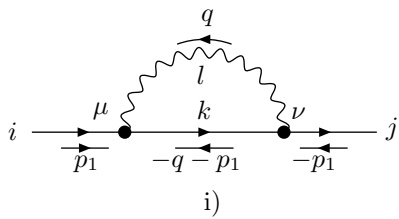
\begin{figure}[!h]
\vspace*{-21mm}
\[
  \vcenter{ \hbox{
  \begin{picture}(150,80)(0,0)

  \ArrowLine(11 ,-21)(29 ,-21)
  \ArrowLine(119,-21)(101,-21)

  \Line(0 ,-15)(65 ,-15)
  \Line(65,-15)(130,-15)
  \ArrowLine(19 ,-15)(20 ,-15)
  \ArrowLine(64 ,-15)(66 ,-15)
  \ArrowLine(76 ,-21)(54 ,-21)
  \ArrowLine(110,-15)(111,-15)

\Text( -9 ,-18)[lb]{$ i $}
\Text( 134,-18)[lb]{$ j $}

\Text( 16,-30)[lb]{$  p_1 $}
\Text( 98,-30)[lb]{$ -p_1 $}
\Text( 45,-30)[lb]{$ -q-p_1 $}

 \Text(23, -10 )[lb]{$ \mu $}
 \Text(98, -10 )[lb]{$ \nu $}

\Text( 59, 3)[lb]{ $ l $}
\Text( 59,-10)[lb]{ $ k $}

 \ArrowArc(65,-10)(30,70,110)
 \Text( 64,25 )[lb]{$ q $}
 \Vertex(35,-15){2.5}
 \Vertex(95,-15){2.5}

  \PhotonArc(65,-15)(30,0,180){2}{12}
\Text(65,-47)[lb]{i)}
  \end{picture} } }
\]
\vspace*{12mm}
\caption[Fermionic self energy, two point diagrams]
        {Two point diagrams.}
\label{TwoPointFermionicDiagrams}
\vspace*{-5mm}
\end{figure}

\begin{figure}[!t]
\vspace*{-5mm}
\[
\begin{array}{ccc}
  \vcenter{ \hbox{
  \begin{picture}(150,80)(0,0)

  \ArrowArc(65,50)(  30,70,110)

  \ArrowLine(11 ,-20)(29 ,-20)
  \ArrowLine(119,-20)(101,-20)

  \PhotonArc(65,50)(25,0,180){2}{12}
  \PhotonArc(65,50)(25,-180,0){2}{12}

  \Line(0 ,-15)(65 ,-15)
  \Line(65,-15)(130,-15)
  \ArrowLine(19 ,-15)(20 ,-15)
  \ArrowLine(110,-15)(111,-15)

 \Vertex(67, 21){2.5}
 \Vertex(67,-15){2.5}
 \Line(67,-12)(67,-7)
 \Line(67,-2)(67,3)
 \Line(67,8  )(67,12)
 \Line(67,17 )(67,21)

\Text( 72,33 )[lb]{$\beta  $}
\Text( 54,33 )[lb]{$\alpha $}

\Text( -9 ,-18)[lb]{$ i $}
\Text( 134,-18)[lb]{$ j $}

\Text( 16,-30)[lb]{$ p_1 $}
\Text( 98,-30)[lb]{$-p_1 $}

\Text( 64,62 )[lb]{$ l $}
\Text( 64,85 )[lb]{$ q $}

\Text(50, 12 )[lb]{$(\delta ) $}
\Text(50,-10 )[lb]{$(\lambda) $}
\Text(15,-50)[lb]{ii) Bosonic component}
  \end{picture} } }
&&
  \vcenter{ \hbox{
  \begin{picture}(150,80)(0,0)

  \ArrowArc(65,50)(  30,70,110)

  \ArrowLine(11 ,-21)(29 ,-21)
  \ArrowLine(119,-21)(101,-21)

   \ArrowArc(65,50)(  25,270,269.999)

 \Vertex(67, 24){2.5}
 \Line(67,-12)(67,-7)
 \Line(67,-2)(67,3)
 \Line(67,8  )(67,12)
 \Line(67,17 )(67,21)

 \Vertex(67,-15){2.5}
  \Line(0 ,-15)(65 ,-15)
  \Line(65,-15)(130,-15)
  \ArrowLine(19 ,-15)(20 ,-15)
  \ArrowLine(110,-15)(111,-15)

\Text( -9 ,-18)[lb]{$ i $}
\Text( 134,-18)[lb]{$ j $}

\Text( 16,-30)[lb]{$ p_1 $}
\Text( 98,-30)[lb]{$-p_1 $}

\Text( 64,62 )[lb]{$ l $}
\Text( 64,85 )[lb]{$ q $}

\Text(50, 12 )[lb]{$(\delta )$}
\Text(50,-10 )[lb]{$(\lambda)$}
\Text(15,-50)[lb]{iii) Fermionic component}
  \end{picture} } }
\end{array}
\]
\vspace*{15mm}
\caption[Fermionic self energy, tadpoles] 
        {Tadpoles.}
\label{FerSETadpoles}
\end{figure}
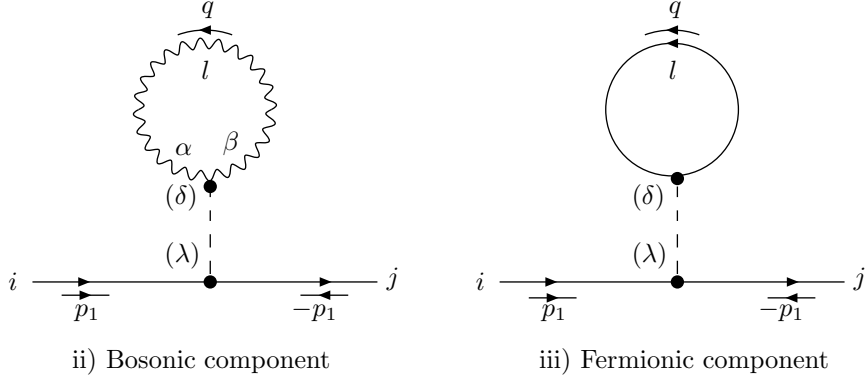

There are two topologies of fermion self energy diagrams 
(two point diagrams Fig.~\ref{TwoPointFermionicDiagrams} and tadpoles Fig.~\ref{FerSETadpoles}).

\clearpage

These diagrams are precomputed and stored in the {\tt FS.sav} file. 
They may be recomputed by a FORM program accessible via the menu sequence\\[.5mm]
{\bf EW $\to$ Precomputation $\to$ Self $\to$ Fermion $\to$ Fermion Self}.\\[.5mm]
The three types of diagrams are defined in the specific procedure {\sf Diagf(i,j)}.

\noindent \underline{Structure of the program {\bf Fermion Self}:}\\
The calculation is done by 12 calls to specific procedure {\sf CalcFer}(i,j), with 
$i=j=11,\,12\,\ldots,\,22$, corresponding to all fermions of four generations, respectively.

Procedure {\sf CalcFer(i,i)} calls  specific procedure {\sf Diagf(i,j)} and
seven intrinsic procedures (i) {\sf FeynmanRules}, (ii) {\sf GammaRight}, (iii) {\sf Diracizing}, 
(iv) {\sf GammaTrace}, (v) {\sf Reduction}, (vi) {\sf Sing} and (vii) {\sf Scalarizing},
see Section~\ref{pintr}.

At each call to procedure {\sf CalcFer}(i,i), the diagrams
Figs. \ref{TwoPointFermionicDiagrams} and \ref{FerSETadpoles}
are calculated for twelve virtual fermions $i=11,\,12\,\ldots,\,22$
and only six virtual bosons $l=1,\,2,\,\ldots,\,6$, corresponding to
$\gamma$, $Z$, $W^\pm$, $h$, $\phi^0$ and $\phi^\pm$, respectively, since Faddeev--Popov ghosts
do not contribute here. 

The topology of the self-energy diagrams is specified in procedure {\sf Diagf(i,j)}
in terms of the vertices and propagators of the diagrams as well as {\it sign} and 
{\it combinatorial} factors ({\sf sign(`l')} and {\sf cft(`l')}).

There is one more FORM program, accessible via the menu sequence \\[1mm]
{\bf QED $\to$ Precomputation $\to$ Self $\to$ Lepton $\to$ Lepton Self} \\[1mm] 
whose structure is very similar to program Fermion Self. It computes the leptonic
self energy in the QED tree of {\tt SANC}.
The results are stored in {\tt LSqed.sav} file.
Precomputed fermionic self-energies are used by FORM programs which calculate fermionic
counterterms  \\[1mm]
{\bf EW $\to$ Precomputation $\to$ Self $\to$ Fermion $\to$ Ren Blocks $\to$ CalcFerRenConst}\\[1mm]
and leptonic counterterms\\[1mm]
{\bf QED $\to$ Precomputation $\to$ Self $\to$ Lepton $\to$ CalcLepRenConst}. \\[1mm]
Both fermionic self-energy diagrams and fermionic counterterms
are used by FORM programs which compute the renormalized fermionic (leptonic in the QED branch) 
propagators \\[1mm]
{\bf EW $\to$ Precomputation $\to$ Self $\to$ Fermion $\to$ Ren Blocks $\to$ CalcFerRenProp} 
and \\[1mm] 
{\bf QED $\to$ Precomputation $\to$ Self $\to$ Lepton $\to$ CalcLepRenProp}, respectively.

In the FORM code {\bf CalcLepRenProp} we show the main steps of the calculations.
The two latter codes end up demonstrating the vanishing of the renormalized propagator
on the corresponding fermionic mass shell, as is required by the on-mass-shell (OMS)
renormalization scheme, see Section 2.2 of~\cite{Andonov:2002xc}.

\subsubsection{Fermionic self energy for $ffbb$ processes\label{fse-ffbb}}
Moving to precomputation of the building blocks for  $ffbb$ processes, we change our conventions.
Now any object: self energy, vertex and box are considered to be 4-legs, rather than 2-, 3- and 
4-legs respectively, as we did before. The main motivation for this change is our observation that
vertices with off-shell fermions are inconvenient to treat and the resulting expressions are more 
compact if we consider 4-legs on-mass-shell building blocks instead of 3-legs off-shell.
Accepting this convention for vertices, it is natural to treat self-energies also like 4-leg
objects shown in Fig.~\ref{ffbbSE}:

\begin{figure}[!h]
\vspace{-1mm}
\[
\begin{array}{ccc}
  \vcenter{\hbox{
\begin{picture}(132,132)(0,0)

\Vertex(0,63){6}

 \ArrowLine(55,115)(33,115)
 \ArrowLine(55,17)(33,17)
  
 \Photon(0,110)(55,110){2}{12}

 \Vertex(0,110){2.5}
 \ArrowLine(0,110)(-22,132)
 \Line(0,22)(0,110)
 \ArrowLine(0,35)(0,50)
 \ArrowLine(0,85)(0,90)

 \Vertex(0,22){2.5}
 \ArrowLine(-22,0)(0,22)
 \Photon(0,22)(55,22){2}{12}

 \ArrowLine(-16,0)(0,16)
 \ArrowLine(-16,132)(0,116) 

 \Text(-32,132)[lt]{\sf ii}
 \Text(42,118)[lb]{$p$}    
 \Text(37,3)[lb]{$Q$}
 \Text(-15,106)[lb]{$\nu$}
 \Text(60,105)[lb]{vu=typeFU}
 \Text(60,18)[lb]{vd=typeFD}
 \Text(-42,137)[lb]{fu=typeIU}
 \Text(-42,-12)[lb]{fd=typeID}
 \Text(-15,17)[lb]{$\mu $}
 \Text(-4,123)[lb]{$p_1$}
 \Text(-4,3)[lb]{$p_2$}
 \Text(35,-12)[lb]{a)}
\end{picture}}}
& \qquad &
  \vcenter{\hbox{
\begin{picture}(132,132)(0,0)
 \Vertex(0,63){6}

 \ArrowLine(55,115)(33,115)
 \ArrowLine(55,17)(33,17)
  
 \Photon(0,110)(55,110){2}{12}
 \Vertex(0,110){2.5}
 \ArrowLine(0,110)(-22,132)
 \Line(0,22)(0,110)
 \ArrowLine(0,35)(0,50)
 \ArrowLine(0,85)(0,90)

 \Vertex(0,22){2.5}
 \ArrowLine(-22,0)(0,22)
 \Photon(0,22)(55,22){2}{12}

 \ArrowLine(-16,0)(0,16)
 \ArrowLine(-16,132)(0,116) 

 \Text(-32,132)[lt]{\sf ii}
 \Text(42,118)[lb]{$Q$}    
 \Text(37,6)[lb]{$p$}
 \Text(-15,105)[lb]{$\mu$}
 \Text(60,18)[lb]{vu=typeFU}
 \Text(60,105)[lb]{vd=typeFD}
 \Text(-42,137)[lb]{fu=typeIU}
 \Text(-42,-12)[lb]{fd=typeID}
 \Text(-15,18)[lb]{$\nu $}
 \Text(-4,123)[lb]{$p_1$}
 \Text(-4,3)[lb]{$p_2$}
 \Text(35,-12)[lb]{b)}
\end{picture}}}
\end{array}
\]
\caption[Self energy $ffbb$ diagrams]
        {Self energy $ffbb$ diagrams.}
\label{ffbbSE}
\end{figure}
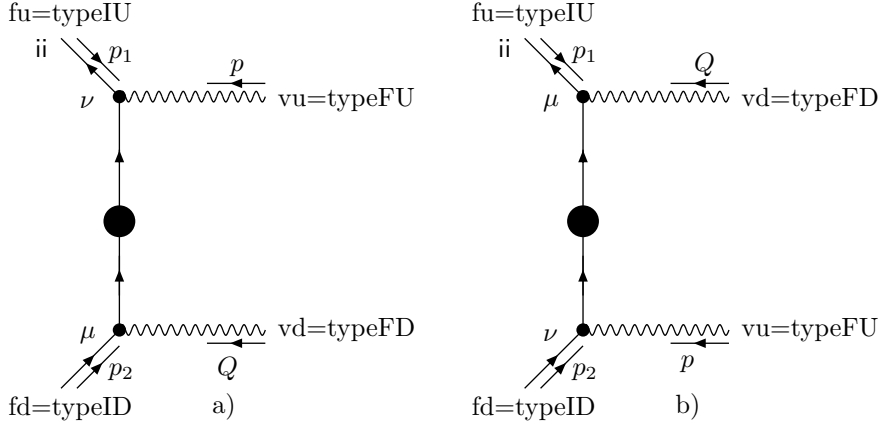
\vspace*{-1mm}

Here the blob at the fermion propagator denotes the sum of all self-energy diagrams
described in Section~\ref{fse}.
These self energies are precomputed by a FORM program accessible via the menu sequence \\
{\bf EW $\to$ Precomputation $\to$ Self $\to$ Fermion $\to$ Ren Blocks $\to$ ffbb RenSelf}.

The required CPU time is still very short, and at user request they may be re-computed. 
The two types of self energies are defined in the specific procedures 
{\sf CalcFerSEt(fu,fd,vd,vu)} and {\sf CalcFerSEu(fu,fd,vu,vd)}, 
which use renormalized fermionic propagators precomputed by {\bf CalcFerRenProp}.
Note that the labels `t' and `u' are associated with the Mandelstam
variables $t,u$, see Section~\ref{proc-ffff}.

Both $t$ and $u$ channel procedures call intrinsic procedures 
{\sf FeynmanRules}, {\sf GammaRight} and {\sf p2I} where the latter, together with several 
{\sf id's} in between, performs obvious identities and change of variables.

Computed diagrams {\sf FSEt`fu'`fd'`vd'`vu'`jl'`k1'} and 
{\sf FSEu`fu'`fd'`vu'`vd'`jl'`k1'} are stored in the file\\
{\sf ffbbSelfxi`xi'`fu'`fd'`vu'`vd'.sav},
where the predefined parameter `xi' has the following meaning:
\vspace*{-3mm}

\begin{verbatim}
#define xi "0" / * .eq.0 to test gauge invariance / * .eq.1 to work in xi=1 gauge
\end{verbatim}
\vspace*{-3mm}

\noindent
This option is introduced mostly to save CPU time since calculation in the {\sf xi=1} gauge 
are much faster than in the $R_\xi$ gauge. However, it is always appealing to see the explicit 
cancellation of gauge parameters. That is why we try to maintain  option {\sf xi "0"}
as long as possible even though in some cases it is extremely time consuming.

There is a similar FORM program in the QED part, accessible via the menu sequence \\
{\bf QED $\to$ Precomputation $\to$ Self $\to$ Lepton $\to$ Ren Blocks $\to$ llAA RenSelf}.

Furthermore, the code, accessible via menu sequence:\\
{\bf EW $\to$ Precomputation $\to$ Self $\to$ Fermion $\to$ Ren Blocks $\to$ ffbb Bornlikect} \\
computes contributions from four diagrams similar to Fig.~\ref{ffbbSE}, in which the self energy 
blob at the fermion propagator is replaced by counterterm `crosses' (one for each of the
four diagrams at each vertex with indices $\mu$ and $\nu$, similar to Fig.~\ref{ffbbvert}).
The result is stored in the file {\sf ffbbBornlikectxi`xi'`fu'`fd'`vu'`vd'.sav}.

The file from
{\bf QED $\to$ Precomputation $\to$ Self $\to$ Lepton $\to$ Ren Blocks $\to$ ffAA Bornlikect}
does the same job in the QED part.
\vspace{-5mm}

\clearpage

\subsection{One-loop vertices}
The current {\tt SANC} version has all $bf\bar{f}$ and $bbb$ 3-leg SM vertices.
\subsubsection{$bf\bar{f}$ vertices\label{vert-bff}}
There are two topologies of $bf\bar{f}$ vertices: {\sf FBF} and {\sf BFB}:
\begin{figure}[!h]
\vspace*{-1mm}
\[
\begin{array}{ccc}
 \vcenter{\hbox{
\begin{picture}(132,132)(20,0)

 \Photon(-20,66)(44,66){2}{7}
    \Vertex(44,66){2.5}
\Text(-72, 64)[lb]{$i=$typeB}
 \Text(-22, 76)[lb]{$Q=-p_1-p_2$}

\Text(61, 75)[lb]{$ k_3 $}
\Text(61, 52)[lb]{$ k_1 $}
\Text(75, 66)[lb]{$ k_2 $}

\Text( 35,55 )[lb]{ $\mu    $}
\Text( 90,22 )[lb]{ $\beta  $}
\Text( 90,105)[lb]{ $\alpha $}

   \ArrowLine(0,71)(21,71) 

   \ArrowLine(44,  66)(88,22)
   \ArrowLine(104,132)(88,116)

   \Text(43,37)[lb]{$-q$}  
   \Text(3,93)[lb]{$q+p_1+p_2$}  

   \ArrowLine(70,34)(54,50)
   \ArrowLine(71,99)(55,83)

\ArrowLine(88,110)(44,66)
    \Vertex(88,110){2.5}
 \Photon(88,110)(88,22){2}{10}

      \ArrowLine(88,22)(110,0)
      \ArrowLine(104,0)(88,16)
      \ArrowLine(94,55)(94,77)

 \ArrowLine(110,132)(88,110)
 \Text(114,132)[lb]{$j=$typeU}
 \Text(114,0)[lb]{$l=$typeD}
 \Text(82,128)[lb]{$p_2$}
 \Text(82,3)[lb]{$p_1$}
\Text(98,66)[lb]{$q+p_1$}
 \Vertex(88,22){2.5}
\Text(0,-10)[lb]{a) Topology FBF}
\end{picture}}}
&\vspace*{5cm}&
  \vcenter{\hbox{
\begin{picture}(132,132)(-40,0)

 \Photon(-20,66)(44,66){2}{7}
    \Vertex(44,66){2.5}
\Text(-72, 64)[lb]{$i=$typeB}
 \Text(5, 76)[lb]{$Q$}

\Text(61, 75)[lb]{$ k_3 $}
\Text(61, 52)[lb]{$ k_1 $}
\Text(75, 66)[lb]{$ k_2 $}

\Text(30,57 )[lb]{$\mu  $}
\Text(39,78 )[lb]{$\mu_3  $}
\Text(39,50 )[lb]{$\mu_1  $}
\Text(93,22 )[lb]{$\beta$}
\Text(93,105)[lb]{$\alpha$}

   \ArrowLine(0,71)(21,71) 

 \Photon(44,66)(88,22){2}{8}
   \ArrowLine(104,132)(88,116)

   \Text(43,37)[lb]{$-q$}  
   \Text(3,93)[lb]{$q+p_1+p_2$}  

   \ArrowLine(70,34)(54,50)
   \ArrowLine(70,98)(54,82)
 \Photon(88,110)(44,66){2}{8}
    \Vertex(88,110){2.5}
\ArrowLine(88,110)(88,22)
      \ArrowLine(88,22)(110,0)
      \ArrowLine(104,0)(88,16)
      \ArrowLine(94,55)(94,77)

 \ArrowLine(110,132)(88,110)
 \Text(114,132)[lb]{$j=$typeU}
 \Text(114,0)[lb]{$l=$typeD}
 \Text(82,128)[lb]{$p_2$}
 \Text(82,3)[lb]{$p_1$}
\Text(98,66)[lb]{$q+p_1$}
 \Vertex(88,22){2.5}
\Text(0,-10)[lb]{b) Topology BFB}
\end{picture}}}
\end{array}
\]
\vspace*{-55mm}
\caption[$bff$ vertices]{$bff$ vertices.}
\label{fig:vffvert}
\end{figure}
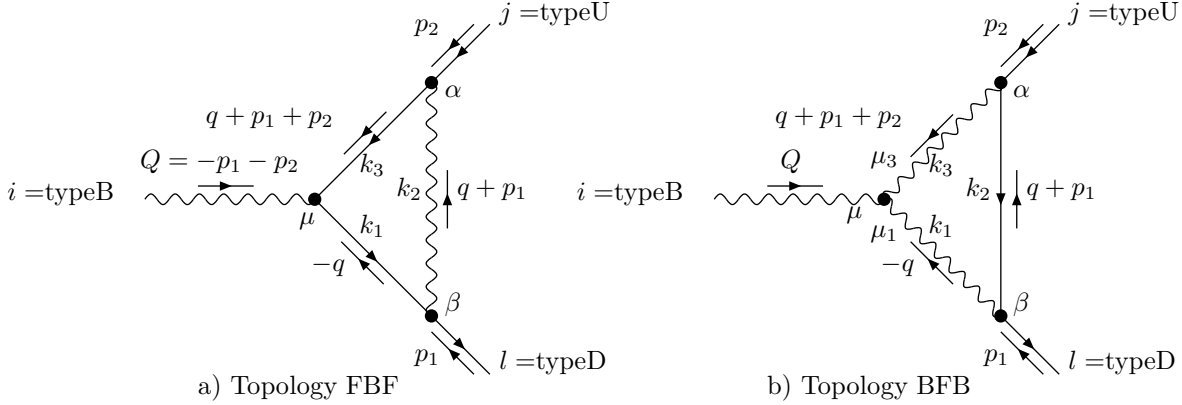

These vertex diagrams are precomputed by a FORM program accessible via menu sequence \\[1mm]
{\bf EW $\to$ Precomputation $\to$ Vertex $\to$ bff $\to$ bff Vertex}.

At user request they may be recomputed. 
The two types of vertices are defined in the specific procedures {\sf FBF(i,j,l)} and
{\sf BFB(i,j,l)}, where the topologies of all vertices of
the type of Fig.~\ref{fig:vffvert} are specified in nested loops over all allowed
field indices of the virtual particles.

\noindent
\underline{Structure of program {\bf bff Vertex}}\\
The calculation is done by a single call to specific procedure
{\sf CalcVertex(`typeB',`typeU',`typeD')}, 
followed by a call to intrinsic procedure {\sf p2D}.

The arguments {\sf typeB}, {\sf typeU}, and {\sf typeD} are the field indices of the
external particles, set in the command that starts this program:
{\sf typeB} is the incoming boson, {\sf typeU} is the antifermion
with incoming 4-momentum $p_2$, and {\sf typeD} is the fermion
with incoming 4-momentum $p_1$.
For instance, to calculate the vertex for $Z\to u\bar{u}$ decay, {\sf typeB},
{\sf typeU}, and {\sf typeD} are set equal to $2,\, 13,\, 13$, respectively.
The results are stored in files {\tt V`typeB'`typeU'`typeD'.sav}.

Procedure {\sf CalcVertex(i,j,l)} calls specific procedures {\sf FBF(i,j,l)} and
{\sf BFB(i,j,l)} and the intrinsic procedures in the following sequence:
{\sf FeynmanRules},  {\sf GammaRight},  {\sf Diracizing}\footnote{For $Wff$ vertices, this
procedure is called exceptionally in the so-called {\em special use mode}, 
see Section~\ref{pintr}. This is due to a delicate treatment of the auxiliary Passarino--Veltman
functions originating from a double pole $1/(p^2)^2$ in the photonic propagator in the $R_\xi$
gauge.}, {\sf Reduction},  {\sf Diracizing}, {\sf Pulling}, {\sf Diraceq}, {\sf Sing}, 
{\sf Scalarizing}, {\sf 2Qs}, {\sf Masshell} and {\sf Symmetrize}.
The intrinsic procedures are described in Section~\ref{pintr}.

In the QED tree there is a similar entry \\[1mm]
{\bf QED $\to$ Precomputation $\to$ Vertex $\to$ Aff $\to$ Aff Vertex}, \\[1mm]
which computes $\gamma l\bar{l}$ vertices for a lepton of a kind {\sf j=typeU=12,16,20}. 
These vertices are defined in the specific procedure {\sf DiagVert(j)}.
The results are stored in files {\tt Vqed`typeU'.sav}.

\subsubsection{$bbb$ vertices\label{vert-bbb}}
The {\em bosonic} component of three-boson vertices has four topologies 
shown in the following diagrams:
\begin{figure}[!h]
\vspace*{-1mm}
\[
\begin{picture}(132,132)(0,0)
 \Photon(-20,66)(44,66){2}{7}
    \Vertex(44,66){2.5}
\Text(-72, 64)[lb]{$i=$typeB}
\Text(-72, 55)[lb]{neutral}
\Text(-22, 76)[lb]{$Q=-p_1-p_2$}
\Text(64, 77)[lb]{$ k_3 $}
\Text(64, 48)[lb]{$ k_1 $}
\Text(75, 63)[lb]{$ k_2 $}
\Text(55, 67)[lb]{$ \beta $}
\Text(55, 55)[lb]{$ \gamma $}
\Text(-25,55 )[lb]{ $\alpha   $}
\Text( 90,28 )[lb]{ $\nu_2  $}
\Text( 65,22 )[lb]{ $\nu_1  $}
\Text( 90,97 )[lb]{ $\mu_1 $}
\Text( 60,103)[lb]{ $\mu_2 $}
   \Text(43,37)[lb]{$-q$}  
   \Text(3,93)[lb]{$q+p_1+p_2$}  
   \ArrowLine(0,71)(21,71) 
\DashLine(44,66)(88,22){3}
\DashLine(88,110)(44,66){3}
\DashLine(88,110)(88,22){3}
   \ArrowLine(104,132)(88,116)
   \ArrowLine(70,34)(54,50)
   \ArrowLine(71,99)(55,83)
    \Vertex(88,110){2.5}
 \Photon(88,22)(110,0){2}{5}
      \ArrowLine(104,0)(88,16)
      \ArrowLine(94,55)(94,77)
 \Photon(110,132)(88,110){2}{5}
 \Text(114,123)[lb]{$\mu$}
 \Text(114,137)[lb]{$j=$typeU}
 \Text(114,-12)[lb]{$l=$typeD}
 \Text(114,3)[lb]{$\nu$}
 \Text(82,123)[lb]{$p_2$}
 \Text(82,3  )[lb]{$p_1$}
 \Text(98,66 )[lb]{$q+p_1$}
 \Vertex(88,22){2.5}
\Text(30,-18)[lb]{Topology T1}
\end{picture}
\]
\vspace*{-5mm}
\[
\begin{picture}(132,132)(0,0)
 \Photon(-20,66)(44,66){2}{7}
   \ArrowLine(0,71)(21,71) 
    \Vertex(44,66){2.5}
\DashCArc(66,66)(22,0,180){3}
\DashCArc(66,66)(22,180,0){3}
\ArrowArc(66,66)( 25,70,110)
\ArrowArc(66,66)(-25,70,110)

 \Vertex(88,66){2.5}
 \Photon(110,88)(88,66){2}{5}
 \ArrowLine(104,88)(89,73)
 \Photon(88,66)(110,44){2}{5}
 \ArrowLine(104,44)(89,59)

 \Text(-72, 64)[lb]{$i=$typeB}
\Text(-72, 55)[lb]{neutral}
\Text(-25, 76)[lb]{$Q=-p_1-p_2$}
\Text(50, 67)[lb]{$ \beta $}
\Text(50, 55)[lb]{$ \gamma $}
\Text(-25,55 )[lb]{ $\alpha   $}
\Text( 68,55 )[lb]{ $\nu_1  $}
\Text( 68,67)[lb]{ $\mu_2 $}
\Text(58,30)[lb]{$-q$}  
\Text(12,94)[lb]{$q+p_1+p_2$}  
\Text( 114,77 )[lb]{$\mu$}
\Text(114,87)[lb]{$j=$typeU}
\Text(114,33)[lb]{$l=$typeD}
\Text(114,45)[lb]{$\nu$}
\Text(91,88)[lb]{$p_2$}
\Text(91,40  )[lb]{$p_1$}
\Text(30,14)[lb]{Topology T2}
\end{picture}
\]
\vspace*{-1.7cm}
\[
\begin{picture}(132,132)(0,0)
 \Photon(-20,66)(44,66){2}{7}
    \Vertex(44,66){2.5}
   \ArrowLine(0,71)(21,71) 
\DashCArc(66,66)(22,0,180){3}
\DashCArc(66,66)(22,180,0){3}
\ArrowArc(66,66)( 25,70,110)
\ArrowArc(66,66)(-25,70,110)

 \Vertex(88,66){2.5}
 \Photon(110,66)(88,66){2}{5}
 \ArrowLine(108,71)(89,71)

\PhotonArc(102,64)(60,-180,-85){2}{20}
\ArrowArc(102,64)(-55,75,95)

\Text(-72, 64)[lb]{$i=$typeB}
\Text(-72, 55)[lb]{neutral}
\Text(-25, 76)[lb]{$Q=-p_1-p_2$}
\Text(50, 67)[lb]{$ \beta $}
\Text(50, 55)[lb]{$ \gamma $}
\Text(-25,55 )[lb]{ $\alpha   $}
\Text( 68,55 )[lb]{ $\nu_1  $}
\Text( 68,67)[lb]{ $\mu_2 $}
\Text(58,30)[lb]{$-q$}   
\Text(63,95)[lb]{$q+p_2$}  
\Text( 114,70 )[lb]{$\mu$}
\Text(114,62)[lb]{$j=$typeU}
\Text(91,76)[lb]{$p_2$}
\Text(91,15  )[lb]{$p_1$}
 \Text(114,-2)[lb]{$l=$typeD}
 \Text(114,10)[lb]{$\nu$}
\Text(30,-13)[lb]{Topology T3}
\end{picture}
\]
\vspace*{1mm}
\[
\begin{picture}(132,132)(0,0)
 \Photon(-20,66)(44,66){2}{7}
    \Vertex(44,66){2.5}
   \ArrowLine(0,71)(21,71) 
\DashCArc(66,66)(22,0,180){3}
\DashCArc(66,66)(22,180,0){3}
\ArrowArc(66,66)( 25,70,110)
\ArrowArc(66,66)(-25,70,110)

 \Vertex(88,66){2.5}
 \Photon(110,66)(88,66){2}{5}
 \ArrowLine(108,71)(89,71)

\PhotonArc(102,64)(60,85,180){2}{20}
\ArrowArc(102,64)(55,85,105)

\Text(-72, 64)[lb]{$i=$typeB}
\Text(-72, 55)[lb]{neutral}
\Text(-30, 76)[lb]{$Q=-p_1-p_2$}
\Text(50, 67)[lb]{$ \beta $}
\Text(50, 55)[lb]{$ \gamma $}
\Text(-25,55 )[lb]{ $\alpha   $}
\Text( 68,55 )[lb]{ $\nu_1  $}
\Text( 68,67)[lb]{ $\mu_2 $}
\Text(58,30)[lb]{$-q$}  
\Text(63,95)[lb]{$q+p_2$}  
\Text( 114,70 )[lb]{$\mu$}
\Text(114,62)[lb]{$j=$typeU}
\Text(91,76)[lb]{$p_2$}
\Text(91,109  )[lb]{$p_1$}
\Text(114,122)[lb]{$l=$typeD}
\Text(114,117)[lb]{$\nu$}
\Text(30,14)[lb]{Topology T4}
\end{picture}
\]
\vspace*{-13mm}
\caption[$bbb$ vertices]
        {Four topologies for three-boson vertices.}
\label{fig:bbb_ve}
\vspace*{-15mm}
\end{figure}
\clearpage

These vertex diagrams are precomputed by a FORM program accessible through the chain of clicks\\[1mm] 
{\bf EW $\to$ Precomputation $\to$ Vertex $\to$ bbb $\to$ Boson $\to$ bvv Vertex}.\\[1mm] 
They also may be recomputed at user request. The four topologies are defined in the 
specific procedures {\sf TribosVertT1(i,l,j)}.

\noindent
\underline{Structure of program {\bf bvv Vertex}}\\
The calculation is done by a single call to specific procedure
{\sf CalcTribosVertT14(`typeB',`typeD',`typeU')} which calls first of all four topologies
in Figs.~\ref{fig:bbb_ve} defined in specific procedures {\sf TribosVertTk(i,l,j)},
with $k=1,2,3,4$.

Just after that four specific procedures {\sf ClTribosVertTk(i,l,j)} perform 
{\em clusterizing} of the computed diagrams. One should note that {\em clusters} have different 
meanings in {\tt SANC}. Here clusterizing means nothing but summation over all virtual field indices,
resulting in the dependence of cluster names only on external field indices.

After clusterizing, the usual calls of the intrinsic procedures follow:
{\sf FeynmanRules}, {\sf 2Qs}, {\sf Reduction}, {\sf 2Qs}, {\sf Sing} and {\sf Scalarizing}.

The {\em fermionic} component of three-boson vertices has only one topology T1.
The diagrams are defined in the specific procedure {\sf TribosVertf(i,l,j)}.

\noindent
\underline{Structure of program {\bf Fermion $\to$ bbb Vertex}}\\
The calculation is done by a single call to specific procedure
{\sf CalcTribosVert(`typeB',`typeD',`typeU')} which at first calls the
topology defined in the specific procedure {\sf TribosVertf(i,l,j)}.
Next, specific procedure {\sf ClT1fer(i,l,j)} performs {\em clusterizing}. 

The calculation of the cluster is followed by calls of the intrinsic procedures:
{\sf FeynmanRules}, {\sf Gamma\-Left}, {\sf Diracizing}, {\sf GammaTrace},
{\sf Reduction}, {\sf Scalarizing}, {\sf Sing}, {\sf Scalprod} and {\sf DivisionGramDet}.

One may access a similar FORM code to precompute the HWW, ZWW and AWW vertices
via sequence
{\bf EW $\to$ Precomputation $\to$ Vertex $\to$ bbb $\to$ Boson $\to$ bWW Vertex}.

\subsubsection{Vertices for $ffbb$ processes\label{vert-ffbb}}
There are four blocks of vertices met in $ffbb$ processes where only diagrams with fermion
exchange contribute at the Born level (we have also $ffZH$ process where there is the Born 
diagram with boson exchange, so-called Higgsstrahlung, but this process is not added to 
{\tt SANC v.1.00}):
\begin{figure}[!h]
\[
\begin{array}{ccc}
  \vcenter{\hbox{
\begin{picture}(132,132)(0,0)

 \ArrowLine(55,115)(33,115)
 \ArrowLine(55,17)(33,17)
  
 \Photon(0,110)(55,110){2}{12}

 \Vertex(0,110){6}
 \ArrowLine(0,110)(-22,132)
 \Line(0,22)(0,110)
 \ArrowLine(0,64)(0,68)

 \Vertex(0,22){2.5}
 \ArrowLine(-22,0)(0,22)
 \Photon(0,22)(55,22){2}{12}

 \ArrowLine(-16,0)(0,16)
 \ArrowLine(-16,132)(0,116) 

 \Text(-32,132)[lt]{\sf ii}
 \Text(42,118)[lb]{$p$}    
 \Text(37,3)[lb]{$Q$}
 \Text(-15,106)[lb]{$\nu$}
 \Text(60,105)[lb]{vd=typeFD}
 \Text(60,18)[lb]{vu=typeFU}
 \Text(-42,137)[lb]{fu=typeIU}
 \Text(-42,-12)[lb]{fd=typeID}
 \Text(-15,17)[lb]{$\mu $}
 \Text(-4,123)[lb]{$p_1$}
 \Text(-4,3)[lb]{$p_2$}
 \Text(35,-25)[lb]{a)}
\end{picture}}}
& \qquad &
  \vcenter{\hbox{
\begin{picture}(132,132)(0,0)
 \ArrowLine(55,115)(33,115)
 \ArrowLine(55,17)(33,17)
  
 \Photon(0,110)(55,110){2}{12}
 \Vertex(0,110){6}
 \ArrowLine(0,110)(-22,132)
 \Line(0,22)(0,110)
 \ArrowLine(0,64)(0,68)

 \Vertex(0,22){2.5}
 \ArrowLine(-22,0)(0,22)
 \Photon(0,22)(55,22){2}{12}

 \ArrowLine(-16,0)(0,16)
 \ArrowLine(-16,132)(0,116) 

 \Text(-32,132)[lt]{\sf ii}
 \Text(42,118)[lb]{$Q$}    
 \Text(37,6)[lb]{$p$}
 \Text(-15,105)[lb]{$\mu$}
 \Text(60,18)[lb]{vd=typeFD}
 \Text(60,105)[lb]{vu=typeFU}
 \Text(-42,137)[lb]{fu=typeIU}
 \Text(-42,-12)[lb]{fd=typeID}
 \Text(-15,18)[lb]{$\nu $}
 \Text(-4,123)[lb]{$p_1$}
 \Text(-4,3)[lb]{$p_2$}
 \Text(35,-25)[lb]{b)}
\end{picture}}}
\\[40mm]  
  \vcenter{\hbox{
\begin{picture}(132,132)(0,0)
 \ArrowLine(55,115)(33,115)
 \ArrowLine(55,17)(33,17)
  
 \Photon(0,110)(55,110){2}{12}

 \Vertex(0,110){2.5}
 \ArrowLine(0,110)(-22,132)
 \Line(0,22)(0,110)
 \ArrowLine(0,64)(0,68)

 \Vertex(0,22){6}
 \ArrowLine(-22,0)(0,22)
 \Photon(0,22)(55,22){2}{12}

 \ArrowLine(-16,0)(0,16)
 \ArrowLine(-16,132)(0,116) 

 \Text(-32,132)[lt]{\sf ii}
 \Text(42,118)[lb]{$p$}    
 \Text(37,3)[lb]{$Q$}
 \Text(-15,106)[lb]{$\nu$}
 \Text(60,105)[lb]{vd=typeFD}
 \Text(60,18)[lb]{vu=typeFU}
 \Text(-42,137)[lb]{fu=typeIU}
 \Text(-42,-12)[lb]{fd=typeID}
 \Text(-15,17)[lb]{$\mu $}
 \Text(-4,123)[lb]{$p_1$}
 \Text(-4,3)[lb]{$p_2$}
 \Text(35,-25)[lb]{c)}
\end{picture}}}
& \qquad &
  \vcenter{\hbox{
\begin{picture}(132,132)(0,0)
 \ArrowLine(55,115)(33,115)
 \ArrowLine(55,17)(33,17)
  
 \Photon(0,110)(55,110){2}{12}
 \Vertex(0,110){2.5}
 \ArrowLine(0,110)(-22,132)
 \Line(0,22)(0,110)
 \ArrowLine(0,64)(0,68)

 \Vertex(0,22){6}
 \ArrowLine(-22,0)(0,22)
 \Photon(0,22)(55,22){2}{12}

 \ArrowLine(-16,0)(0,16)
 \ArrowLine(-16,132)(0,116) 

 \Text(-32,132)[lt]{\sf ii}
 \Text(42,118)[lb]{$Q$}    
 \Text(37,6)[lb]{$p$}
 \Text(-15,105)[lb]{$\mu$}
 \Text(60,18)[lb]{vd=typeFD}
 \Text(60,105)[lb]{vu=typeFU}
 \Text(-42,137)[lb]{fu=typeIU}
 \Text(-42,-12)[lb]{fd=typeID}
 \Text(-15,18)[lb]{$\nu $}
 \Text(-4,123)[lb]{$p_1$}
 \Text(-4,3)[lb]{$p_2$}
 \Text(35,-25)[lb]{d)}
\end{picture}}}
\end{array}
\]
\vspace*{3mm}
\caption[$ffbb$ vertices]
        {Four $ffbb$ vertex diagrams.}
\label{ffbbvert}
\end{figure}
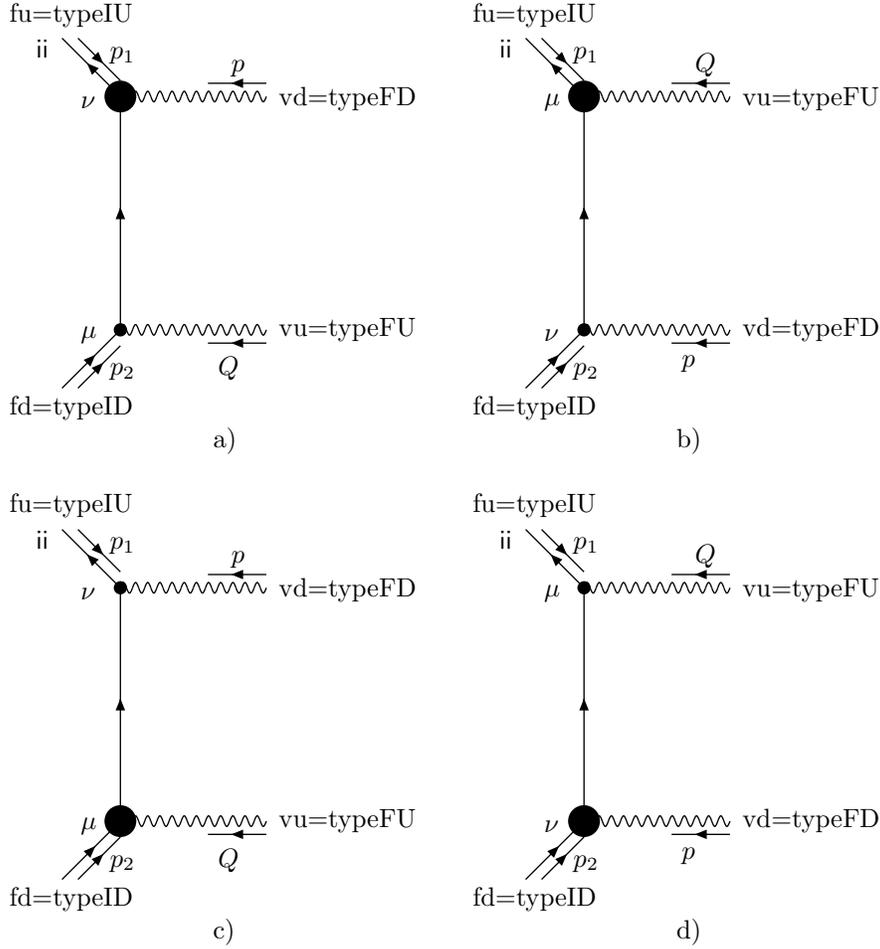

The four building blocks are precomputed by a FORM program accessible via menu sequence: \\[1mm]
{\bf EW $\to$ Precomputation $\to$ Vertex $\to$ bff $\to$ ffbb Vertex}.

\noindent
\underline{Structure of program {\bf ffbb Vertex}}\\
The calculation is done by $2\otimes3$ calls to the following specific procedures:\\
{\sf CalcVert(`typeFU',`typeID',`typeIU',mu)}~~~~~~~~~~~~~~~~~ --- computes vertex blob with index 
$\mu$;\\
{\sf CalcVertmut(`typeIU',`typeID',`typeFD',`typeFU',mu)} --- computes diagram b); \\
{\sf CalcVertmuu(`typeIU',`typeID',`typeFU',`typeFD',mu)} --- computes diagram c); \\
{\sf CalcVert(`typeFD',`typeID',`typeIU',nu)}~~~~~~~~~~~~~~~~~~ --- computes vertex blob with index 
$\nu$;\\
{\sf CalcVertnut(`typeIU',`typeID',`typeFD',`typeFU',nu)}~~ --- computes diagram d); \\
{\sf CalcVertnuu(`typeIU',`typeID',`typeFU',`typeFD',nu)}~~ --- computes diagram a).


The procedure {\sf CalcVert} recomputes $bff$ vertices 
for given set {\sf `typeFU',`typeID',`typeIU'} and creates expressions
{\sf vfbf`vu'`fd'`fu'`mu'`k1'`k2'`k3'} and {\sf vbfb`vu'`fd'`fu'`mu'`k1'`k2'`k3'} similar
to those described in Section~\ref{vert-bff}, but not applying intrinsic procedure
{\sf Masshell} since in the case under consideration one of fermions is off-shell.
The four procedures {\sf CalcVertmut, CalcVertmuu, CalcVertnut} and {\sf CalcVertnuu}
have very similar structures. 

In each of four  procedures the intrinsic procedures {\sf FeynmanRules, GammaRight,
bpIdentities, p2m, p2I} and {\sf p2p} are called. Next follows an important shift of the
virtual fermion 4-momenta to `real' four momenta $p_1, p_2, p, Q$, 
and finally {\sf ExtMomentumWI} is applied.

All the building blocks are clusterized by summation over virtual momenta in vertex loops
(see discussion of clusterization in Section~\ref{ffbb-box}.).

Besides {\sf xi} there are three more internally defined options: {\sf on}
(see Section~\ref{ffbb-box} for more discussion about option {\sf on})
and {\sf mf, mp}, of which the latter two allow to neglect fermion masses.
\vspace*{-2mm}

\begin{verbatim}
#define on "1"
* .eq.0 photons are off mass-shell; .eq.1 photons are on  mass-shell
#define mf "0"
* .eq.0 fermion mass, pm(`typeID')=0;  .eq.1 it is not neglected
#define mp "0"
* .eq.0 partner mass, pm(`typeIDp')=0; .eq.1 it is not neglected
\end{verbatim}
\vspace*{-2mm}

\noindent
The result of this calculation is stored in the file \\
{\sf ffbbVertxi`xi'on`on'mf`mf'mp`mp' `typeIU'`typeID'`typeFU'`typeFD'.sav}.
\vspace*{-5mm}

\clearpage

\subsection{One-loop boxes}
Approaching the description of precomputation of boxes, one should note that the world of boxes 
is much more rich and complex compared to self-energies and vertices. If for the latter case we
still could profess the idea of allowing  recomputation of all one-loop vertices needed for the 
process
under consideration, then in the case of boxes we must change our strategy if the user wants to 
carry out some recomputation. 

For boxes, the idea of precomputation becomes vitally important for realization of {\tt SANC} 
project.
As will be explained below, calculation of some boxes for some particular processes takes so much 
time that an external user should refrain from repeating precomputation. Furthermore, the richness 
of boxes requires more classification. Depending on the type of external lines (fermion or boson), 
we will distinguish three large classes of boxes: {\sf ffff}, {\sf ffbb} and {\sf bbbb}. 

\subsubsection{Boxes for $ffff$ processes\label{ffff-box}}
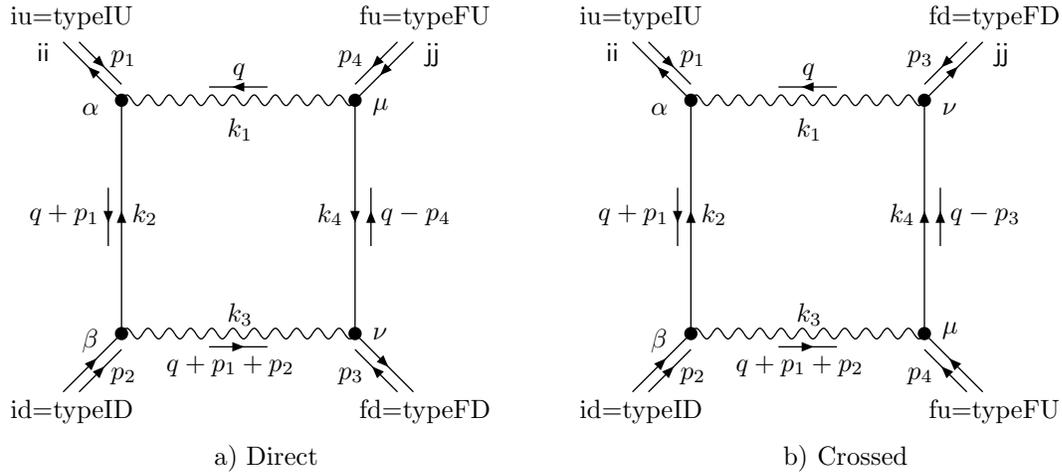
\begin{figure}[!h]
\vspace*{3mm}
\[
\begin{array}{ccc}
\vcenter{\hbox{
\begin{picture}(132,132)(0,0)

\Text(-42,137)[lb]{iu=typeIU}
\Text(-32,132)[lt]{\sf ii}
\Text(-15,105)[lb]{$ \alpha $}
\Text(-4,123)[lb]{$ p_1 $}
\ArrowLine(-16,132)(0,116)
\ArrowLine(0,110)(-22,132)

\Text(4,63)[lb]{$ k_2 $}
\Text(-35,63)[lb]{$ q+p_1 $}
\ArrowLine(-5,77)(-5,55)
\ArrowLine(0,22)(0,110)
\Vertex(0,22){2.5}
\Vertex(0,110){2.5}

\Text(-42,-12)[lb]{id=typeID}
\Text(-15,15)[lb]{$ \beta $}
\Text(-4,3)[lb]{$ p_2 $}
\ArrowLine(-16,0)(0,16)
\ArrowLine(-22,0)(0,22)

\Text(42,118)[lb]{$ q $} 
\ArrowLine(55,115)(33,115)
\Photon(0,110)(88,110){2}{11}
\Text(40,95)[lb]{$ k_1 $}

\Text(40,25)[lb]{$ k_3 $}
\Photon(0,22)(88,22){2}{11}
\ArrowLine(33,17)(55,17)
\Text(17,6)[lb]{$ q+p_1+p_2 $}

\Text(90,137)[lb]{fu=typeFU}
\Text(115,132)[lt]{\sf jj}
\Text(82,123)[lb]{$ p_4$ }
\Text(95,105)[lb]{$ \mu $}
\ArrowLine(110,132)(88,110)
\ArrowLine(104,132)(88,116)

\Text(98,63)[lb]{$ q-p_4 $}
\ArrowLine(94,55)(94,77)
\Text(74,63)[lb]{$ k_4 $}
\ArrowLine(88,110)(88,22)
\Vertex(88,22){2.5}
\Vertex(88,110){2.5}
 
\Text(90,-12)[lb]{fd=typeFD}
\Text(95,20)[lb]{$ \nu$ }
\Text(82,3  )[lb]{$ p_3$ }
\ArrowLine(88,22)(110,0)
\ArrowLine(104,0)(88,16)
 
\Text(35,-30)[lb]{a) Direct}
\end{picture}}}
& \qquad \qquad \qquad &
\vcenter{\hbox{
\begin{picture}(132,132)(0,0)

\Text(-42,137)[lb]{iu=typeIU}
\Text(-32,132)[lt]{\sf ii}
\Text(-15,105)[lb]{$ \alpha $}
\Text(-4,123)[lb]{$ p_1 $}
\ArrowLine(-16,132)(0,116)
\ArrowLine(0,110)(-22,132) 

\Text(4,63)[lb]{$ k_2 $}
\Text(-35,63)[lb]{$ q+p_1 $}
\ArrowLine(-5,77)(-5,55)
\ArrowLine(0,22)(0,110)
\Vertex(0,22){2.5}
\Vertex(0,110){2.5}

\Text(-42,-12)[lb]{id=typeID}
\Text(-15,15)[lb]{$ \beta $}
\Text(-4,3)[lb]{$ p_2 $}
\ArrowLine(-16,0)(0,16)
\ArrowLine(-22,0)(0,22)

\Text(42,118)[lb]{$ q $}
\ArrowLine(55,115)(33,115)
\Photon(0,110)(88,110){2}{11}
\Text(40,95)[lb]{$ k_1 $}

\Text(40,25)[lb]{$ k_3 $}
\Photon(0,22)(88,22){2}{11}
\ArrowLine(33,17)(55,17)
\Text(17,6)[lb]{$ q+p_1+p_2 $}

\Text(90,137)[lb]{fd=typeFD}
\Text(115,132)[lt]{\sf jj}
\Text(82,123)[lb]{$ p_3 $}
\Text(95,105)[lb]{$\nu$}
\ArrowLine(88,110)(110,132)
\ArrowLine(104,132)(88,116)

\Text(98,63)[lb]{$q-p_3$}
\ArrowLine(94,55)(94,77)
\Text(75,63)[lb]{$ k_4 $}
\ArrowLine(88,22)(88,110)
\Vertex(88,22){2.5}
\Vertex(88,110){2.5}

\Text(90,-12)[lb]{fu=typeFU}
\Text(95,20)[lb]{$ \mu $}
\Text(82,3)[lb]{$p_4$}
\ArrowLine(110,0)(88,22)
\ArrowLine(104,0)(88,16)

\Text(35,-30)[lb]{b) Crossed}
\end{picture}}}
\end{array}
\]
\vspace*{.6cm}
\caption[$ffff$ boxes, direct and crossed]
        {Direct and crossed boxes for $ffff$ processes.}
\label{ffffboxesT2T4}
\end{figure}

The {\sf ffff} boxes, which are met in the description of 4f processes, are of two topologies,
{\em direct} and {\em crossed}, and are characterized by two fermionic currents {\sf ii} and 
{\sf jj}, coupled by two bosons with field indices {\sf k1} and {\sf k3};
see Fig.~\ref{ffffboxesT2T4} showing all the field indices (for external and virtual fields), 
types of external fermions {\sf typeXX}, Lorenz indices and momentum flows for the direct and 
crossed topologies.

These diagrams describe both {\em neutral current} (NC) and {\em charged current} (CC) boxes
both for {\tt 1$\to$ 3} decays and {\tt 2$\to$ 2} reactions in any channel $(s,t,u)$.
In the general case, virtual boson field indices run from 1 to 6, and virtual fermion indices run
over doublets {\sf \{typeID,typeIDp\}} and {\sf \{typeFU,typeFUp\}}, where {\sf typeIDp} and
{\sf typeFUp} are isospin partners of {\sf typeID} and {\sf typeFU}, respectively.

The calculation of direct and crossed boxes is realized in two specific procedures
{\sf direct(iu,id,fu,fd)} and {\sf crossed(iu,id,fu,fd)} in nested loops over all allowed
field indices of the virtual particles.

Such a realization may apparently take into account all the external fermion masses. However, 
in practical applications we usually treat fermion masses of one of two currents, {\sf ii} or
{\sf jj}, as being massless, which is true for the processes of the kind $f_1+f_1+f+f\to 0$,
upon which we concentrate. Here $f_1$ denotes a massless first generation fermion or any neutrino. 

Various FORM codes computing these boxes are accessible via chains of clicks \\[1mm]
{\bf EW $\to$ Box $\to$ ffff $\to$ Neutral Current $\to$ NC ZZ(ZA,WW) Box} and \\[1mm]
{\bf EW $\to$ Box $\to$ ffff $\to$ Charged Current $\to$ CC INI(FIN) Box}.\\[1mm]
Here INI(FIN) means which of the masses, initial fermions {\sf ii} or final fermions
{\sf jj} are kept non-zero.
All the codes have very similar structures with minor modifications.
The calculation is done by a single call to specific procedure 
{\sf CalcBoxNC(`typeIU',`typeID',`typeFU',`typeFD')} which calls first the two
specific procedures --- {\sf direct(iu,id,fu,fd)} and {\sf crossed(iu,id,fu,fd)}.
In these procedures all the box diagrams are defined in terms of vertices, propagators and external
spinors {\sf tlo(p1), tro(p2), tle(p3)} and {\sf tre(p4)}. 

The subsequent calculations are very similar for all 4f-boxes, both NC and CC.

In the beginning of each {\sf NC(CC) XX Box} program, described in this Section, 
there is an internal definition {\sf \#define xi "0"}
which chooses among two internal options with an obvious meaning described in Section 
\ref{fse-ffbb}.

Note, that for 4f boxes we
gain only little CPU time choosing {\sf xi "1"} option, since action of 
{\sf BoxPrereductionNC(CC)} results in nearly complete cancellation of $\xi$ already
before {\sf Scalarizing}.
For example, the program {\sf NC ZA Box} needs about 3 minutes CPU time at a 1.6 GHz computer for
both options. A similar picture is valid also for the program {\sf CC FIN Box}.

Actually, 4f-boxes are the last precomputation programs which still may be recomputed at user
request. However, we do not recommended to recompute them, because 3 min is already a noticeable 
time.

The results of calculations of NC boxes are stored in the files
{\sf ffffZZxi`xi'`typeIU'`typeID'`typeFU'`typeFD' .sav} and 
{\sf ffffZAxi`xi'`typeIU'`typeID'`typeFU'`typeFD'.sav} and are 
loaded by a FORM program \\[1mm]
{\bf EW $\to$ Box $\to$ ffff $\to$ Neutral Current $\to$ NC FF Box} \\[1mm]
which constructs box form factors stored in the files 
{\sf ffffNCFF`typeIU'`typeID'`typeFU'`typeFD'.sav}.
This special FORM program will be described elsewhere.


\subsubsection{Boxes for $ffbb$ processes\label{ffbb-box}}
There are seven topologies of boxes which are met in the description of 2f2b processes.
Their enume\-ration is borrowed from Ref.~\cite{Bardin:1999ak}. 
All these boxes have apparently only one fermionic current conventionally marked by the current 
index {\sf ii} but different numbers of internal bosonic lines. 
\vspace*{1mm}

\leftline{\bf Topologies T2, T4}

\begin{figure}[!t]
\[
\begin{array}{ccc}
  \vcenter{\hbox{
\begin{picture}(132,132)(0,0)

\Text(-32,132)[lt]{\sf ii}
\Text(40,25)[lb]{$ k_3 $}
\Text(40,95)[lb]{$ k_1 $}

\Text(4,63 )[lb]{$ k_2 $}
\Text(74,63)[lb]{$ k_4 $}

\Text(42,118)[lb]{$q$}    
\Text(17,6)[lb]{$q+p_1+p_2$}

 \ArrowLine(55,115)(33,115)
 \ArrowLine(33,17)(55,17)
  
 \ArrowLine(88,22)(88,110)
 \Vertex(88,22){2.5}
 \Photon(88,22)(110,0){2}{5}

 \ArrowLine(88,110)(0,110)

 \Vertex(0,110){2.5}
 \ArrowLine(0,110)(-22,132)
 \Photon(0,110)(0,22){2}{11}

 \Vertex(0,22){2.5}
 \ArrowLine(-22,0)(0,22)
 \ArrowLine(0,22)(88,22)

 \ArrowLine(104,132)(88,116)
 \Vertex(88,110){2.5}
 \Photon(110,132)(88,110){2}{5}

 \ArrowLine(104,0)(88,16)
 \ArrowLine(94,55)(94,77)

 \ArrowLine(-16,0)(0,16)
 \ArrowLine(-16,132)(0,116) 

 \Text(-15,105)[lb]{$\alpha$}
 \Text(95,105)[lb]{$\mu$}
 \Text(90,137)[lb]{vu=typeFU}
 \Text(90,-12)[lb]{vd=typeFD}
 \Text(-42,137)[lb]{fu=typeIU}
 \Text(-42,-12)[lb]{fd=typeID}
 \Text(95,20)[lb]{$\nu$}
 \Text(-15,15)[lb]{$\beta $}
 \Text(-4,123)[lb]{$p_1$}
 \Text(-4,3)[lb]{$p_2$}
 \Text(82,123)[lb]{$p_4$}
 \Text(82,3 )[lb]{$p_3$}
 \Text(98,63)[lb]{$q-p_4$}
 \Text(-35,63)[lb]{$ q+p_1 $}
 \ArrowLine(-5,77)(-5,55)
 
\Text(35,-15)[lb]{a) T2}
\end{picture}}}
& \qquad \qquad \qquad &
  \vcenter{\hbox{
\begin{picture}(132,132)(0,0)

\Text(-32,132)[lt]{\sf ii}
\Text(40,25)[lb]{$ k_3 $}
\Text(40,95)[lb]{$ k_1 $}

\Text(4,63 )[lb]{$ k_2 $}
\Text(75,63)[lb]{$ k_4 $}

\Text(42,118)[lb]{$q$}    
\Text(17,6)[lb]{$q+p_1+p_2$}

 \ArrowLine(55,115)(33,115)
 \ArrowLine(33,17)(55,17)
  
 \ArrowLine(88,22)(88,110)
 \Vertex(88,22){2.5}
 \Photon(88,22)(110,0){2}{5}

 \ArrowLine(88,110)(0,110)

 \Vertex(0,110){2.5}
 \ArrowLine(0,110)(-22,132)
 \Photon(0,110)(0,22){2}{11}

 \Vertex(0,22){2.5}
 \ArrowLine(-22,0)(0,22)
 \ArrowLine(0,22)(88,22)

 \ArrowLine(104,132)(88,116)

 \Vertex(88,110){2.5}
 \Photon(110,132)(88,110){2}{5}

 \ArrowLine(104,0)(88,16)
 \ArrowLine(94,55)(94,77)

 \ArrowLine(-16,0)(0,16)
 \ArrowLine(-16,132)(0,116) 

 \Text(-15,105)[lb]{$\alpha$}
 \Text(95,105)[lb]{$\nu$}
 \Text(90,137)[lb]{vd=typeFD}
 \Text(90,-12)[lb]{vu=typeFU}
 \Text(-42,137)[lb]{fu=typeIU}
 \Text(-42,-12)[lb]{fd=typeID}
 \Text(95,20)[lb]{$\mu$}
 \Text(-15,15)[lb]{$\beta $}
 \Text(-4,123)[lb]{$p_1$}
 \Text(-4,3)[lb]{$p_2$}
 \Text(82,123)[lb]{$p_3$}
 \Text(82,3 )[lb]{$p_4$}
 \Text(98,63)[lb]{$q-p_3$}
 \Text(-35,63)[lb]{$ q+p_1 $}
 \ArrowLine(-5,77)(-5,55)
 
\Text(35,-15)[lb]{b) T4}
\end{picture}}}
\end{array}
\]
\caption[$ffbb$ boxes, topologies T2 and T4]
        {Boxes for $ffbb$ processes, topologies T2 and T4.}
\label{ffbbboxesT2T4}
\end{figure}
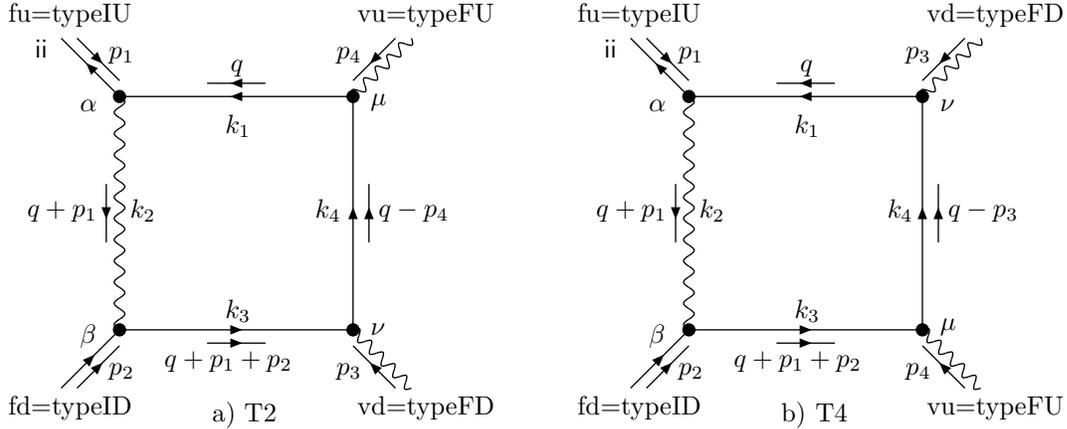

\noindent We begin with the simplest case of topologies having only one virtual bosonic line, see
Fig.~\ref{ffbbboxesT2T4} showing all the field, type and Lorentz indices and momentum flows.
These two topologies are actually of the type of direct and crossed ones considered in the previous 
section. However, their structure is quite different.

These diagrams also describe both NC and CC boxes in any channel ({\sf s,t,u}), both decays and 
reactions. The virtual boson field indices run from 1 to 6, and virtual fermion indices run
over doublets {\sf \{typeID,typeIDp\}}, where {\sf typeIDp} is the isospin partner of {\sf typeID}.
It is foreseen to cover CC processes in the future; currently we have only NC processes.

The calculation of T2 and T4 topologies is realized in two specific procedures
{\sf boxT2(fu,fd,vd,vu)} and {\sf boxT4(fu,fd,vu,vd)} in nested loops over all allowed
field indices of the virtual particles.
Contrary to the 4f-case, we do take into account the external fermion masses.

The calculation starts by two calls to specific procedures
{\sf CalcBoxT2(`typeIU',`typeID',`typeFD',`typeFU')} and
{\sf CalcBoxT4(`typeIU',`typeID',`typeFU',`typeFD')} which call 
two specific procedures shown above.

Then, for each topology, the calculation continues by clustering boxes calling
the procedure {\sf ClusterboxT2(4)(`fu',`fd',`vd',`vu')} which creates four clusters, 
{\sf Cl`k'T2(4)`fu'`fd'`vd'`vu'} for {\sf k=1,4}, depending
on $\xi_{\sss A}$, $\xi_{\sss Z}$, $\xi_{\sss W}$, and independent of any $\xi$, respectively.
Next follow calls to the intrinsic procedures
{\sf FeynmanRules, GammaRight, Diracizing, Diraceq, Reduction, Pulling, Diraceq, PullingOrder,
Scalprod, Sing, ExtMomentumWI} described in Section~\ref{pintr}.

Then follows a do-loop over {\sf k} making separate scalarizations of four clusters inside which
six more intrinsic procedures are called: {\sf Scalarizingdp, Scalarizing, DivisionGramDet,
bpIdentities, p2m, p2p}.
\vspace*{3mm}

In the beginning of the program there are four usual definitions:
\vspace*{-2mm}

\begin{verbatim}
#define xi "0" / #define on "0" / #define mf "1" / #define mp "1".
\end{verbatim}
\vspace*{-2mm}
the meaning of three of which was explained in the previous sections.
\vspace*{3mm}

Note that for the $2f2b$ processes where one (or two) bosons are photons one may, 
of course, choose {\sf on "1"} which greatly saves CPU time
(see tables of CPU time below in this section).
Option {\sf on "0"} is basically foreseen for $2\to 3$ 
processes where these $2f2b$ boxes with off shell boson(s) are building blocks.

The results of calculations of these boxes are stored in the files\\
{\sf ffbbT2xi`xi'on`on'mf`mf'mp`mp'`fu'`fd'`vd'`vu'.sav} and 
{\sf ffbbT4xi`xi'on`on'mf`mf'mp`mp'`fu'`fd'`vu'`vd'.sav}.

The calculation of $2f2b$ boxes takes a lot of CPU time. 
The {\sf NC T2T4 Box} is the fastest, but even this program should not be recomputed by users.
The other $ffbb$ box topologies, except {\sf T7}, take much more CPU time.

In conclusion of this section we present an instructive example of how much CPU time is needed 
to compute T2+T4 topologies for the $d\bar{d}\to\gamma\gamma$ process at a 3 GHz PC running
Linux:\\[1mm]
{\sf xi "0", on "0", mf "1", mp "0"} --- ~5 hours,   \\
{\sf xi "0", on "1", mf "1", mp "1"} --- 70 minutes, \\
{\sf xi "1", on "1", mf "1", mp "1"} --- 60 minutes. \\
This is why a recomputation of these boxes is not allowed.
\vspace*{1mm}

\begin{figure}[!h]
\[
\begin{array}{ccc}
  \vcenter{\hbox{
\begin{picture}(132,132)(0,0)

\Text(-32,132)[lt]{\sf ii}
\Text(40,25)[lb]{$ k_3 $}
\Text(40,95)[lb]{$ k_1 $}

\Text(4,63 )[lb]{$ k_2 $}
\Text(74,63)[lb]{$ k_4 $}

\Text(42,118)[lb]{$q$}    
\Text(17,6)[lb]{$q+p_1+p_2$}

      \ArrowLine(55,115)(33,115)
      \ArrowLine(33,17)(55,17)
  
\Photon(88,110)(88,22){2}{11}
\Vertex(88,22){2.5}
\Photon(88,22)(110,0){2}{5}
\Photon(88,110)(0,110){2}{11}

\Vertex(0,110){2.5}
\ArrowLine(0,110)(-22,132)
\ArrowLine(0,22)(0,110)

\Vertex(0,22){2.5}
\ArrowLine(-22,0)(0,22)
\Photon(0,22)(88,22){2}{11}

\ArrowLine(104,132)(88,116)

\Vertex(88,110){2.5}
\Photon(110,132)(88,110){2}{5}

\ArrowLine(104,0)(88,16)
\ArrowLine(94,55)(94,77)

\ArrowLine(-16,0)(0,16)
\ArrowLine(-16,132)(0,116) 

 \Text(-15,105)[lb]{$\alpha$}
 \Text(95,105)[lb]{$\mu$}
 \Text(90,137)[lb]{vu=typeFU}
 \Text(90,-12)[lb]{vd=typeFD}
 \Text(-42,137)[lb]{fu=typeIU}
 \Text(-42,-12)[lb]{fd=typeID}
 \Text(95,20)[lb]{$\nu$}
 \Text(-15,15)[lb]{$\beta $}
 \Text(-4,123)[lb]{$p_1$}
 \Text(-4,3)[lb]{$p_2$}
 \Text(82,123)[lb]{$p_4$}
 \Text(82,3 )[lb]{$p_3$}
 \Text(98,63)[lb]{$q-p_4$}
 \Text(-35,63)[lb]{$ q+p_1 $}
 \ArrowLine(-5,77)(-5,55)
 
\Text(35,-15)[lb]{a) T1}
\end{picture}}}
&  \qquad \qquad \qquad &
  \vcenter{\hbox{
\begin{picture}(132,132)(0,0)

\Text(-32,132)[lt]{\sf ii}
\Text(40,25)[lb]{$ k_3 $}
\Text(40,95)[lb]{$ k_1 $}

\Text(4,63 )[lb]{$ k_2 $}
\Text(75,63)[lb]{$ k_4 $}

\Text(42,118)[lb]{$q$}    
\Text(17,6)[lb]{$q+p_1+p_2$}

\ArrowLine(55,115)(33,115)
\ArrowLine(33,17)(55,17)
\ArrowLine(0,22)(0,110)  
\Photon(88,110)(88,22){2}{11}
\Vertex(88,22){2.5}
\Photon(88,22)(110,0){2}{5}

\Photon(88,110)(0,110){2}{11}

\Vertex(0,110){2.5}
\ArrowLine(0,110)(-22,132)

\Vertex(0,22){2.5}
\ArrowLine(-22,0)(0,22)
\Photon(0,22)(88,22){2}{11}

\ArrowLine(104,132)(88,116)

\Vertex(88,110){2.5}
 \Photon(110,132)(88,110){2}{5}

      \ArrowLine(104,0)(88,16)
      \ArrowLine(94,55)(94,77)

      \ArrowLine(-16,0)(0,16)
      \ArrowLine(-16,132)(0,116) 

 \Text(-15,105)[lb]{$\alpha$}
 \Text(95,105)[lb]{$\nu$}
 \Text(90,137)[lb]{vd=typeFD}
 \Text(90,-12)[lb]{vu=typeFU}
 \Text(-42,137)[lb]{fu=typeIU}
 \Text(-42,-12)[lb]{fd=typeID}
 \Text(95,20)[lb]{$\mu$}
 \Text(-15,15)[lb]{$\beta $}
 \Text(-4,123)[lb]{$p_1$}
 \Text(-4,3)[lb]{$p_2$}
 \Text(82,123)[lb]{$p_3$}
 \Text(82,3  )[lb]{$p_4$}
 \Text(98,66 )[lb]{$q-p_3$}
 \Text(-35,63)[lb]{$ q+p_1 $}
 \ArrowLine(-5,77)(-5,55)
 
\Text(35,-15)[lb]{b) T3}
\end{picture}}}
\end{array}
\]
\caption[$ffbb$ boxes, topologies T1 and T3]
        {Boxes for $ffbb$ processes, topologies T1 and T3.}
\label{ffbbboxesT1T3}
\end{figure}
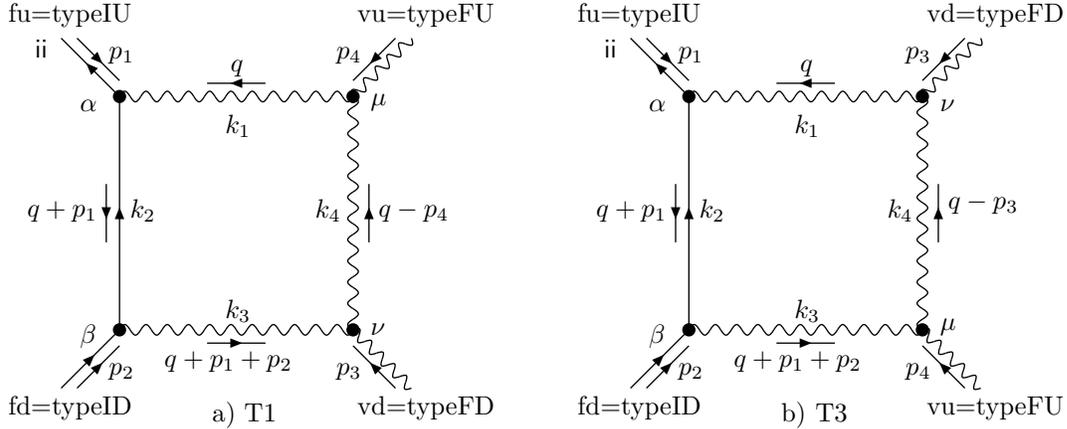

\leftline{\bf Topologies T1, T3}
\noindent
We jump now to the most complex case of topologies having three internal bosonic lines, 
see Fig.~\ref{ffbbboxesT1T3} showing all the field, type and Lorentz indices and momentum flows.
These two topologies are also of the type of direct and crossed ones,
however their structure is much more complex than anything considered so far.

The paragraph after Fig.~\ref{ffbbboxesT2T4} applies also in this case.

The calculation of T1 and T3 topologies is realized in two specific procedures
{\sf boxT1(fu,fd,vd,vu)} and {\sf boxT3(fu,fd,vu,vd)} in nested loops over all allowed
field indices of the virtual particles.
As before, we take into account the external fermion masses.

So far we have considered $2f2b$ processes for the case when one of the bosons is a photon. 
In such a case the internal bosons must be charged (with field indices $\pm 3$ and $\pm 6$). 

The calculation starts by two calls to specific procedures
{\sf CalcBoxT1(`typeIU',`typeID',`typeFD',`typeFU')} and
{\sf CalcBoxT3(`typeIU',`typeID',`typeFU',`typeFD')} which call
the two specific procedures mentioned above.

Then, for each topology, the calculation continues by calling
the procedure {\sf ClusterboxT1(3)(fu,fd,vd,vu)} which creates in this case only two clusters, 
{\sf Cl`k'T1(3)`fu'`fd'`vd'`vu'} for {\sf k=2,3}, where {\sf k=2} collects all neutral virtual
bosons and {\sf k=3} all charged bosons.
Next follow standard calls to the intrinsic procedures
{\sf FeynmanRules, GammaRight, Diracizing, Diraceq, Reduction, Pulling, Diraceq, PullingOrder,
Scalprod, Sing, ExtMomentumWI} described in Section~\ref{pintr}.

Then follows a call to the intrinsic procedure {\sf ScalarizingProj} which scalarizes two clusters,
{\sf k=2,3}, splitting them first into as many Dirac--Lorentz structures as it has.
This is done in order to avoid limitations of {\tt FORM v3.0} which cannot handle internal files
of length greater than 1.6 Gb or so. This might be circumvented by switching to {\tt FORM v3.1},
however, we did not manage to switch to this version so far.
 
Inside itself procedure {\sf ScalarizingProj} creates many intermediate expressions to which
intrinsic procedures {\sf Scalarizing, DivisionGramDet, p2m} are applied. At the end,
{\sf ScalarizingProj} collects these pieces together again.

The intrinsic procedure {\sf p2p} is called at the end.

The results of calculations of boxes of T1 and T3 topologies are stored in the files\\
{\sf ffbb`k'T1xi`xi'on`on'mf`mf'mp`mp'`fu'`fd'`vd'`vu'.sav} and \\
{\sf ffbb`k'T3xi`xi'on`on'mf`mf'mp`mp'`fu'`fd'`vu'`vd'.sav}.

The label `k' stands for two clusters: $=2$ with all {\em neutral} virtual bosons, and 
$=3$ with all charged virtual bosons {\sf `k1',`k3',`k4'}. 
\vspace*{1mm}

In the beginning of the program there are four usual definitions:
\vspace*{-2mm}
\begin{verbatim}
#define xi "0" / #define on "0" / #define mf "1" / #define mp "1"
\end{verbatim}
\vspace*{-2mm}
\noindent
having the same meaning as explained in the previous section.
\vspace*{1mm}

A table of CPU times for the $d\bar{d}\to\gamma\gamma$ boxes of topologies T1+T3 looks as follows:\\
{\sf xi "0", on "0", mf "1", mp "0"} --- 90 hours,   \\
{\sf xi "0", on "1", mf "1", mp "1"} --- ~7 hours,   \\
{\sf xi "1", on "1", mf "1", mp "1"} --- 14 minutes. \\
Of course, a recomputation of these boxes is not allowed.
\vspace*{-5mm}

\clearpage

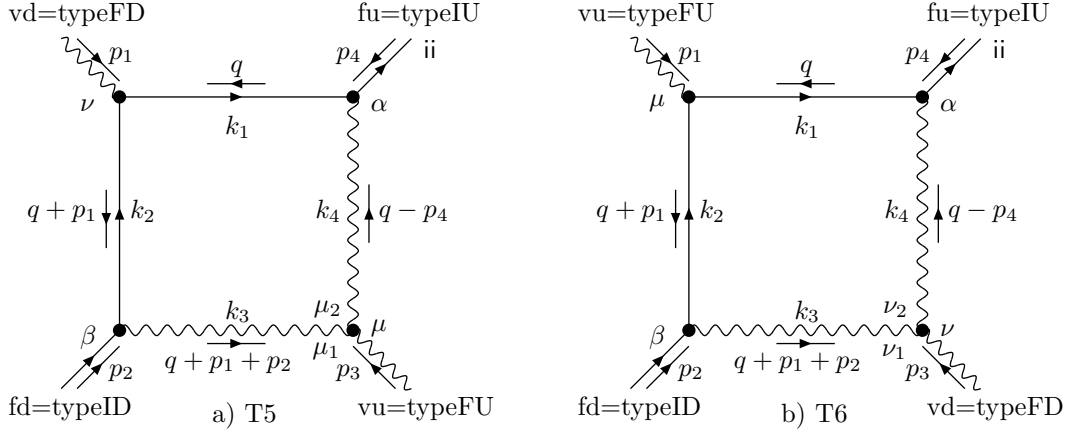
\begin{figure}[!h]
\[
\begin{array}{ccc}
  \vcenter{\hbox{
 \begin{picture}(132,132)(0,0)

\Text(115,132)[lt]{\sf ii}
\Text(40,25)[lb]{$ k_3 $}
\Text(40,95)[lb]{$ k_1 $}

\Text(4,63 )[lb]{$ k_2 $}
\Text(74,63)[lb]{$ k_4 $}

\Text(42,118)[lb]{$q$}    
\Text(17,6)[lb]{$q+p_1+p_2$}

      \ArrowLine(55,115)(33,115)
      \ArrowLine(33,17)(55,17)
  
\Photon(88,110)(88,22){2}{11}
\Vertex(88,22){2.5}
\Photon(88,22)(110,0){2}{5}
\ArrowLine(0,110)(88,110)

\Vertex(0,110){2.5}
\Photon(0,110)(-22,132){2}{5}
\ArrowLine(0,22)(0,110)

\Vertex(0,22){2.5}
\ArrowLine(-22,0)(0,22)
\Photon(0,22)(88,22){2}{11}

\ArrowLine(104,132)(88,116)

\Vertex(88,110){2.5}
\ArrowLine(88,110)(110,132)

\ArrowLine(104,0)(88,16)
\ArrowLine(94,55)(94,77)

\ArrowLine(-16,0)(0,16)
\ArrowLine(-16,132)(0,116) 

 \Text(-15,105)[lb]{$\nu$}
 \Text(95,105)[lb]{$\alpha$}

 \Text(90,137)[lb]{fu=typeIU}
 \Text(90,-12)[lb]{vu=typeFU}
 \Text(-42,137)[lb]{vd=typeFD}
 \Text(-42,-12)[lb]{fd=typeID}
 \Text(-15,15)[lb]{$\beta $}
 \Text(-4,123)[lb]{$p_1$}
 \Text(-4,3)[lb]{$p_2$}
 \Text(82,123)[lb]{$p_4$}
 \Text(82,3 )[lb]{$p_3$}
 \Text(98,63)[lb]{$q-p_4$}

 \Text(-35,63)[lb]{$q+p_1$}
\ArrowLine(-5,75)(-5,53)
 
 \Text(95,20)[lb]{$\mu$}
 \Text(73,11)[lb]{$\mu_1$}
 \Text(73.5,27)[lb]{$\mu_2$}

\Text(35,-15)[lb]{a) T5}
\end{picture}}}
& \qquad \qquad \qquad &
  \vcenter{\hbox{
 \begin{picture}(132,132)(0,0)

\Text(115,132)[lt]{\sf ii}
\Text(40,25)[lb]{$ k_3 $}
\Text(40,95)[lb]{$ k_1 $}

\Text(4,63 )[lb]{$ k_2 $}
\Text(74,63)[lb]{$ k_4 $}

\Text(42,118)[lb]{$q$}    
\Text(17,6)[lb]{$q+p_1+p_2$}

      \ArrowLine(55,115)(33,115)
      \ArrowLine(33,17)(55,17)
  
\Photon(88,110)(88,22){2}{11}
\Vertex(88,22){2.5}
\Photon(88,22)(110,0){2}{5}
\ArrowLine(0,110)(88,110)

\Vertex(0,110){2.5}
\Photon(0,110)(-22,132){2}{5}
\ArrowLine(0,22)(0,110)

\Vertex(0,22){2.5}
\ArrowLine(-22,0)(0,22)
\Photon(0,22)(88,22){2}{11}

\ArrowLine(104,132)(88,116)

\Vertex(88,110){2.5}
\ArrowLine(88,110)(110,132)

\ArrowLine(104,0)(88,16)
\ArrowLine(94,55)(94,77)

\ArrowLine(-16,0)(0,16)
\ArrowLine(-16,132)(0,116) 

 \Text(-15,105)[lb]{$\mu$}
 \Text(95,105)[lb]{$\alpha$}

 \Text(90,137)[lb]{fu=typeIU}
 \Text(90,-12)[lb]{vd=typeFD}
 \Text(-42,-12)[lb]{fd=typeID}
 \Text(-42,137)[lb]{vu=typeFU}

 \Text(-15,15)[lb]{$\beta $}
 \Text(-4,123)[lb]{$p_1$}
 \Text(-4,3)[lb]{$p_2$}
 \Text(82,123)[lb]{$p_4$}
 \Text(82,3 )[lb]{$p_3$}
 \Text(98,63)[lb]{$q-p_4$}

 \Text(-35,63)[lb]{$q+p_1$}
 \ArrowLine(-5,75)(-5,53)
 
 \Text(95,20)[lb]{$\nu$}
 \Text(73,11)[lb]{$\nu_1$}
 \Text(73.5,27)[lb]{$\nu_2$}

\Text(35,-15)[lb]{b) T6}
\end{picture}}}

\end{array}
\]
\caption[$ffbb$ boxes, topologies T5 and T6]
        {Boxes for $ffbb$ processes, topologies T5 and T6.}
\label{ffbbboxesT5T6}
\end{figure}

\leftline{\bf Topologies T5, T6}
\noindent Now we switch to the intermediate case of topologies having two internal bosonic lines, 
see Fig.~\ref{ffbbboxesT5T6} showing all the field, type and Lorentz indices and momentum flows.
These two topologies are {\em not} the usual pair of direct and crossed ones.
Note that they are both drawn as two different direct boxes.

The paragraph after Fig.~\ref{ffbbboxesT2T4} is applicable here too.
For these topologies, contrary to all previous cases, the leptonic current flows through the 
diagram: from lower--left (p2) to upper--right corner (p4), rather then from lower--left (p2) 
to upper--left corner (p1).
This forces us to introduce the notion of {\em in the sense of Reduction,} i.e. to perform
the calculation of a diagram denoting momentum flows as is suitable for Reduction and at the end
come back to the {\em Real} notation for external momenta and Mandelstam invariants.
This is why the corresponding procedure is called {\sf isoR2Real}, see Section~\ref{pintr}.

The calculation of T5 and T6 topologies is realized in two specific procedures
{\sf boxT5(vd,fd,vu,fu)} and {\sf boxT6(vu,fd,vd,fu)} in nested loops over all allowed
field indices of the virtual particles.

Even though we consider the processes with a photon, for this case internal bosons 
can be charged or neutral if the photon is coupled to the fermion line.
The virtual fermion index runs over a doublet.

The calculation starts by several calls to specific procedure
{\sf CalcBoxT5(`typeFD',`typeID',`typeFU',`typeIU', k3min,k3max,k4min,k4max)} and
by a single call to
{\sf CalcBoxT6(`typeFU',`typeID',`typeFD',`typeIU',k3min,k3max, k4min,k4max)}.
They, in turn, call the two specific procedures shown above.

Such an asymmetry is due to the fact that we assume {\sf `typeFD'} to be a photon,
and therefore the virtual bosons
in diagram T6 must be charged: {\sf k3min=3, k3max=6, k4min=3, k4max=6}. Note, that
{\sf k3=\{k3min,k3max\}} and {\sf k4=\{k4min,k4max\}}.
\vspace*{1mm}

For diagrams with two virtual bosons the clustering is performed in the following way:\\[.5mm]
--- Cluster 22, {\sf k3=\{2,5\}, k4=\{2,5\}}\\
--- Cluster 33, {\sf k3=\{3,6\}, k4=\{3,6\}}\\
--- Cluster 42, {\sf k3=\{4,4\}, k4=\{2,5\}}\\
--- Cluster 24, {\sf k3=\{2,5\}, k4=\{4,4\}}\\
--- Cluster 44, {\sf k3=\{4,4\}, k4=\{4,4\}}
\vspace*{1mm}

Then the calculation continues for each topology by calls to the intrinsic procedures
{\sf FeynmanRules, GammaRight, Diracizing, Reduction, Pulling, Diraceq, PullingOrder,
Scalprod, Sing, ExtMomentumWI} and then to {\sf ScalarizingProj} as described in the previous 
section.

The intrinsic procedures {\sf p2p} and {\sf isoR2Real} are called at the end.

The results of calculations of boxes of T5 and T6 topologies are stored in the files\\
{\sf ffbb`k3min'`k4min'T5xi`xi'on`on'mp`mp'`vd'`fd'`vu'`fu'.sav} and \\
{\sf ffbb`k3min'`k4min'T6xi`xi'on`on'mp`mp'`vu'`fd'`vd'`fu'.sav}.
\vspace*{2mm}

In the beginning of the program there are usual definitions:
\vspace*{-2mm}

\begin{verbatim}
#define xi "0" / #define on "0" / #define mf "1" /#define mp "1"
\end{verbatim}
\vspace*{-2mm}

A table of  CPU times for the $d\bar{d}\to\gamma\gamma$ boxes of topologies T5+T6 reads:\\[.5mm]
{\sf xi "0", on "0", mf "1", mp "0"} --- ~32 hours,   \\
{\sf xi "0", on "1", mf "1", mp "1"} --- 4.5 hours,   \\
{\sf xi "1", on "1", mf "1", mp "1"} --- ~19 minutes. \\
A recomputation of these boxes is not allowed either.
\vspace*{2mm}

\leftline{\bf Topology T7}

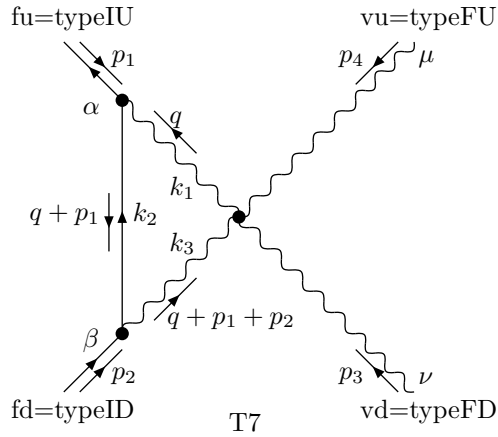
\begin{figure}[!h]
\[
  \vcenter{\hbox{
\begin{picture}(132,132)(0,0)

\Text(18,52)[lb]{$ k_3 $}
\Text(18,73)[lb]{$ k_1 $}
\Text(4,63 )[lb]{$ k_2 $}
\Text(18,100)[lb]{$ q $}    

\Photon(88,22)(110,0){2}{3}
\Photon(88,22)(0,110){2}{11}

\Vertex(44,66){2.5}

\Vertex(0,110){2.5}
\ArrowLine(0,110)(-22,132)
\ArrowLine(0,22)(0,110)

\Vertex(0,22){2.5}
\ArrowLine(-22,0)(0,22)
\Photon(0,22)(88,110){2}{11}

\ArrowLine(104,132)(88,116)

\Photon(110,132)(88,110){2}{3}

\ArrowLine(104,0)(88,16)

\ArrowLine(-16,0)(0,16)
\ArrowLine(-16,132)(0,116) 

\ArrowLine(28,90)(12,106) 
\ArrowLine(-5,75)(-5,53)

\ArrowLine(12,27)(28,43) 
\Text(17,24)[lb]{$q+p_1+p_2$}
\Text(-35,63)[lb]{$q+p_1$}
 \Text(-15,105)[lb]{$\alpha$}
 \Text(112,123)[lb]{$\mu$}
 \Text(90,137)[lb]{vu=typeFU}
 \Text(90,-12)[lb]{vd=typeFD}
 \Text(-42,137)[lb]{fu=typeIU}
 \Text(-42,-12)[lb]{fd=typeID}
 \Text(112,3)[lb]{$\nu$}
 \Text(-15,15)[lb]{$\beta $}
 \Text(-4,123)[lb]{$p_1$}
 \Text(-4,3)[lb]{$p_2$}
 \Text(82,123)[lb]{$p_4$}
 \Text(82,3  )[lb]{$p_3$}
 
\Text(37,-15)[lb]{ T7}

\end{picture}}}
\]
\vspace*{1mm}
\caption[$ffbb$ boxes, topology T7]
        {Boxes for $ffbb$ processes, topology T7.}
\label{ffbbboxesT7}
\end{figure}

\noindent
Topology T7 also has two internal bosonic lines, see Fig.~\ref{ffbbboxesT7}, however it is rather
a pinch of topologies T1 and T4 (bosonic line with field index k4 is pinched out).

The calculation of the T7 topology is realized in the specific procedure
{\sf boxT7(fu,fd,vd,vu)} in nested loops over field indices.
If there is only one photon in the final state, the virtual bosons must be charged.
This is why there is only one cluster in this case.

This is a vertex-like diagram and for this reason it is much simpler then all the others.

The calculation starts by a single call to specific procedure
{\sf CalcBoxT7(`typeIU',`typeID',`typeFD',`typeFU')} which calls 
the procedure {\sf boxT7} shown above.

The calculation continues by calls to the intrinsic procedures
{\sf FeynmanRules}, {\sf GammaRight}, {\sf Diracizing}, {\sf Diraceq},
{\sf Reduction}, {\sf Pulling}, {\sf Diraceq}, {\sf Scalprod}, {\sf Sing},
{\sf ExtMomentumWI}, {\sf Scalarizing, p2m, DivisionGramDet}.

The results are stored in the file {\sf ffbbT7xi`xi'on`on'mp`mp'`fu'`fd'`vd'`vu'.sav}.\\

In the beginning of the program there are four usual definitions:
\begin{verbatim}
#define xi "0" / #define on "0" / #define mf "1" / #define mp "1".
\end{verbatim}

The topology T7 needs only a few seconds of CPU time.

\clearpage

\section{SANC Procedures}
\subsection{Introduction}
At \underline{Level 1} the {\tt SANC} database contains FORM {\em programs} and 
{\em procedures}.
All FORM programs are accessible for the user via a sequence of clicks on {\tt SANC} tree,
when we reach a file and open it. The situation is very different for the procedures 
which are of three kinds: {\em specific, special} and {\em intrinsic}.

A procedure is called specific if it is used only in one particular program. Normally, specific
procedures are included in the corresponding FORM program body and are also open 
for the user. Moreover, they are very easy to read and describe.

Special procedures are usually used by only a limited number of FORM programs and, similar to
specific procedures, they do a job which is relevant for these programs only. However, contrary to
specific procedures we do not open them for the users, since their content is not so transparent
to be read and described.
It is envisaged to upgrade special procedures to the level of intrinsic procedures
in future versions of {\tt SANC}.

Intrinsic procedures are used by many FORM programs and should be easily used in any new
program. Their functions are totally determined by the list of their arguments which are of two
types: genuine arguments (FORM variables), denoted below as {\bf AVALUE}, and {\em options}, 
usually integer numbers, denoted as {\bf IVALUE}, which are used as switches governing calculation 
flow inside the procedure. Sometimes an {\bf IVALUE} (see below) stands for a {\em special use},
i.e. a special flow inside an intrinsic procedure.

\subsection{Intrinsic procedures\label{pintr}}
Below we give a list of intrinsic procedures which are mostly met in the
{\bf Precomputation} trees of SANC. In this list, which is not complete by far,
the procedures are presented in alphabetic order, and the `treatment' 
of their options and arguments is explained.

\def\phIV{\phantom{ IVALUE =}}
\def\phAV{\phantom{ AVALUE =}}
\newcommand{\qhat}{\hat{q}}

\begin{description}
 \item[\underline{\sf a2b}]: replaces symbol ``a'' to symbol ``b''.
Possible arguments: $\stw\to\ctw$, $\mw\to\mz$, $\gamma_6\to\gamma_5$, $\gamma_7\to\gamma_5$,
                    $\sigma_{i}\to\sigma_{j}$, $\delta_{i}\to\delta_{j}$ {\it etc.} and 
{\it vice versa.}
\begin{description}
 \item[AVALUE = ($\stw,\ctw$)], for example.
\end{description}
\end{description}

\begin{description}
\item [\underline{\sf bpIdentities}]:
applies identities to the so-called auxiliary 
Passarino--Veltman functions (bp0=$b_0$ and bp1=$b_1$~\cite{Bardin:1999ak}), attempts to exclude 
bp0 (and bp1).
\begin{description} 
 \item[IVALUE = (I)]
 \item[\phIV I=0] --- to exclude bp0        
 \item[\phIV I=1] --- to exclude bp0 and bp1.
\end{description}
\end{description}

\begin{description} 
\item[\underline{\sf Diraceq}]: applies Dirac equations
to expressions preliminary simplified with the aid of {\underline{\sf Pulling}}.
\begin{description} 
 \item[AVALUE = (i,j,k,l)] 
 \item[\phAV i] for spinor tlo, Dirac equation:         $ {\rm tlo}\,\sla{p_1} =-i m_i \,{\rm tlo}$
 \item[\phAV j] for spinor tro, \phantom{\,Dirac equation:}$\sla{p_2} \,{\rm tro} = i m_j \,{\rm tro}$
 \item[\phAV k] for spinor tre, \phantom{Dirac equation:}$ {\rm tle}\, \sla{p_3}=-i m_k \,{\rm tle}$
 \item[\phAV i] for spinor tle, \phantom{~Dirac equation:}$\sla{p_4} \,{\rm tre} = i m_l \,{\rm tre}$
\end{description}
where (i,j,k,l) are field indices.
\end{description}

\begin{description} 
\item[\underline{\sf Diracizing}]: expressions of the form $\qhat\Gamma\qhat$ and
$\gamma_\alpha\Gamma\gamma_\alpha$ are simplified; here $\qhat\equiv\gamma\cdot q$,
and $\Gamma$ is a string of up to five $\gamma$ matrices.
The final step consists of setting $\qhat\cdot\qhat=q^2$ and
$\gamma_\alpha\gamma_\alpha=n$, where $n=4-\varepsilon$ is the dimension of momentum space.
\vspace*{-1mm}
\begin{description} 
\item[IVALUE = (I)]
\item[\phAV I=0] normal use
\item[\phAV I=1] a special use in Wff vertices.
\end{description}
\end{description}

\begin{description} 
\item[\underline{\sf DirectProdSumm}]: performs summation in direct products of $\gamma$ matrices
such as in well known identity
$$
 \gamma_\mu\gamma_\alpha\gamma_\nu\gamma_6\otimes
 \gamma_\nu\gamma_\beta \gamma_\mu\gamma_6=
4\gamma_\beta\gamma_6\otimes\gamma_\alpha\gamma_6;
$$
this procedure knows 76 identities of such a kind.
\begin{description} 
 \item[AVALUE = (i,j,k,l)], the same arguments as in {\underline{\sf Diraceq}}.
\end{description}
\end{description}

\begin{description} 
\item[\underline{\sf DivisionGramDet}]:
realizes various possibilities to use the algebra of Gram determinants to simplify raw
expressions.
\vspace*{-1mm}
\begin{description} 
 \item[IVALUE = (I)]  
 \item[\phAV I=0] division of det3i (active in all subsequent options) 
 \item[\phAV I=1] division of det4i for $O_s,T_s$ topology, for example T1,T2
 \item[\phAV I=2] division of det4i for $O_s,U_s$ topology, for example T3,T4
 \item[\phAV I=3] division of det4i for $T_s,U_s$ topology, for example T5
 \item[\phAV I=4] division of det4i for $U_s,T_s$ topology, for example T6.
\end{description}
\end{description}

\begin{description} 
\item[\underline{\sf Expansions}]:
expands $B_0$ and $B^F_0$ functions for small values of some of its arguments.  
\vspace*{-1mm}
\begin{description}
\item[AVALUE,IVALUE = (FI)]
 \item[\phAV FI] field index
\end{description}
\end{description}

\begin{description} 
\item[\underline{\sf ExpansionPhotMassShell}]:
puts an external bosonic momentum in $ffbb$ processes to the corresponding mass shell.
For example
in the process $ff\to\gamma B$ for $p^2_\gamma$ or $p^2_{\sss B}$ its action means
\vspace*{-1mm}
\begin{description}
\item[IVALUE,AVALUE = (I,J,mp,pGs,pBs)]
 \item[\phAV I=1,~J=1] $pGs=0,\qquad\;pBs=0$
 \item[\phAV I=1,~J=2] $pGs=0,\qquad\;pBs=-mp^2$
 \item[\phAV I=2,~J=1] $pGs=-mp^2,\;  pBs=0$
 \item[\phAV I=2,~J=2] $pGs=-mp^2,\;  pBs=-mp^2$.
\end{description}
\end{description}

\begin{description}
\item [\underline{\sf ExtMomentumWI}]: applies Ward identities for external vector boson momenta,
i.e. sets $(p_I)_\mu=0$ and $(p_I)_\nu=-(p_J)_\nu-(p_K)_\nu$.
\vspace*{-1mm}
\begin{description}
 \item[AVALUE = (I,mu,J,K,nu)]
\end{description}
\end{description}

\begin{description} 
\item[\underline{\sf f2f}]:
realizes the possibility (in particular cases for certain arguments of PV functions)
of replacing $B_0\to A_0$, $B_0^F\to B_0^F$, $B_0\to b_{1}$ and {\it vice versa},
if concrete arguments of the PV functions allow such replacements.
\vspace*{-1mm}
\begin{description} 
 \item[AVALUE = (b0,a0)], for example.
\end{description}
\end{description}

\clearpage

\begin{description} 
\item[\underline{\sf FeynmanRules}]: applies Feynman rules for propagators and vertices,
see Section~\ref{frules}.
\begin{description} 
 \item[IVALUE = (I)] 
 \item[\phAV I=0] for QED part
 \item[\phAV I=1] for EW and QCD parts.
\end{description}
\end{description}

\begin{description} 
\item[\underline{\sf GammaLeft}]: all Dirac matrices $\gamma_5$,
$\gamma_6=  1+\gamma_5$ and $\gamma_7=1-\gamma_5$ are moved to the left and
the expression 
is simplified using identities $(\gamma_5)^2=1$, $\gamma_6\gamma_7=0$, etc.
\end{description}

\begin{description} 
\item[\underline{\sf GammaRight}]: the same as in {\sf GammaLeft}, but all matrices 
$\gamma_i $ are moved to the right.
\end{description}

\begin{description} 
\item[\underline{\sf GammaTrace}]: the traces of products of $\gamma$ matrices are evaluated 
in $n$ dimensional space. 
\end{description}

\begin{description} 
 \item[\underline{\sf Globals}]: performs global declarations by FORM Tables 
of particle names, particle masses, electric charges, ghost charges, 
mass ratios, coupling constants, weak isospins, gauge parameters, combinatorial factors etc.
\end{description}

\begin{description} 
\item[\underline{\sf isoR2Real}]:
realizes the ideology of shifting from the level {\em in the sense of reduction $p_i$}
to the real 4-momenta $p_i$, $(p_i)_{input} \to (p_i)_{output}$ and 
{\sf (Invariants)}$_{\rm input} \to$ {\sf (Invariants)}$_{\rm output}$, see
item {\bf Topologies T5, T6}.
\begin{description} 
  \item[AVALUE = (p1out,p2out,p3out,p4out,Qsout,Tsout,Usout)]
\end{description}
\end{description}
Only {\sf output} values appear in the argument list; the {\sf input} string is assumed to be 
{\sf p1,p2,p3,p4,Qs,Ts,Us}.

\begin{description} 
\item[\underline{\sf m2zero}]: sets a mass $mp$ to $0$ in expressions and in the arguments of all
functions.
\begin{description}
\item[AVALUE = (mp)]
\end{description}
\end{description}

\begin{description} 
\item[\underline{\sf Masshell}]:
This procedure has four arguments, which must be fermionic field 
indices; a field, whose index is an argument of {\sf Masshell}, is put on its mass shell.
Thus the command \verb+#call Masshell(`iu',,,)+ sets \verb+p(`iu')^2+ equal to \verb+-pm(`iu')^2+.
\begin{description}
 \item[AVALUE = (iu,id,fu,fd)], the same list as in {\underline{\sf Diraceq}}.
\end{description}
\end{description}

\begin{description} 
\item[\underline{\sf open}]: opens a symbol, e.g. substitutes a combination of coupling constants
like \verb+vmaen=ven-aen+.
\begin{description}
 \item[AVALUE = (a)]
\end{description}
\end{description}

\begin{description} 
\item[\underline{\sf openall}]: opens all coupling constants, charges, etc. and then
{\underline{\sf substitute}}s them.
\end{description}

\begin{description} 
\item[\underline{\sf opensymbol}]: opens all symbols as {\underline{\sf openall}} but only in terms
containing symbol `a'.
\begin{description}
 \item[AVALUE = (a)]
\end{description}
\end{description}

\begin{description} 
\item[\underline{\sf p2D}]:
replaces $p_i$ and $p_j$ by vectors Q and D: $Q =- (p_i+p_j)$ and $D = p_i-p_j$.
\begin{description} 
\item[AVALUE = (i,j)]
\end{description}
\end{description}

\begin{description} 
\item[\underline{\sf p2I}]: changes a $p^2$ to an invariant $I$.
\begin{description} 
\item[AVALUE = (p,I)]
\end{description}
\end{description}

\begin{description} 
\item[\underline{\sf p2m}]: 
puts a 4-momentum $p$ to its mass shell, $p^2=-mp^2$,
in the expressions and in the arguments of all functions.
\begin{description} 
\item[AVALUE = (p,mp)]
\end{description}
\end{description}

\begin{description} 
\item[\underline{\sf p2p}]: changes a $p^2$ to $P^2$ and $\hat p \to \hat P $ 
in the string of gamma matrices.
\begin{description} 
\item[IVALUE,AVALUE = (I,p,P).]
\item[\phAV$\qquad\qquad$ I=0] $p^2$ changes to $P^2$                                        
\item[\phAV$\qquad\qquad$ I=1] $p^2$ changes to $P^2$ and $\hat p \to \hat P $
\end{description}
\end{description}

\begin{description} 
\item[\underline{\sf p2Qs}]: 
expresses all scalar products $p_i\cdot p_j$ in terms of
$p_1^2$, $p_2^2$ and $Q^2=(p_1+p_2)^2$ for a three point function with $Q+p_1+p_2=0$.
\begin{description} 
\item[AVALUE = (p,mp)]
\end{description}
\end{description}
                    
\begin{description}                           
\item[\underline{\sf PoleSep}]:
separates the PV functions explicitly into their $1/\bar\varepsilon$ pole parts
and finite parts $A_0^{F}$ and $B_0^F$.
\end{description}

\begin{description}                           
\item[\underline{\sf Pulling}]: is applied to expressions of the form
$$
\bar{u}(p_1)\left(\gamma_\alpha\hat{p}_1\hat{p}_2\gamma_\beta\ldots\right)u(p_2)
$$
with the result that $\hat{p}_1$ is placed next to $\bar{u}(p_1)$ and
$\hat{p}_2$ is placed in front of ${u}(p_2)$, after which the expression is
simplified using the Dirac equation by a call to procedure {\underline{\sf Diraceq}}.
\begin{description} 
\item[IVALUE = (I)] 
\item[\phIV I=0] main option, is used in all programs up to ffbb boxes; 
\item[\phIV $\quad\;\;$] eliminates $p_4$ in {\sf ii} current and $p_2$ in {\sf jj} current
\item[\phIV 1,2,3,4] is used in ffbb boxes;
\item[\phIV $\quad\;\;$] eliminates $p_1$ or $p_2$ or $p_3$ or $p_4$ in {\sf ii} current.
\footnote{See examples of assignment of current labels {\sf ii} in Figs.~\ref{ffffboxesT2T4} 
and \ref{ffbbboxesT2T4}.}
\end{description}
\end{description}

\begin{description} 
\item[\underline{\sf PullingOrder}]: this procedure orders
$\gamma$ strings containing $\hat{p}$, $\gamma_\mu$ and $\gamma_\nu$ into
$\gamma_\mu \hat{p} \gamma_\nu$ with one of the three factors possibly missing.
Thus the surviving expressions are $\gamma_\mu \hat{p} \gamma_\nu$,
$\gamma_\mu \hat{p}$, $\gamma_\mu\gamma_\nu$, and $\hat{p} \gamma_\nu$.
\begin{description}
\item[AVALUE = ($p,\mu,\nu$)] 
\end{description}
\end{description}

\begin{description} 
\item[\underline{\sf Reduction}]: it has options {\sf I=0, 1}: if {\sf I=0}, then
the user can perform a {\it prereduction}\footnote{
The prereduction consists of simplifications, such as the replacement
of $q^2/(q^2+m^2)$ by $1-m^2/(q^2+m^2)$.} ``{\it by hand}'';\\
if {\sf I=1}, the {\it standard prereduction} is done automatically.\\
After a prereduction is done, the reduction is performed on integrals of the form
\bqa
\int \frac{d^nq\{1,\,q_\mu,\,q_\mu q_\nu,\dots \}}{d_0d_1d_2d_3}
\label{def:int}
\eqa
where $\{1,\,q_\mu,\,q_\mu q_\nu,\dots \}$ means one of the expressions:
scalar, vector, tensor~\footnote{In the present version of
SANC we have tensors of up to the 4th rank and N-point functions for N up to 4.}
and $d_i$ are given by
\bqa
d_0 &=& q^2+m_1^2,               \nonumber\\
d_1 &=& (q+p_1)^2+m_2^2,         \nonumber\\
d_2 &=& (q+p_1+p_2)^2+m_3^2,     \nonumber\\
d_3 &=& (q+p_1+p_2+p_3)^2+m_4^2. \nonumber
\eqa

\noindent
The {\sf N-point function}, i.e. a one-loop diagram 
with {\sf N} external legs, is defined by the following diagram:

\vspace*{-2mm}
\[
\begin{array}{c}
\begin{picture}(100,100)(0,-50)
\SetScale{1}
\DashArrowLine(0,0)(30,0){3}
\DashArrowLine(50,57)(50,27){3}
\ArrowLine(80,27)(50,27)
\ArrowLine(50,27)(30,0)
\ArrowLine(30,0)(50,-27)
\ArrowLine(50,-27)(80,-27)
\DashArrowLine(50,-57)(50,-27){3}
\SetScale{1}
\Text(36,40)[lb]{$p_1$}
\Text(5.5,15)[lb]{$q+p_1$}
\Text(-25,-20)[lb]{$q+p_1+p_2$}
\Text(62.5,-39)[lb]{$q+p_1+p_2+p_3$}
\Text(3,3)[lb]{$p_2$}
\Text(36,-45)[lb]{$p_3$}
\Text(62.5,30)[lb]{$q$}
\Text(62.5,14)[lb]{$d_0$}
\Text(45,10)[lb]{$d_1$}
\Text(45,-15)[lb]{$d_2$}
\Text(62.5,-22.5)[lb]{$d_3$}
\end{picture}
\end{array}
\]
\vspace*{.2cm}

As a result the integrals (\ref{def:int}) are replaced by linear combinations of
Passarino--Veltman functions $(PV)_{ij}\in\{A_{ij},\,B_{ij},\,C_{ij},\,D_{ij}\}$
where $i=1,\,2,\,3$ or $4$ for vector, 2nd, 3rd or 4th rank tensor, respectively,
and $j$ is a sequential number. 

\begin{description}
\item[IVALUE = (I)] 
\item[\phIV I=0] works without internal Prereduction
\item[\phIV I=1] the standard internal Prereduction is performed
\end{description}
\end{description}

\begin{description} 
\item[\underline{\sf substitute}]: substitutes an argument ``a'' 
(charge, isospin or coupling constant of a particle).
For example, to substitute the isospin of the particle `typeID'=11 we call
{\sf substitute(i3(`typeID'))} with the result {\sf i3(`typeID')=1/2}.
\begin{description}
\item[AVALUE = (a)]
\end{description}
\end{description}

\begin{description} 
\item[\underline{\sf Scalarizing}]: expresses all (PV)$_{ij}$ functions in terms of scalars PV$_0$.
Option {\bf I=0} acts differently for boxes and N=2,3 point functions;
for 4-point functions the masses of the particles with momenta $p_1$and $p_2$ are set equal to zero,
for 2,3-point functions Scalarizing is exact in masses. Four digit options are applied only 
for 4-point functions as explained below. In all cases Scalarizing is exact in masses.
Various options have been introduced to save CPU time.
\begin{description}
\item[IVALUE = (I)]
\item[\phAV I=1234] all masses are different
\item[\phAV I=1134] masses $m_1$ and $m_2$ are equal
\item[\phAV I=1133] masses $m_1,m_2$ and $m_3,m_4$ are equal
\item[\phAV I=1232] masses $m_2$ and $m_4$ are equal (for boxes T5, T6)
\item[\phAV I=0   ] masses $m_1$ and $m_2$ are equal to zero (for boxes)
\item[] The latter option must be used also for self energies and vertices.
\end{description}
\end{description}

\begin{description} 
\item[\underline{\sf Scalarizingdp}]: Scalarizing of dp$_{ij}$ PV series, 
see Ref.~\cite{Bardin:1999ak} for definitions of $d\equiv$dp functions.
\begin{description} 
 \item[IVALUE = (I)]
 \item[\phIV I=0] scalarizing of {\sf dp} series setting masses $m_1$ and $m_2$ equal to zero
 \item[\phIV I=1] scalarizing of {\sf dp} series  exact in all masses
\end{description}
\end{description}

\begin{description}
\item[\underline{\sf ScalarizingProj}]: acts in $ffbb$ boxes of topologies T1, T3, T5 and T6.
At first it projects the Input expression into a number of terms according to different
structures with corresponding coefficients. Next it starts to apply the procedure 
{\it Scalarizing} to each term.
The choice of index of {\sf Scalarizing(I)} depends of the type of box topology.
Finally, it forms Output expression by summing up all terms.
\begin{description}
\item[AVALUE = (k3min,k4min,Topology,NameInput,p,mu,nu,p1,p2,I,NameOutput)]
\item[\phAV k3min,k4min] the indices, defining a cluster, see items {\bf Topologies T1,T3}\linebreak
$\,$\hspace*{13mm} and {\bf Topologies T5,T6}
\item[\phAV Topology] number of the topology (1,3,5,6)
\item[\phAV NameInput] name of input expression
\item[\phAV p,mu,nu] the same arguments as in {\sf PullingOrder}
\item[\phAV p1,p2] fermionic 4-momenta defining basis of structures
\item[\phAV I] index of procedure {\it Scalarizing}
\item[\phAV NameOutput] name of output expression
\end{description}
\end{description}

\begin{description} 
\item[\underline{\sf Scalprod}]: calculates scalar products for 4-point function with 
4-momenta satisfying $p_1+p_2+p_3+p_4=0$.
\begin{description}
\item[AVALUE = (p)]
\item[\phAV p] scalar products $p_i \cdot p_j$
\item[\phAV K] scalar products $K_i \cdot K_j$, etc.
\end{description}
\end{description}

\begin{description} 
\item[\underline{\sf Sing}]: In procedure {\sf Sing}, the dimension $n$ is set equal 
to $4-\eps$, and then the PV functions, multiplied by $\eps$, are analyzed: if a PV function has 
a pole, then the product $\eps\times$PV
is replaced by its residue, and finally $\eps$ is set equal to zero.
\end{description}

\begin{description} 
\item[\underline{\sf Symmetrize}]: symmetrizes Passarino-Veltman functions.
Thus $B_0$ is symmetrized using the symmetry property
${\ds B_0(Q^2,\,m_1,\,m_2)= B_0(Q^2,\,m_2,\,m_1)}$.
\begin{description}
\item[IVALUE = (I)]                 
\item[\phAV I=0] symmetrizes $B_0$ functions
\item[\phAV I=1] symmetrizes $B_0$ and $C_0$ functions.
\end{description}
\end{description}

\begin{description} 
\item[\underline{\sf Xi1}]: sets all gauge parameters $\xi,\xi_{\sss A},\xi_{\sss Z}$ 
equal to one. These parameters are present in all intermediate
contributions in $R_\xi$ gauge but cancel in gauge invariant physical observables.
\end{description}

\subsection{Feynman rules\label{frules}}
The {\tt SANC} collection of Feynman rules is based on the Standard Model Lagrangian in
$R_\xi$ gauge with three gauge fixing parameters $\xi$, $\xi_{\sss A}$, 
and $\xi_{\sss Z}$~\cite{Bardin:1999ak}.\\
\underline{Propagators}\\
Every propagator should be multiplied by a factor of
${\ds{{1}/{\left( 2\pi\right)^4 i}}}\;.$  \\
The propagator of a {\bf fermion} $f$ is a non-commuting function.
It is defined by the following {\tt SANC} command:
\vspace*{-2mm}
\[
\begin{array}{cc}
\mbox{pr(k,p,ii)}, \qquad 
&
\begin{array}{c}
\vcenter{\hbox{
  \begin{picture}(60,20)(3,5)
  \SetScale{2.}
    \ArrowLine(0,5)(30,5)
  \end{picture}}}
\\ f
\end{array}
\label{fer_prop}
\end{array}
\vspace*{-2mm}
\]
where $k$ is the field index, $p$ is the fermionic 4-momentum and $ii$ is the fermionic current 
label.

The {\bf vector boson} propagator is a commuting function:
\[
\begin{array}{cc}
 \mbox{pr(k,$\mu$,$\nu$,p)}, \qquad
&
\vcenter{\hbox{
  \begin{picture}(60,20)(3,0)
  \SetScale{2.}
    \Photon(0,5)(30,5){2}{4}
   \Text(-10,5)[lb]{$\Large\mu$}
   \Text(65,5)[lb]{$\Large\nu$}
  \end{picture}}}
\end{array}
\]
where $k$ is the field index, $\mu, \nu$ are the corresponding Lorentz indices and
$p$ is the bosonic 4-momentum. \\
\underline{Vertices}

In the presently available class of diagrams, {\it vertices} are of three kinds:
boson-fermion-fermion (bff), three-boson (bbb) vertices and four-boson (bbbb) vertices.
A {\it vertex} is a non-commuting function.
Every vertex should be multiplied by a factor of $\left( 2\pi\right)^4 i\;.$ 

\underline{{\sf	bff} vertices}

The {\tt SANC} command  for this type of vertex is:
\vspace*{2mm}
\[
\begin{array}{cc}
\hspace*{-6cm}
\vcenter{\hbox{
\mbox{vert(i,l,-j,$\alpha$,ii)}}},
&
\hspace*{-4cm}
\vcenter{\hbox{
\begin{picture}(45,52)(-190,0)
  \Photon(0,26)(30,26){3}{5}
    \Text(0,31)[lb]{$i,\alpha$}
  \ArrowLine(45,52)(30,26)
    \Text(50,52)[lb]{$l$}
  \ArrowLine(30,26)(45,0)
    \Text(50,0)[lt]{$j$}
\end{picture}
}}
\end{array}
\vspace*{2mm}
\]
where $i$, $j$ and $l$ are field indices, $\alpha$ is a Lorentz label and {\sf ii} is a fermionic 
current index. The first field index refers to a boson; the other field indices refer to
the incoming $l$ and outgoing $-j$ fermion fields. 

\underline{{\sf bbb} vertices}

The {\tt SANC} command for trilinear vector boson vertices is of the following form:

\[
\begin{array}{cc}
\hspace*{-6cm}
\vcenter{\hbox{
\mbox{vert(i,-j,l,$\alpha$,$\mu$,$\nu$,$Q$,$p_1$,$p_2$)}}},
&
\hspace*{-6cm}
\vcenter{\hbox{
\begin{picture}(45,52)(-200,5)
  \Photon(0,26)(30,26){3}{5}
    \Text(0,31)[lb]{$i,\alpha,Q$}
  \Photon(45,52)(30,26){3}{5}
  \ArrowLine(37.6,13.6)(37,12.8)
    \Text(50,52)[lb]{${l,\nu,p_2}$}
  \Photon(30,26)(45,0){3}{5}
  \ArrowLine(34,38)(38.5,33.5)
    \Text(50,0)[lt]{$j,\mu,p_1$}
\end{picture}
}}
\end{array}
\vspace*{5mm}
\]
where $i$, $j$ and $l$ are boson field indices, $\alpha, \mu, \nu$ are Lorentz labels and 
$Q,p_1,p_2$ are incoming momenta such that $Q+p_1+p_2=0$.
Significant is that the triplets of arguments, \{i,-j,l\}, \{$\alpha$, $\mu$, $\nu$\}
and \{$Q$, $p_1$, $p_2$\} are written in the same order according to the rule:
``{\em from ingoing neutral to ingoing negative charge flow \ }'', where  the positive
charge flow is shown by the arrows in the diagram.

In the vertex diagrams involving a Higgs boson or scalar unphysical fields
(Higgs-Kibble ghosts), the Lorentz indices $\mu$ are shown in brackets since they are
{\it dummy} or {\it silent} indices, kept in the \verb+vert+ command for formal
reasons, whereas these vertices do not depend on $\mu$.


In the book~\cite{Bardin:1999ak} all tri-linear bosonic vertices together with their 
Feynman rules, involving Higgs bosons, scalar unphysical fields $\phi^0$ and $\phi^\pm$
and Faddeev-Popov ghosts are shown. Many of these diagrams carry silent (or dummy) Lorentz
indices as discussed above in connection with $bff$ vertices, and also dummy 4-momenta.

\underline{{\sf bbbb} vertices}

The {\tt SANC} command to define this class of vertices is (see the generic diagram)

\[
\begin{array}{ccc}
\hspace*{-6cm}
\vcenter{\hbox{
\mbox{vert(i,\,j,\,k,\,l,\,$\alpha$,\,$\beta$,\,$\mu$,\,$\nu$)}}},
&
\hspace*{-4cm}
\vcenter{\hbox{
\begin{picture}(45,52)(-200,0)
  \Photon(10,52)(30,26){3}{5}
  \Photon(10,0)(30,26){3}{5}
  \Photon(30,26)(50,52){3}{5}
  \Photon(30,26)(50,0){3}{5}
    \Text(60,48)[lb]{${l,\nu}$}
    \Text(-15,50)[lb]{${i,\alpha}$}
    \Text(60,5)[lt]{${k,\mu}$}
    \Text(-15,5)[lt]{${j,\beta}$}
\end{picture}
}}
\end{array}
\]
where $i,\,j,\,k,\,l$ are field indices and $\alpha$, $\beta$, $\mu$, $\nu$  are
corresponding Lorentz indices.

In the book~\cite{Bardin:1999ak} all quadri-linear bosonic vertices together with 
their Feynman rules are presented.

\clearpage

\section{Processes, available in SANC v.1.00\label{processes}}
In this section we briefly discuss the available in QED and EW branches processes.
In this paper we do not discuss processes of QCD branch which are scarce.

\begin{figure}[!h]
\[
\begin{array}{ccc}
\vcenter{\hbox{
\begin{picture}(100,100)(112,430)
\includegraphics[width=80mm,height=190mm]{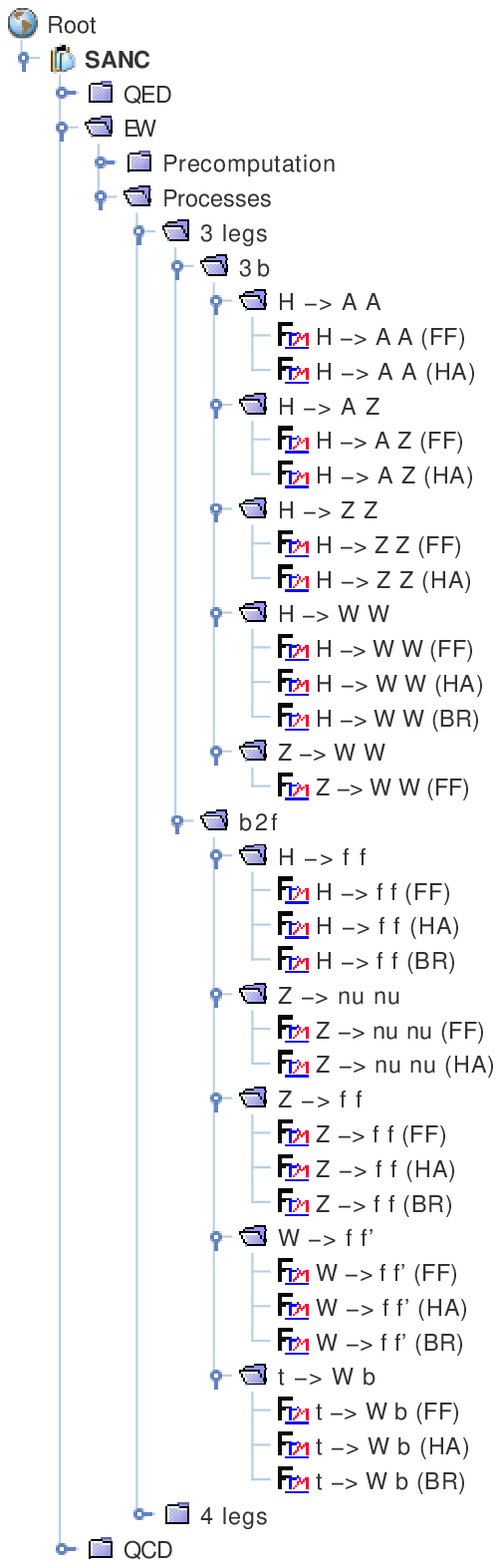}
\end{picture}}}
&\qquad&
\vcenter{\hbox{
\begin{picture}(100,100)(20,430)
\includegraphics[width=80mm,height=190mm]{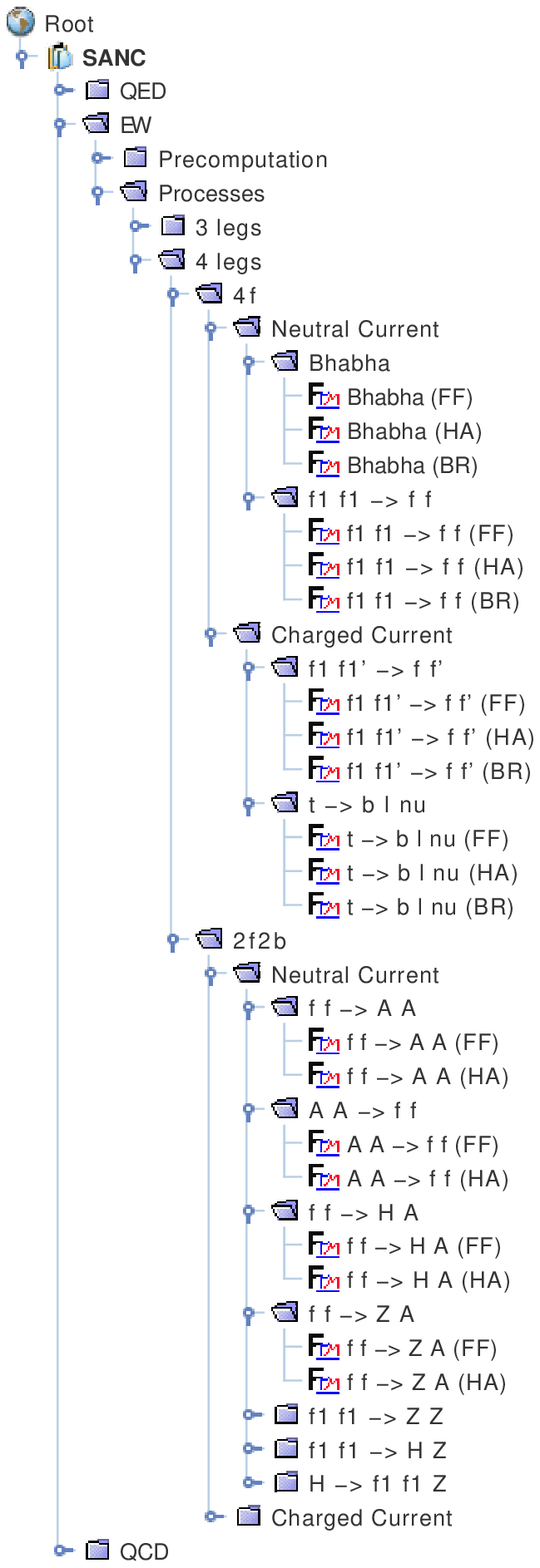}
\end{picture}}}
\end{array}
\]
\vspace*{148mm}
\caption[Available processes in EW part]
        {Available processes in EW part.}\label{ProcEW}
\label{fig111}
\end{figure}

The Fig.~\ref{ProcQED} shows the fully open menu for ``Processes'' in the QED branch of 
{\tt SANC} whose structure we briefly describe.
 The Figures~\ref{ProcEW} show all available 3-leg and 4-leg EW processes. 
We will not describe them in this part of the description since after explaining
the QED branch, they may be easily interpreted.
 Moreover, the QED branch has mostly a pedagogical purpose.
This is why it is worth devoting some time to it already in this first part mostly dealing
with Precomputation. However, we will not describe here the structure of corresponding modules,
leaving this for a second part of the {\tt SANC} description.
QED processes are presented by three classes: 1) a heavy photon decay; 2) $e^+e^-$ annihilation 
into a lepton pair (including Bhabha scattering); 3) Compton-like processes, i.e. 
$e^+e^- \to \gamma\gamma$ or some other cross channel.
Note that by our convention QED contains massless photons and three generations of leptons. 
We consider, nevertheless, the decay of a heavy photons for pedagogical reasons.

When we arrive via a menu sequence, {\it e.g.} {\bf QED $\to$ Processes $\to$ A $\to$ ll Decay},
we normally see three modules {\sf FF, HA and BR}. Modules {\sf FF} compute the scalar form 
factors of a given process. 
As was already stressed, they are channel independent, modulo a crossing transformation. 
Then a FORTRAN code to compute them can be automatically generated as described 
in Section~\ref{Uguide}. 
Modules {\sf HA} compute channel dependent HAs. The channel is evident for 
examples 1) and 2); for example 3) we have for the time being HAs for the
annihilation channel $e^+e^- \to \gamma\gamma$ and for the inverse channel 
$\gamma\gamma  \to e^+e^-$. 

\begin{floatingfigure}{78mm}
\hspace*{-6mm}
\vspace*{-10mm}
\includegraphics[width=78mm,height=95mm]{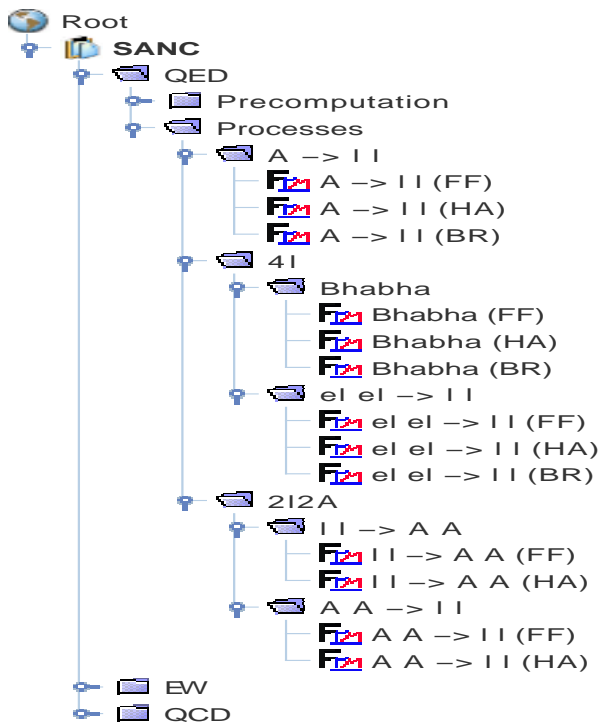}
\vspace*{5mm}
\caption[Available processes in QED part]
{Available processes in QED part.}\label{ProcQED}
\end{floatingfigure}

Both in QED and EW branches, modules {\sf BR} compute analytically the contributions 
due to accompanying Bremsstrahlung.
In this connection, present implementation of Brem\-sstrah\-lung into \linebreak
{\tt SANC} is not homogeneous.
As a rule, for all $1\to 2$ decays we have both Soft and Hard photon contributions. 
If only neutral particles are involved, for example $Z\to \nu\nu$, the module BR is not present
in the tree.
There is one exception: for $Z\to W^{+}W^{-}$ decay we have only one module {\sf FF},
since it is unphysical and we implemented {\sf FF} for future use as a building block for more
complicated processes. As a rule, for $4f$ processes we also have both Soft and Hard photon 
Bremsstrahlung with exception of Bhabha process where we have only Soft contribution.
For CC $2f_1\to 2f$ processes where we have realized quite involved calculations 
of Hard Bremsstrahlung with a possibility to impose simple cuts, see also~\cite{Arbuzov:2005dd}.
For the tree body decays $t\to b l\nu$ we have implemented both Soft and Hard contributions. 
For $2f2b$ processes we had so far no {\sf BR} and no {\tt s2n} calculations.

This work is being started in {\tt version 1.10}, where
we have implemented {\sf FF} and {\sf HA} modules for three more $ffbb$ 
processes: $f_1\bar{f_1}\to ZZ$, $f_1\bar{f_1}\to ZH$ and $H\to f_1\bar{f_1}Z$.
We recall, that $f_1$ stands for a massless fermion (its mass is retained only in the arguments 
of logs, if necessary). For the decay channel $H\to f_1\bar{f_1}Z$ and annihilation process
$f_1\bar{f_1}\to ZH$ we have implemented ({\bf BR}) modules and
{\tt s2n} calculations, 
see also~\cite{Bardin:2005dp} which contains an extensive presentation of numerical results.
\vspace*{-10mm}

\clearpage

\clearpage

\section{User Guide}\label{Uguide}
\subsection{Getting started}\label{Uguide:Intro}
\subsubsection{SANC installation}
To work with {\tt SANC}, one must install a {\tt SANC} client on ones computer.
The {\tt SANC} client can be downloaded from the {\tt SANC} project homepage
{\it http://sanc.jinr.ru} or {\it http://pcphsanc.cern.ch}.
 On the homepage select {\bf Download}, then download the client appropriate
for your operational system (Linux, Windows), save it to your home directory
and follow the instructions.
\footnote{To install and run {\tt SANC} client one should
have the Java Runtime Environment (JRE) at least version
5.0 Update 5 installed, see section {\bf Minimum System Requirements}
of the {\bf Download} page at the {\tt SANC} project homepage.}

In Linux, opening the *.tgz file creates the directory
{\bf /home/$<${\it user}$>$/sanc\_installer}.
Change to that directory, read the {\tt Readme} and execute {\tt install.sh}.
{\tt SANC} is then installed in {\bf /home/$<${\it user}$>$/sanc\_1.00.00}.
To start a {\tt SANC} session, go to directory
{\bf /home/$<${\it user}$>$/sanc\_1.00.00/bin} and execute {\bf sanc}.

In Windows, start {\tt sanc\_installer.exe} program and follow
the instructions, restart computer. To start a {\tt SANC} session,
click {\tt SANC} client icon.

\subsubsection{SANC windows}
\begin{figure}[h]
\centering
\setlength{\unitlength}{1mm}
\begin{picture}(180,100)
\put(0,-98){\makebox(0,0)[lb]{\psfig{file=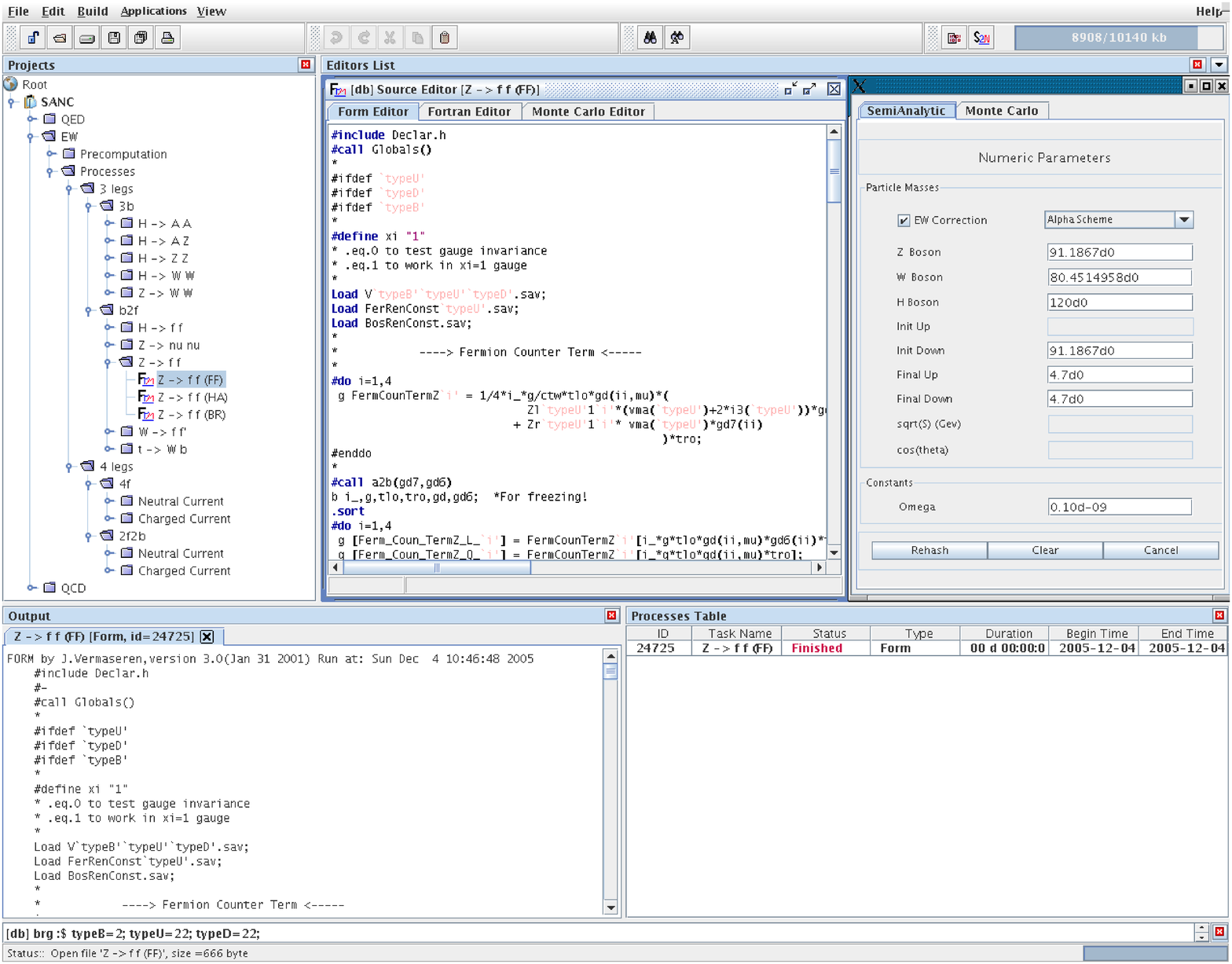,width=200mm,height=160mm,angle=90}}}
\end{picture}
\vspace*{93mm}
\caption[Main SANC window]
        {Main SANC window}
\label{fig:fullshot}
\vspace*{-10cm}
\end{figure}

At the beginning of a client session the main {\tt SANC} window opens,
see Fig.~\ref{fig:fullshot},\footnote{In the figure the windows are shown
after several of the steps described below.}
with several {\bf toolbars} and {\bf windows} or {\bf fields}:\\
{$\bullet$} on top is the {\bf Menu bar} with menus {\it File},
{\it Edit}, {\it Build}, {\it Applications} and {\it View};\\
{$\bullet$} underneath is a row of three {\bf Toolbars}:
{\tt File}, {\tt Edit} and {\tt Build}\\
{$\bullet$} underneath that on the left is the {\bf SANC tree} field, 
and to the right of it the {\bf Editors List} window;
{$\bullet$} underneath is the {\bf Output} window and underneath that
is the {\bf Console};\\
{$\bullet$} below, at the bottom, lies the {\bf Status} bar.\\
Other fields do arise in the course of the work.

The five menus have the options shown in Table \ref{options}.
Menus with $\to$ have further extensions. For example, {\it Toolbars} has four
options; they duplicate the {\it File}, {\it Edit} and {\it Build} toolbars,
which are activated by default, and a latent option {\it Memory}.
When the latter option is activated, two numbers are displayed: the first one
is the current usage of the Java Virtual Machine (JVM) memory, and the second one
is the total size of the JVM memory.
All options can be unchecked in menu {\it View} $\to$ {\it Toolbars} $\to$.

\begin{table}[tbh]
\begin{center}
\caption{The {\tt SANC} Menus and their options.}\label{options}
\vspace*{2mm}
\begin{tabular}{||l|l|l|l|l||}
\hline\hline
{\it File}              &{\it Edit}   &{\it Build}  &{\it Applications} &{\it View}           \\
\hline\hline
{\it Login ...}         &{\it Undo}   &{\it Compile}&{\it Editor Form}  &{\it Toolbars} $\to$ \\
{\it Open Project ...}  &{\it Redo}   &{\it Run S2N}&{\it Numeric Form} &{\it Projects}       \\
{\it Mount Filesystem} $\to$&{\it Cut}&             &{\it Graphics Form}&{\it Editors List}   \\
{\it Unmount Filesystem}&{\it Copy}   &             &                   &{\it Processes Table}\\
{\it Save}              &{\it Paste}  &             &                   &{\it Console}        \\
{\it Save All}          &{\it Find}   &             &                   &{\it Output}         \\
{\it Print} ...  $\to$  &{\it Replace}&             &                   &{\it Status Bar}     \\
{\it Exit}              &{\it Settings}&            &                   &{\it ProgressBar}    \\
                        &             &             &                   &{\it Full Screen}    \\
                        &             &             &                   &{\it Look And Feel} $\to$ \\
                        &             &             &                   &{\it Suggestions}    \\
\hline\hline
\end{tabular}
\end{center}\end{table}


\subsubsection{Login procedure}
{$\bullet$} To log in, click the {\bf Login} icon (the first icon of the File toolbar).
The Login panel opens with a choice of {\tt SANC} servers:
{\it local}, {\it sanc.jinr.ru} and {\it pcphsanc.cern.ch};
choose one of the latter ones
(the {\it local} server is for PCs which have the server itself installed),
then enter the login name {\em guest} and password {\em guest}.\\
{$\bullet$} Click the {\bf Open Project} icon (the second icon of the File toolbar).
This opens the {\bf Open Project} panel.
There are two projects: {\tt Lessons} and {\tt SANC}.
Select project {\tt SANC} and press {\bf OK};
then the {\tt SANC} tree appears in the {\tt SANC} tree field.

\subsubsection{The SANC tree}
The {\tt SANC} tree has three options: {\bf QED, EW} and {\bf QCD}.
Selection of one of these opens the next level of options:
{\bf Precomputation} and {\bf Processes}. 

Here we describe the sequence of steps for option {\bf EW $>$ Processes}.
The use of the {\bf Precomputation} branch was described to an extent 
in Section~\ref{precomputation}.

The available processes are subdivided into {\bf 3legs} and {\bf 4legs}.
The two branches of {\bf 3legs} are {\bf 3b} and {\bf b2f} decays,
and those of {\bf 4legs} are {\bf 4f} and {\bf 2f2b} processes;
here {\bf b} and {\bf f} denote any {\it boson} and {\it fermion}, respectively.
For each of the latter two there is a branch for {\bf Neutral Current} and 
a branch for {\bf Charged Current} processes. 
The next branching is into the available processes of that class.

\subsubsection{Naming conventions}   
In {\tt SANC} we use naming conventions for fields (or particles) shown in
Table~\ref{listfilds} where N is the field index, and in the columns
headed ``name'' we show the names used internally in {\tt SANC}. All associated 
parameter symbols are derived from these names. Thus the mass, charge
and weak isospin of the electron are denoted {\tt mel}, {\tt qel} and
{\tt i3el}, respectively, also the vector and axial vector coupling constants 
({\tt vel, ael}) and their sum ({\tt vpael}) and difference ({\tt vmael}).

\begin{table}[!h]
\caption[List of fields.]
        {List of fields.}
\label{listfilds}
\vspace*{3mm}
\begin{tabular}{||r|c|c|c|c|c|c|c|c|c|c|c|c|c|c||}
\hline
\hline
\multicolumn{3}{||c}{bosons}
&
\multicolumn{9}{|c|}{fermions}
&
\multicolumn{3}{c||}{QCD}\\
\hline    
\multicolumn{3}{||c}{}
&
\multicolumn{3}{|c}{1st generation}
&
\multicolumn{3}{|c}{2nd generation}
&
\multicolumn{3}{|c}{3rd generation}
&
\multicolumn{3}{|c||}{}\\
\hline
   $N$ &  field  &name& $N$ &field&name& $N$ &field &name& $N$&field &name& $N$&field &name \\
\hline
     1 &$A$      & gm & 11 &$\nu_e$&en&15&$\nu_\mu$ & mn & 19 &$\nu_\tau$&tn& 23 &  g  & gn \\
     2 &$Z$      &  z & 12 & $e^-$& el & 16 &$\mu^-$& mo & 20 &$\tau^-$& ta & 24 &$Y_g$&  - \\
$\pm$3 &$W^{\pm}$&  w & 13 &  $u$ & up & 17 &  $c$  & ch & 21 &  $t$   & tp &    &     &    \\
     4 &$H$      &  h & 14 &  $d$ & dn & 18 &  $s$  & st & 22 &  $b$   & bt &    &     &    \\
     5 &$\phi^{0}$&  -&    &      &    &    &       &    &    &        &    &    &     &    \\
$\pm$6 &$\phi^{\pm}$&-&    &      &    &    &       &    &    &        &    &    &     &    \\
     7 &$X^+$    &   -&    &      &    &    &       &    &    &        &    &    &     &    \\
     8 &$X^-$    &   -&    &      &    &    &       &    &    &        &    &    &     &    \\
     9&$Y_{\sss{Z}}$&-&    &      &    &    &       &    &    &        &    &    &     &    \\
    10&$Y_{\sss{A}}$&-&    &      &    &    &       &    &    &        &    &    &     &    \\
\hline
\hline
\end{tabular}
\end{table}
\vspace*{-20mm}

\clearpage

\clearpage

\subsection{Benchmark case 1: $b\to ff$ decays}\label{Uguide:decay}
\subsubsection{Semianalytical calculation}
Consider the $Z\to b\bar{b}$ decay.
First we open the relevant branch of the {\tt SANC} tree:
\vspace*{2mm}

\centerline{\bf EW $\to$ Processes $\to$ b2f $\to$ Z $\to$ ff}

\vspace*{2mm}
There are three FORM programs: ({\bf{FF}}) {\it Form Factor},
({\bf{HA}}) {\it Helicity Amplitudes}, and ({\bf{BR}}) {\it Bremsstrah\-lung}.

Select ({\bf{FF}}) by a click with the right mouse button, this also
pulls down a menu. On the menu
left-click on {\bf Open}. A {\bf Source Editor} window opens
with three tags: {\bf Form Editor}, {\bf Fortran Editor}, and
{\bf Monte Carlo Editor}. The first of these is activated by default and the
FORM source code is displayed in the field.

The particle indices can be seen in the {\bf Console} field;
by default they are:
{\tt typeB = 2} ($Z$ boson), {\tt typeU = 22}
and {\tt typeD = 22} ($b$ quarks).
To change the final state fermions, their particle numbers can be changed
by editing them
in the {\bf Console} field and pressing {\tt Enter}.\footnote{This
need be done only once for a particular choice; to repeat, put
the cursor at the end of the {\bf Console} field, press the
down-arrow key, select the required line of particle numbers using the
up- and down-arrow keys and confirm by pressing {\tt Enter}.}

After choosing the process, open the {\bf Numeric Form} panel from the
{\it Application} menu.
In this panel the particle masses and other relevant information are displayed.

Next the FORM code is compiled by clicking on the {\bf Compile} button --- the
first icon in the {\tt Build} toolbar (or by pressing the {\bf F7} function key). 
After compilation the FORM {\it log file} is shown in the {\bf Output} field.

Clicking on the {\bf Run S2N} button generates the  FORTRAN code;
the FORTRAN code can be seen in the {\bf Output} field.\\
Repeat the sequence of steps for ({\bf{HA}}) and ({\bf{BR}}).

The progress of work can be monitored by activating the {\bf Processes Table}
(see Table~\ref{options}).

The entire {\bf Output} field is arranged in sheets with tags;
for inspection any sheet can be brought to the forground by clicking on its tag.

Once the three FORM codes ({\bf{FF}}), ({\bf{HA}}) and ({\bf{BR}})
have been compiled and tranfered to the FORTRAN codes one can get
the numeric results by the following sequence of operations:
\begin{itemize}
\item[(i)]
open the {\em FORTRAN editor} sheet of the {\tt Editors List}
(belonging to the ({\bf{FF}}) FORM code),
\item[(ii)]
open the  {\bf Numeric Form} panel from the {\tt Applications} menu,
\item[(iii)]
press the {\bf Rehash} button at the bottom of the {\bf Numeric Form} panel,
then the {\bf Compile} button.
\end{itemize}

The answer appears in the {\bf Output} field. 
It consists of a list of {\bf Input parameters} and a set of results:
$\Gamma(\mbox{born})$, the total width [TotalWidth] in Born approximation, 
$\Gamma(\mbox{born+virt+soft})$ and the total width, 
$\Gamma(\mbox{born+virt+soft+hard})$. 
Also shown is the parameter $\omega$, set to $10^{-10}\,\mbox{GeV}$ by default. 
This parameter defines the separation between soft and hard radiation.
It can be modified in the corresponding box of the {\bf Numeric Form} panel. 
Rerunning the program after changing the value of $\omega$
(using the sequence {\bf Rehash $>$ Compile}) gives a result that differs only
in the value of $\Gamma(\mbox{born+virt+soft})$.
The born+virt+soft width is sensitive to parameter $\omega$
and can become unphysical (negative) for very small values of $\omega$.
Increasing $\omega$ and rerunning gives positive values.

\subsubsection{Monte Carlo calculation}
The user can also carry out a Monte Carlo calculation generating
various histograms: {\bf Photon Energy}, {\bf Fermion Energy},
{\bf Photon-Fermion Angular} and {\bf Fermion-antiFermion Angular}.
To do this one must bring the Monte Carlo sheet of the {\bf Numeric Form}
to the foreground, 
check the boxes of the required histograms, then bring the MC sheet of the
{\bf Editors List} to the forground and rerun the program by clicking
the {\bf Compile} button.
After a while the {\bf Histogram Form} is displayed. 
This form has a menu bar; menu {\it Option} allows display of the histogram
statistics.
On the Monte Carlo sheet one can also select the random number 
generator,\footnote{Three random number generators
provided are: Ranlux, Ranmar and Mersenne Twister.} modify the number of MC events
and the range of real photon energies $k_{0\,min}$ and $k_{0\,max}$,
where $k_{0\,min}=\omega$ and $k_{0\,max}$ can be used as an experimental cut.\\
The {\bf Rehash} button must be pressed after each change
in the {\bf Numeric Form}
before clicking on the {\bf Compile} button.
 
The results for the decay rates (in GeV)
of the semianalytical calculation
and of the Monte Carlo calculation for 100\,000 events are summarised
in Table~\ref{benchmk1}. The numerical values are truncated to 6 significant
figures.

\begin{table}
\begin{center} 
\caption{Benchmark Results for $\Gamma(Z\to b\bar{b})$ decay\label{benchmk1}}
\vspace*{2mm}
\begin{tabular}{|cc|c|c|c|}
\hline
 & & $\Gamma_{Born}$ & $\Gamma_{Born+virt+soft}$ & $\Gamma_{Total}$\\
\hline
SA & & 0.356948 & 0.336732 & 0.360224 \\
MC & 100 k &    &          & 0.360229 $\pm$ 0.000721 \\
\hline
\end{tabular} 
\end{center} 
\end{table}

\subsection{Benchmark case 2: the process $2f\to 2f$}\label{Uguide:colln}
Consider the $4f$ {\bf CC} process $f_1 \bar{f}'_1\to f \bar{f}'$.
Implemented are the  processes ${u}\, \bar{d}\to \ell^+\,{\nu}_\ell$,
its charge conjugate and the decay $t\to b\,\ell^+\,\nu_\ell$.
For each process there are three FORM programs: ({\bf{FF}}) {\it Form Factor}, 
({\bf{HA}}) {\it Helicity Amplitudes}, and ({\bf{BR}}) {\it Bremsstrah\-lung}.
Each of these in turn is opened, compiled and run as above in
Section~\ref{Uguide:decay}.

For process  $u\,\bar{d}\to e^+\,{\nu}_e$  we have
in the {\bf Console} window the particle indices shown in Table \ref{ff2ff}.
\vspace*{-2mm}
\begin{table}[!h]
\begin{center}
\caption{Assignment of particle numbers for
 process  ${u}\, \bar{d}\to e^+\,{\nu}_e$}\label{ff2ff}
\vspace*{3mm}
\begin{tabular}{||c|l||} 
\hline\hline
typeIU = 14 & initial Up-type antiparticle ($\bar{d}$ quark)\\
typeID = 13 & initial Down-type particle ($u$ quark)        \\
typeFU = 12 & final Up-type antiparticle (positron)         \\
typeFD = 11 & final Down-type particle (neutrino)           \\ 
\hline\hline
\end{tabular}
\vspace*{2mm}
\end{center}
\end{table}
\vspace{-5mm}
These can be changed to typeIU = 13, typeID = 14, typeFU = 11 and typeFD = 12
for process  $\bar{u}\,{d}\to e^-\,\bar{\nu}_e$
by editing the particle numbers as explained
above\footnote{See Fig.~\ref{ffffboxesT2T4}a)
for definitions of particle types {\sf typeXX}.}.

Next bring the {\it Fortran Editor} sheet of the {\bf Editors List}
and the {\bf Numeric Form} panel to the foreground.
Shown on the {\bf Numeric Parameter}  sheet are the particle masses
in GeV/$c^2$ and the CMS energy in GeV, also the cosine of the CMS
angle between the incident and outgoing particle momenta.

Click on the {\bf Rehash} button at the bottom of the 
{\bf Numeric Form} panel: the main module of FORTRAN code appears in the
{\it Fortran Editor} sheet of the {\bf Editors List}.
Then click on {\bf Compile}. The final answer appears in the {\bf Output} field.
It consists of the parameters used ($\alpha$, $G_F$, particle masses,
the 't Hooft scale $\mu$ and the
Mandelstam variables), and the resulting differential cross sections
$d\,\sigma/d\,\cos\,\theta$ in picobarns in the {\sf Born} approximation
and  {\sf Born+one-loop}. 
The results for the default parameters and for several scattering angles
are summarised in Table~\ref{benchmk2}.
The numerical values are truncated to 6 figures.

\begin{table}[!h]
\begin{center}
\caption{CMS differential cross sections in pb for $u\bar{d}\to e^+\nu_e$}
\label{benchmk2}
\vspace*{3mm}
\begin{tabular}{|c|c|ccc|} 
\hline
             & &\multicolumn{3}{|c|}{$\sqrt{s}$ GeV} \\
$\cos\theta$ & & 40 & 80  & 120 \\ \hline
-0.9  & Born            & 3.58202 &  10818.4 &  12.0561 \\
      & Born + one-loop & 3.53427 &  9990.97 &  28.1226 \\
\hline
-0.5  & Born            & 2.23256 &  6742.78 &  7.51423 \\
      & Born + one-loop & 2.18961 &  6226.00 &  12.8563 \\
\hline
 0.0  & Born            & 0.99225 &  2996.79 &  3.33966 \\
      & Born + one-loop & 0.97192 &  2769.11 &  5.12160 \\
\hline
 0.5  & Born            & 0.24806 &  749.198 &  0.83491 \\
      & Born + one-loop & 0.24453 &  695.224 &  1.47999 \\
\hline
 0.9  & Born            & 0.00992 &  29.9679 &  0.03339 \\
      & Born + one-loop & 0.01072 &  30.2878 &  0.09277 \\
\hline
\end{tabular}\end{center}\end{table}


Here the one-loop corrections are purely weak and QED corrections 
comprise one-loop virtual QED corrections and soft and hard radiations.

The {\sf Born+one-loop} cross section is insensitive to the 't Hooft scale parameter 
$\mu$ which cancels between one-loop electroweak
and the QED part of virtual corrections.

Input parameters can be changed by editing the appropriate field of
the  {\bf Numeric Form} panel and pressing the {\bf Rehash} button.
Again the {\bf Rehash} button must be pressed before pressing {\bf Compile}.
\footnote{To produce whole Table~\ref{benchmk2} one can set flag
{\tt tbprint = 1} in the {\it Fortran Editors} sheet.
After editing the code one has not need press the {\bf Rehash} button, but
just {\bf Compile}.}

In the NC sector there are many more processes.
Here $f_1$ is a {\it massless} fermion of the {\it first
generation}\footnote{The masses of first generation fermions
are retained only in $\log$s to regulate collinear singularities.}
or {\it any} neutrino, and $f$ is {\it any} fermion.
All procedures described above for the CC processes apply also in this case. 

Monte Carlo calculations are not yet implemented for $2\to 2$ processes.




\vspace*{5mm}

\def\href#1#2{#2}
\addcontentsline{toc}{section}{Acknowledgments}

\noindent{\bf Acknowledgments}

The authors are very much indebted to G.~Passarino for critical reading
of the manuscript and useful comments.
We are grateful to S.~Jadach, W.~Placzek, F.~Tkachov, B.~Ward, and Z.~Was 
for numerous discussions. We thank V.~Kolesnikov and E.~Uglov for
managing and supporting the SANC computer system.

\clearpage

\def\href#1#2{#2}
\addcontentsline{toc}{section}{References}


\end{document}